\documentclass[a4paper,fleqn,usenatbib]{mnras}

\usepackage{newtxtext,newtxmath} 

\usepackage[T1]{fontenc} 
\usepackage{ae,aecompl}

\usepackage{graphicx}	
\usepackage{amsmath}	
\usepackage{amssymb}	

\newcommand{\elecd}{$n_{\rm e}$} 
\newcommand{\elect}{$T_{\rm e}$} 
\newcommand{\tf}{$t^{2}$} 
\newcommand{\hb}{H$\beta$}

\newcommand{\foii}{[O\thinspace{\sc ii}]} 
\newcommand{\foiii}{[O\thinspace{\sc iii}]} 
\newcommand{\fsii}{[S\thinspace{\sc ii}]}

\newcommand{\fnii}{[N\thinspace{\sc ii}]}

\newcommand{\fcliii}{[Cl\thinspace{\sc iii}]}

\newcommand{\ffeiii}{[Fe\thinspace{\sc iii}]}

\newcommand{\oiii}{O\thinspace{\sc iii}} 
 
\newcommand{\nii}{N\thinspace{\sc ii}} 
\newcommand{\niii}{N\thinspace{\sc iii}}

\newcommand{\oii}{O\thinspace{\sc ii}} 
 
\newcommand{\cii}{C\thinspace{\sc ii}} 
\newcommand{\ciii}{C\thinspace{\sc iii}} 
\newcommand{\civ}{C\thinspace{\sc iv}}

\newcommand{\fariii}{[Ar\thinspace{\sc iii}]}

\newcommand{\hi}{H\,{\sc i}} 
\newcommand{\hii}{H\thinspace{\sc ii}} 
 
\newcommand{\hei}{He\thinspace{\sc i}} 
\newcommand{\heii}{He\thinspace{\sc ii}}

\newcommand{\ts}{\emph{$t^2$}} 
 
\newcommand{\adfo}{${\rm ADF(O^{2+})}$} 
 
\newcommand\ionic[2]{${\rm #1^{#2}}$}           

\newcommand{\cmc}{{\rm cm$^{-3}$}}

\title[C, N and O abundance gradients in M101 and M31]{Carbon, nitrogen and oxygen abundance gradients in M101 and M31}

\author[C.~Esteban et al. ]{ C.~Esteban,$^{1, 2}$\thanks{E-mail: cel@iac.es (CE)} F.~Bresolin,$^{3}$ J. Garc{\'{\i}}a-Rojas,$^{1, 2}$ and L.~Toribio San Cipriano$^{1, 2}$
\\ 
$^{1}$Instituto de Astrof\'isica de Canarias, E-38200 La Laguna, Tenerife, Spain\\ 
$^{2}$Departamento de Astrof\'isica, Universidad de La Laguna, E-38206, La Laguna, Tenerife, Spain\\ 
$^{3}$Institute for Astronomy, 2680 Woodlawn Drive, Honolulu, HI 96822, USA\\
}
\date{Accepted XXX. Received YYY; in original form ZZZ}

\pubyear{2019}

\begin{document} 
\label{firstpage} 
\pagerange{\pageref{firstpage}--\pageref{lastpage}} 
\maketitle


\begin{abstract} We present deep spectrophotometry of 18 {\hii} regions in the nearby massive spiral galaxies M\,101 and M\,31. We have obtained direct 
determinations of electron temperature in all the nebulae. We detect the {\cii} 4267 \AA\ line in several {\hii} regions, permitting to derive the radial gradient of C/H in both galaxies. We also determine the radial gradients of O/H, N/O, Ne/O, S/O, Cl/O and Ar/O ratios. As in other spiral galaxies, the C/H gradients are steeper than those of O/H producing negative slopes of the C/O gradient. The scatter of the abundances of O with respect to the gradient fittings do not support the presence of significant chemical inhomogeneities across the discs of the galaxies, especially in the case of M101. We find trends in the S/O, Cl/O and Ar/O ratios as a function of O/H in M101 
that can be reduced using {\elect} indicators different from the standard ones for calculating some ionic abundances. The distribution of the N/O ratio with respect to O/H is rather flat in 
M31, similarly to previous findings for the Milky Way. Using the disc effective radius -- $R_{\mathrm e}$ -- as a normalization parameter for comparing gradients, we find that the latest estimates of $R_{\mathrm e}$ for the Milky Way provide an excess of metallicity in apparent contradiction with the mass-metallicity relation; a value about two times larger might solve the problem. Finally, using different abundance ratios diagrams we find that the enrichment timescales of C and N 
result to be fairly similar despite their different  nucleosynthetic origin.
\end{abstract}





\section{Introduction} 
\label{sec:intro} 
The analysis of collisionally excited lines (hereafter CELs) in the spectra of {\hii} regions allows the derivation of the chemical abundances of elements such as N, O, Ne, S, Cl, Ar and Fe in the ionized gas phase of the interstellar medium (ISM). These abundances trace the present-day chemical composition of the host galaxies at the locations where the {\hii} regions lie. Spiral galaxies show radial variations of chemical abundances along their discs -- radial gradients -- that are essential observational constraints for chemical evolution models. The form of such gradients reflects the action of nuclear processes in stellar interiors, and the effects of the star formation history and gas flows in the chemical evolution of galaxies. 

C is the second most abundant heavy element in the Universe after O, and is a major constituent of interstellar dust and organic molecules with an indisputable biogenic importance.  None of the ionic species of C that can be present in 
{\hii} regions emit CELs in the optical range of the spectrum and the determination of its abundance is far more difficult than for O or N, for example. Despite its importance, there are rather limited determinations of the C abundance in extragalactic \hii\ regions. Most of these C abundances are derived from the {\ciii}] 1909 \AA\  and {\cii}] 2326 \AA\  CELs in the UV, which can only be observed from space and can be strongly affected by the uncertainty in the choice of UV reddening function \citep{garnettetal95, garnettetal99}. A more robust determination can be obtained from the \ionic{C}{2+}/\ionic{O}{2+} ratio derived making use of UV CELs, 
whose uncertainty due to reddening is minimal due to the closeness in wavelength between the \ciii] and \oiii] lines used to derive such ionic abundances \citep[e.g.][]{bergetal16, bergetal19}. However, there is another possibility to derive C abundances in ionized nebulae: the use of the faint recombination line (hereafter RL) of {\cii} 
4267~{\AA}, which has also the advantage of lying in a spectral zone free of blending with other emission lines. There are several works reporting the detection of {\cii} 4267~{\AA} and determination of C abundance in extragalactic {\hii} regions \citep{estebanetal02, apeimbert03, lopezsanchezetal07, bresolin07, estebanetal09, estebanetal14, toribiosanciprianoetal16, toribiosanciprianoetal17}. In particular, we obtained the first determination 
of the C/H and C/O radial gradients of the ionized gas in M101 \citep{estebanetal09}, M33 and NGC~300 \citep{toribiosanciprianoetal16}, as well as  the Milky Way \citep{estebanetal05, estebanetal13}.  

M101 (NGC~5457, Pinwheel Galaxy) is a face-on SABc spiral galaxy located at a distance of 6.7 Mpc \citep{freedmanetal01}. It contains a large number of bright {\hii} regions. \citet{kennicuttetal03} determined the radial O abundance gradient of this galaxy making use of 
direct determinations of the electron temperature (hereafter {\elect}) for 20 {\hii} regions. \citet{bresolin07} presented spectroscopical results for some {\hii} regions of the inner disc of M101, determining the C abundance from the {\cii} 4267 \AA\ RL for the single region H1013. Deep spectroscopy taken by \cite{estebanetal09} permitted, for the first time, to estimate the radial C/H gradient in M101. \citet{lietal13} observed 28 {\hii} regions and derived {\elect} for 10 of them, finding no evidence for significant large-scale azimuthal variations of the O/H ratio across the disk of M101. \citet{croxalletal16} carried out the most extensive study of the abundance gradients in M101 to date, as part of the CHAOS (CHemical Abundances Of Spirals) program \citep{bergetal15}. \citet{croxalletal16} obtained direct determinations of {\elect} for 74 {\hii} regions and derived the radial gradients of the O/H, N/O, S/O, Ne/O and Ar/O ratios. 
 
M31 (NGC~224, Andromeda Galaxy) is an inclined SAb spiral galaxy, the most massive component of the Local Group, located at a distance of 744 kpc \citep{vilardelletal10}. Despite its closeness and obvious interest, there are only few works devoted to the determination of the 
chemical composition of {\hii} regions in this galaxy. This is perhaps due to the faintness of the {\hii} regions of M31. \citet{blairetal82} determined the radial O abundance gradient but based on empirical strong-line 
methods for estimating the nebular metallicity (no direct determinations of {\elect}). Later on, \citet{galarzaetal99} and \citet{bresolinetal99} were still unable to detect auroral lines for deriving {\elect} and also used 
strong line methods for estimating O abundances. It was not until 2009 when the first determination of {\elect} was finally made for the {\hii} region K932 of M31 from spectra taken with the 10m Keck~I telescope 
\citep{estebanetal09}. Using the same telescope,  \citet{zuritabresolin12} were the first to determine radial abundance gradients based on direct determinations of {\elect} for a sizable number of {\hii} regions of 
M31, in 9 out of their 31 objects. 

In this paper, we present very deep optical spectra of 11 and 7 {\hii} regions of the M101 and M31 galaxies, respectively. The main aim is to determine the radial abundance of C, N, O and other elements in these galaxies 
based on direct determinations of {\elect}. The paper is organized as follows. In Section~\ref{sec:obs} we describe the sample selection, the observations and data reduction procedure. In Section~\ref{sec:lines} we 
describe the emission-line measurements and the reddening correction. In Section~\ref{sec:results} we present the physical conditions and ionic and total abundances determined for the sample objects. In 
Sections~\ref{sec:M101} and \ref{sec:M31} we present and discuss our determinations of the radial abundance gradients in M101 and M31, respectively. In Section~\ref{sec:discussion} we  discuss our results focusing our attention on the C and N abundances and the enrichment of the ISM in these elements. Finally, in Section~\ref{sec:conclusions} we summarize our main conclusions.

\begin{table*} 
\centering \caption{Data of the sample objects and their observations.} 
\label{tab:journal} 
\begin{tabular}{lccccccccc} 
\hline
& R.A.$^{\rm a}$ & Decl.$^{\rm a}$ &  & PA & Extracted area & &  & & Exposure time \\ 
{\hii} region & (J2000) & (J2000) &  $R/R_\mathrm{25}$ &  ($^\circ$) & ($\mathrm{arcsec^2}$) & $S$(H$\alpha$)$^{\rm b}$ & Grating & Airmass & (s) \\ 
\hline 
\\
\multicolumn{10}{c}{M101}\\
\\
NGC~5462 & 14:03:53.11 & 54:22:06.4 & 0.43 & $-$25 & 4.1 $\times$ 0.8 & 2.35 & 1000B & 1.28 & 3 $\times$ 882 \\
& & & & & & & 2500V & 1.13 & 3 $\times$ 882 \\
& & & & & & & 2500U & 1.19 & 6 $\times$ 920 \\
NGC~5455 & 14:03:01.17 & 54:14:29.4 & 0.47 & 160 & 5.1 $\times$ 0.8 & 5.39 & 1000B & 1.14 & 3 $\times$ 881 \\
& & & & & & & 2500V & 1.11 & 3 $\times$ 887 \\
& & & & & & & 2500U & 1.18 & 6 $\times$ 920 \\
H219 & 14:02:46.84 & 54:14:49.9 & 0.50 & 160 & 5.1 $\times$ 0.8 & 0.54 & 1000B & 1.14 & 3 $\times$ 881 \\
& & & & & & & 2500V & 1.11 & 3 $\times$ 887 \\
& & & & & & & 2500U & 1.18 & 6 $\times$ 920 \\
NGC~5447 & 14:02:28.13 & 54:16:26.9 & 0.55 & 42 & 6.1 $\times$ 0.8 & 1.23 & 1000B & 1.15 & 3 $\times$ 882 \\
& & & & & & & 2500U & 1.38 & 6 $\times$ 920 \\
H37 & 14:02:17.66 & 54:22:34.1 & 0.59 & 14 & 5.1 $\times$ 0.8 & 0.37 & 1000B & 1.16 & 3 $\times$ 882 \\
& & & & & & & 2500U & 1.12 & 6 $\times$ 920 \\
H1146 & 14:03:49.08 & 54:28:10.4 & 0.62 & 63 & 6.1 $\times$ 0.8 & 2.83 & 1000B & 1.23 & 3 $\times$ 820 \\
& & & & & & & 2500U & 1.15 & 6 $\times$ 920 \\
H1216 & 14:04:11.34 & 54:25:18.7 & 0.67 & 30 & 3.6 $\times$ 0.8 & 2.06 & 1000B & 1.27 & 3 $\times$ 890 \\
& & & & & & & 2500V & 1.18 & 3 $\times$ 890 \\
& & & & & & & 2500U & 1.18 & 6 $\times$ 920 \\
H1118 & 14:03:43.16 & 54:29:51.2 & 0.69 & 63 & 5.8 $\times$ 0.8 & 1.86 & 1000B & 1.23 & 3 $\times$ 820 \\
& & & & & & & 2500U & 1.15 & 6 $\times$ 920 \\
NGC~5471 & 14:04:28.98 & 54:23:49.0 & 0.81 & 30 & 4.1 $\times$ 0.8 & 3.20 & 1000B & 1.27 & 3 $\times$ 890 \\
& & & & & & & 2500V & 1.18 & 3 $\times$ 890 \\
& & & & & & & 2500U & 1.18 & 6 $\times$ 920 \\
H681 & 14:03:13.31 & 54:35:42.5 & 1.05 & 90 & 7.4 $\times$ 0.8 & 0.15 & 1000B & 1.31 & 3 $\times$ 893 \\
& & & & & & & 2500V & 1.13 & 3 $\times$ 890 \\
& & & & & & & 2500U & 1.11 & 6 $\times$ 920 \\
SDH323 & 14:03:49.64 & 54:38:07.4 & 1.25 & 30 & 11.4 $\times$ 0.8 & 0.10 & 1000B & 1.25 & 3 $\times$ 882 \\
& & & & & & & 2500V & 1.11 & 3 $\times$ 890 \\
& & & & & & & 2500U & 1.14 & 6 $\times$ 920 \\
\\
\multicolumn{10}{c}{M31}\\
\\
BA289 & 00:41:29.20 & 40:51:04.0 & 0.35 & $-$40 & 22.9 $\times$ 0.8 & 0.21 & 1000B & 1.52 & 3 $\times$ 450 \\
& & & & & & & 2500V & 1.48 & 3 $\times$ 1380 \\
& & & & & & & 2500U & 1.68 & 3 $\times$ 900 \\
BA310 & 00:40:30.81 & 40:42:30.2 & 0.44 & 43 & 15.2 $\times$ 0.8 & 0.76 & 1000B & 1.14 & 3 $\times$ 450 \\
& & & & & & & 2500V & 1.37 & 3 $\times$ 1380 \\
& & & & & & & 2500U & 1.21 & 3 $\times$ 900 \\
K703 & 00:45:11.70 & 41:37:02.0 & 0.47 & 43 & 5.1 $\times$ 0.8 & 0.20 & 1000B & 1.22 & 3 $\times$ 450 \\
& & & & & & & 2500V & 1.29 & 3 $\times$ 1380 \\
& & & & & & & 2500U & 1.06 & 3 $\times$ 900 \\
K160 & 00:43:35.00 & 41:09:39.0 & 0.52 & $-$23 & 3.6 $\times$ 0.8 & 2.85 & 1000B & 1.04& 3 $\times$ 450 \\
& & & & & & & 2500V & 1.07 & 3 $\times$ 1380 \\
& & & & & & & 2500U & 1.07& 3 $\times$ 900 \\
BA379 & 00:39:18.77 & 40:21:58.7 & 0.71 & $-$18 & 15.2 $\times$ 0.8 & 0.59 & 1000B & 1.25 & 3 $\times$ 450 \\
& & & & & & & 2500V & 1.06 & 3 $\times$ 1380 \\
& & & & & & & 2500U & 1.02 & 3 $\times$ 900 \\
BA374 & 00:39:33.49 & 40:20:26.1 & 0.75 & $-$71 & 5.1 $\times$ 0.8 & 0.13 & 1000B & 1.20 & 3 $\times$ 450 \\
& & & & & & & 2500V & 1.49 & 3 $\times$ 1380 \\
& & & & & & & 2500U & 1.29 & 3 $\times$ 900 \\
BA371 & 00:39:37.42 & 40:20:11.1 & 0.76 & $-$71 & 2.5 $\times$ 0.8 & 0.52 & 1000B & 1.20 & 3 $\times$ 450 \\
& & & & & & & 2500V & 1.49 & 3 $\times$ 1380 \\
& & & & & & & 2500U & 1.29 & 3 $\times$ 900 \\\hline 
\end{tabular} 
\begin{description} 
\item[$^{\rm a}$] Coordinates of the centre of the extracted aperture. 
\item[$^{\rm b}$] Measured mean H$\alpha$ surface brightness in the areas extracted for spectroscopical analysis in units of 10$^{-14}$ erg cm$^{-2}$ s$^{-1}$ arcsec$^{-2}$.
\end{description} 
\end{table*} 

\section{Sample selection, Observations and Data Reduction} 
\label{sec:obs} 

\begin{figure*} 
\centering 
\includegraphics[scale=0.25]{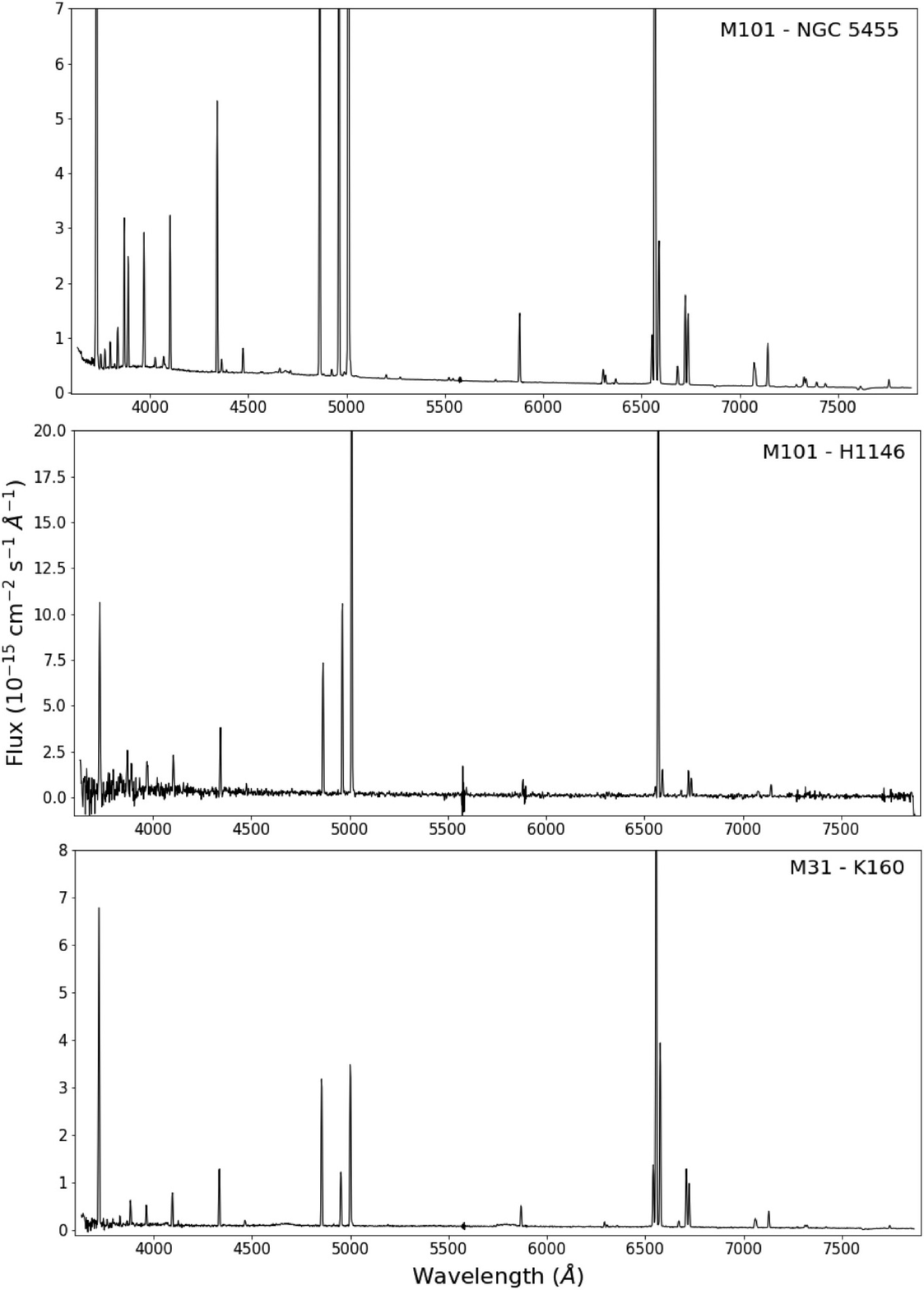} 
\caption{Flux-calibrated GTC OSIRIS spectra of NGC~5455 (upper panel), H1146  (middle panel) and K160 (bottom panel).  NGC~5455 and H1146 correspond to the spectra with the highest and lowest signal-to-noise ratio of the sample, respectively.} 
\label{fig:spectra}
 \end{figure*} 
 
 The observations were performed with the 10.4 m Gran Telescopio Canarias (GTC) at the Observatorio del Roque de los Muchachos (La Palma, Spain). The {\hii} regions of M101 were observed in service mode using one-hour observing blocks distributed in several nights in March 2014, May 2015 and between March and May 2016.  The objects of M31 were also observed in service mode in December 2017. The spectra were taken with the OSIRIS (Optical System for Imaging and low-Intermediate-Resolution Integrated Spectroscopy) spectrograph \citep{cepaetal00, cepaetal03}, which consists of a mosaic of two Marconi CCD42-82 CCDs each with 2048 $\times$ 4096 pixels and a 74 pixel gap between them. Each pixel has a physical size of 15$\mu$m. We used a binning 2 $\times$ 2 for our observations, giving a plate scale of 0.254 arcsec. OSIRIS was used in long-slit mode, centering the objects in CCD2. In some cases, the slit covered two nearby {\hii} regions and eventually one of them fell onto CCD1. The slit length was 7.4 arcmin and its width was set to 0.8 arcsec. We used three grisms -- R1000B, R2500V and R2500U -- for all objects except for NGC~5447, H37, H1118 and H1146. For these four objects we only used R1000B and R2500U. R1000B covers the whole optical spectral range, from 3600 to 7750 \AA\, achieving an effective spectral resolution of 6.52 \AA. This grism was used to determine the reddening coefficient and to measure the lines redder than {\hei} 5876 \AA. R2500V and R2500U give effective spectral resolutions of about 2.46 \AA. R2500V covers from 4430 to 6070 \AA\ and was used to deblend the RLs of multiplet 1 of \oii\ around 4650 \AA\ from the  {\ffeiii} 4658 \AA\ line. The third grism, R2500U, covers from 3430 to 4620 \AA\ and was used to improve the detection and measurement of {\cii} 4267 \AA\ and deblend {\foiii} 4363 \AA\ from nearby sky and nebular emission features.  

The total integration time of the spectra taken with each grism varied from 1350 to 5520 s depending on the object and observing run. We took three or six consecutive exposures in each case.  Additional 60 s exposures were taken with the R1000B grism -- and R2500V for the brightest objects -- in order to avoid problems with the 
possible  saturation of the brightest lines. The position of the centre and position angle (PA) of the slit were selected in order to cover the brightest zones of the nebulae. In some cases, where two nearby {\hii} regions could be observed with the same slit position, the PA was fixed to facilitate the simultaneous observation of both objects. The extension of the extracted area in each bidimensional spectrum optimizes the signal-to-noise ratio of the intensity of several faint lines, especially for {\cii} and {\oii} RLs -- when these lines were detected -- as well as the  {\foiii} 4363 \AA\ and {\fnii} 5755 \AA\ auroral lines, which are necessary for the determination of the {electron temperature {\elect} (see Section~\ref{sec:conditions}). 

Table~\ref{tab:journal} shows the relevant parameters of the observations of {\hii} regions of M101 and M31. The different columns give the coordinates of the centre of the extracted aperture, the fractional galactocentric distance of the object in units of the isophotal radius ($R/R_\mathrm{25}$), the position angle (PA), the aperture size of the extraction,  the mean H$\alpha$ surface brightness -- uncorrected for reddening -- in the extracted area, the gratings used, the mean airmass and the total exposure time. The airmasses of the observations go from 1.11 to 1.38 (mean of 1.18) for the objects of M101 and from 1.02 to 1.68 (mean of 1.26) for M31, so the zenith distance was never larger than 53$^\circ$ and usually about 35$^\circ$. The atmospheric differential refraction is not expected to be a problem for these observations because the objects are several arcsecs extended. 

The fractional galactocentric distance of the objects {$R/R_\mathrm{25}$} (the Galactocentric distance normalized to the isophotal radius of the disk, $R_\mathrm{25}$). has been taken from the literature or recalculated using recent determinations of the parameters of the galaxies. In the case of M31, the {$R/R_\mathrm{25}$} ratios for BA310, BA371, BA374 and BA379 have been taken from \cite{zuritabresolin12}. For the rest of the objects of M31, we have used the galactocentric distance obtained by \cite{blairetal82} and \cite{galarzaetal99}, recomputed assuming the parameters used by \citet{zuritabresolin12}, i.e. an inclination angle of 77$^\circ$ \citep{corbellietal10} and a distance of 744 kpc \citep{vilardelletal10}. The OSIRIS spectra were reduced using {\sc iraf}\footnote{{\sc iraf}, the Image Reduction and Analysis 
Facility, is distributed by the National Optical Astronomy Observatory, 
which is operated by the Association of Universities for Research 
in Astronomy under cooperative agreement with the National Science 
Foundation.} v2.16. Data reduction followed the standard procedure for long-slit 
spectra. The spectrograms were wavelength calibrated with Hg-Ar, Ne and Xe lamps. The absolute flux calibration was achieved by observations of the standard stars Ross 640, G191$-$B2B, Feige 66, Hiltner 600, GD140 and GD153. 
We used polynomials of relatively high order -- 5 to 7 -- to properly fit the sensitivity curves of the R1000B spectra and avoid problems with the fluxes of the {\hi} Balmer lines. Particular care was taken in the background subtraction because the sky background emission is not completely homogeneous along the GTC OSIRIS long slit \citep{fangetal15}. 

\section{Line intensity measurements} 
\label{sec:lines} 

Line fluxes of the spectra of the {\hii} regions of M101 and M31 included in Table~\ref{tab:journal} were measured with the {\sc splot} routine of {\sc iraf} by integrating all the flux in the line between two given limits and over the average local continuum. All line fluxes of a given spectrum have been normalized to H$\beta$ = 100.0. In the case of line blending, we applied a double or multiple Gaussian profile fit procedure using the {\sc splot} routine of {\sc iraf} to measure the individual line intensities. The identification of the lines was made following our previous results on deep spectroscopy of extragalactic \hii\ regions \citep[see][]{estebanetal09, toribiosanciprianoetal16}. The number of lines detected and identified in the objects depends mainly on their surface brightness and varies from 26 in H1146 to 96 in NGC~5455. In Figure~\ref{fig:spectra} we show the spectra of those two {\hii} regions and K160 of M31. 

We detect broad emission features that characterize the spectra of massive Wolf-Rayet (WR) stars in several  {\hii} regions. In NGC~5455, H219 and H1118 of M101 
\citep[the features were previously detected in NGC~5455 by][]{croxalletal16} and in 
K160 of M31. In all the objects we detect both the so-called blue and red WR bumps, indicating the presence of WR stars of the nitrogen and carbon sequences, WN and WC. The blue WR bump consists of the blend of the broad {\heii} 4686 \AA, {\ciii}/{\civ} 4650 \AA\ and {\niii} 4640 \AA\ lines and originate mainly from the emission of WN 
stars. The red bump is produced by the blend of {\ciii} 5698 \AA\ and {\civ} 5808 \AA\ broad emission lines. This last one is the strongest emission line of WC stars, but is barely seen in WN stars. 

The observed line intensities have been corrected for interstellar reddening. This was done using the reddening constant, 
$c$({\hb}), obtained from the intensities of the Balmer lines. However, the fluxes of the {\hi} lines may be also 
affected by underlying stellar absorption. Consequently, we have performed an iterative procedure to derive both 
$c$(H$\beta$) and the equivalent widths of the absorption in the {\hi} lines following 
the procedure outlined by \citet{lopezsanchezetal06}. We have used the reddening function, $f(\lambda)$, normalized to H$\beta$ derived by \cite{cardellietal89} -- assuming 
 $R_V$ = 3.1 -- and the observed H$\alpha$/H$\beta$, H$\gamma$/H$\beta$ and H$\delta$/H$\beta$ line ratios determined  from the spectra obtained with the R1000B grism. We have considered the theoretical line ratios expected for case B recombination given by 
 \citet{storeyhummer95} for the physical conditions derived for the nebulae -- see Section~\ref{sec:conditions}. In tables A1 to A6 -- provided as Supplemental Appendix at MNRAS online -- we include the list of line identifications -- first 3 columns,  the reddening function, $f(\lambda)$ -- fourth column -- and dereddened flux line ratios with respect to H$\beta$ -- remaining columns. The quoted line intensity errors include the estimated flux calibration error ($\sim$ 2 percent), uncertainties in line flux measurement and error propagation in the reddening coefficient. Colons indicate line intensity errors of the order or greater than 40\%. The last two rows of each table include the reddening coefficient and the observed -- uncorrected for reddening -- integrated H$\beta$ flux, $F$(H$\beta$), of the extracted aperture for each object.

\section{Physical Conditions and chemical abundances} 
\label{sec:results} 

For the 18  {\hii} regions of M101 and M31 observed we have determined the physical conditions -- electron temperature, {\elect}, and density, {\elecd} -- and the ionic abundances making use of the version 1.0.26 of {\sc pyneb} \citep{Luridianaetal15} in combination with the atomic data listed in Table~\ref{tab:atomic} and the line-intensity ratios and their uncertainties  given in tables A1 to A6 of the Supplementary Appendix.

\begin{table*} 
\centering 
\caption{Atomic dataset used for collisionally excited lines.} 
\label{tab:atomic} 
\begin{tabular}{lcc} 
\hline 
& Transition probabilities &  \\ Ion & and energy levels & Collisional strengths \\ 
\hline 
N$^+$ & \citet{froesefischertachiev04} & \citet{tayal11} \\ 
O$^+$ & \citet{froesefischertachiev04} & \citet{kisieliusetal09} \\ 
O$^{2+}$ &  \citet{froesefischertachiev04, storeyzeippen00} & \citet{storeyetal14} \\ 
Ne$^{2+}$ & \citet{galavisetal97} & \citet{mclaughlinbell00} \\ 
S$^+$ & \citet{podobedovaetal09} & \citet{tayalzatsarinny10} \\ 
S$^{2+}$ &  \citet{podobedovaetal09} & \citet{tayalgupta99} \\ 
Cl$^{2+}$ & \citet{mendoza83} & \citet{butlerzeippen89} \\ 
Ar$^{2+}$ & \citet{mendoza83, kaufmansugar86} & \citet{galavisetal95} \\ 
Ar$^{3+}$ & \citet{mendozazeippen82a, kaufmansugar86} & \citet{zeippenetal87} \\
Fe$^{2+}$ &  \citet{quinet96, johanssonetal00} & \citet{zhang96} \\ 
\hline 
\end{tabular} 
\end{table*} 

\subsection{Physical Conditions } 
\label{sec:conditions} 

We have derived  {\elecd} using the density-sensitive emission line ratios {\fsii}~6717/6731 and {\foii}~3726/3729 for all the objects, but also {\fcliii}~5518/5538 for some of them. 
The {\fsii} and {\foii} indicators give consistently small densities in all cases and lower than 100 {\cmc} in most of them. Three objects show  {\elecd}({\fcliii}) substantially larger than the other two indicators, but the values have very large uncertainties and have not been considered. {\elect}({\foii}) and {\elect}({\fnii}) have been corrected from the contribution to the intensity of the {\foii} 7319, 7330~{\AA} and {\fnii} 5755~{\AA} lines due to recombination following the formulae derived by \citet{liuetal00} and making a preliminary estimation of the N$^{2+}$ abundance in the nebulae. We have determined {\elect} using the following emission line ratios: {\fnii}~5755/(6548+6584), {\foii}~(7319+7330)/(3726+3729), {\fsii}~(4069+4076)/(6717+6731), {\foiii}~4363/(4959+5007) and {\fariii}~5192/(7136+7751). The physical conditions of all the objects are presented in Table~\ref{tab:physcond}. We assume a two-zone approximation for the nebula, estimating the representative values of the electron temperature for the zones where low- and high-ionization potential ions 
are present, {\elect}(low) and {\elect}(high). Those values will be used for determining ionic abundances except for some particular ions, as will be explained in sections~\ref{sec:abund} and \ref{sec:M101_other}. When two or more indicators of {\elect} for low-ionization potential ions are reported, {\elect}(low) is calculated as the mean of {\elect}({\fnii}), {\elect}({\foii}) and  {\elect}({\fsii}) weighted by their inverse relative errors. The suitability of {\elect}({\fsii}) to calculate the mean may be questionable. \ionic{S}{0} has an ionization potential lower than \ionic{H}{0}, \ionic{O}{0} or \ionic{N}{0} and the zone were \ionic{S}{+} lies may include parts of the nebula outside the ionization front, where the physical conditions of the gas may be different. In any case,  the values of {\elect}({\fsii}) are consistent with those obtained with the other {\elect} indicators for low ionization species within the errors. NGC~5471 is an exception, and shows an abnormally high value of  {\elect}({\fsii}), which has not been considered for determining {\elect}(low). 
The {\elect}(high) was determined as the weighted mean of {\elect}({\foiii}) and {\elect}({\fariii}) when both quantities are obtained in a given nebula. All {\elect} determinations are included in Table~\ref{tab:physcond}. Several objects  lack one of the temperature determinations, {\elect}(low) or  {\elect}(high), in these cases the temperature that cannot not be calculated is estimated using equation 3 of \citet{estebanetal09}, which is based on results for similar {\hii} regions observed in the Milky Way and other local galaxies. 

\begin{table*} 
\centering 
\caption{Physical conditions for the {\hii} regions of M101 and M31.} 
\label{tab:physcond} 
\begin{tabular}{l c c c c c c c} 
\hline 
     Parameter & Lines & NGC~5462 & NGC~5455 &  H1216 & NGC~5471 & H37 & H219 \\  
\hline 
{\elecd} ({\cmc}) & {\fsii} & 110 $\pm$ 50 & 160 $\pm$ 50 & $<$100 &  170 $\pm$ 60 & $<$100 & $<$100 \\
 & {\foii} & $<$100  & 240 $\pm$ 60 & $<$100  & 210 $\pm$ 60 & $<$100 & $<$100  \\
 & {\fcliii} &  $-$ & $<$100  &  $<$100  & $<$100  & $<$100  &  850$^{+1430}_{-850}$ \\
{\elect} (K) &  {\fnii} & 9980 $\pm$ 1490 & 10170  $\pm$  190 & 10840 $\pm$ 320 & 12310 $\pm$ 520 & 11440 $\pm$ 11930 & 10170 $\pm$ 600 \\
 & {\foii} & 11100 $\pm$ 4670 & 11020 $\pm$ 1710 & 12280 $\pm$ 1110 & 12830 $\pm$ 1810 & 9710 $\pm$ 1570 & 11700 $\pm$ 5360 \\
  & {\fsii} & 11130 $\pm$ 1080 & 10910 $\pm$ 690 & 13580 $\pm$ 4190 & 22910 $\pm$ 4700 & 15800 $\pm$ 5180 & 8830 $\pm$ 1900\\
 & {\bf {\elect}(low)} & {\bf 10750 $\pm$ 860} & {\bf 10230 $\pm$ 180} & {\bf 10960 $\pm$ 310} & {\bf 12470 $\pm$ 500} & {\bf 10680 $\pm$ 1180} & {\bf 10060 $\pm$ 570} \\
 & {\foiii} & 9990 $\pm$ 110 & 9640 $\pm$ 130 & 10670 $\pm$ 170 & 13970 $\pm$ 180 & 11050 $\pm$ 210 & 9640 $\pm$ 580\\ 
 & {\fariii} &  $-$ &11570 $\pm$ 920 & $-$ & 14970 $\pm$ 1630 & $-$ & $-$ \\ 
 & {\bf {\elect}(high)} & {\bf 9470 $\pm$ 120} & {\bf 9670 $\pm$ 130} & {\bf 10670 $\pm$ 170} & {\bf 13990 $\pm$ 180} & {\bf 11050 $\pm$ 210} & {\bf 9640 $\pm$ 580} \\
 \\
     & & NGC~5447 & H681 & H1118 & H1146 & SDS323 & BA289 \\
  \\
 {\elecd} ({\cmc}) & {\fsii} & $<$100 & $<$100 & $<$100 &  $<$100 & 120$^{+230}_{-120}$ & $<$100 \\
	& {\foii} & 140 $\pm$ 50 & $<$100  & $<$100 & $<$100 & $<$100  & $<$100  \\
{\elect} (K) &  {\fnii} & $-$ & $-$ & $-$ & $-$ & $-$ & 6930 $\pm$ 460 \\
 & {\fsii} & 11270 $\pm$ 1220 & 13270 : & $-$ & $-$& $-$ & $-$ \\
 & {\bf {\elect}(low)} & {\bf 11270 $\pm$ 1220} & {\bf 12830 $\pm$ 260$^{\rm a}$} & {\bf 12020 $\pm$ 400$^{\rm a}$} & {\bf 11450 $\pm$ 830$^{\rm a}$} & {\bf 13560 $\pm$ 640$^{\rm a}$} & {\bf 6880 $\pm$ 470} \\
 & {\foiii} & 9360 $\pm$ 110 & 13780 $\pm$ 360 & 12640 $\pm$ 560 & 11830 $\pm$ 1170 & 14800 $\pm$ 900 & $-$ \\ 
  & {\bf {\elect}(high)} & {\bf 9360 $\pm$ 110} & {\bf 13780 $\pm$ 360} & {\bf 12640 $\pm$ 560} & {\bf 11830 $\pm$ 1170} & {\bf 14800 $\pm$ 900} & {\bf 5390 $\pm$ 670$^{\rm b}$} \\
\\
     & &K703 & K160 & BA379 & BA371 & BA310 & BA374 \\
\\
{\elecd} ({\cmc}) & {\fsii} &  $<$100  & $<$100& $<$100  & $<$100 & $<$100 & $<$100  \\
 & {\foii} & $<$100  & $<$100 & $<$100  & 210 $\pm$ 60 & $<$100 & $<$100  \\
 & {\fcliii} &  $-$ & 1330 $\pm$ 1300  & $<$100  &  1880 $\pm$ 1540  & $<$100  &  $-$  \\
{\elect} (K) &  {\fnii} & 8720 $\pm$ 480 & 8010 $\pm$ 290 & 9400 $\pm$ 200 & 9740 $\pm$ 510 & 8070 $\pm$ 770 & 11140 $\pm$ 1240 \\
 & {\foii} & 8550 $\pm$ 6650 & 8290 $\pm$ 680 & 9860 $\pm$ 700 & 9980 $\pm$ 810 & $-$ & 9900 $\pm$ 3930 \\
 & {\fsii} & 8490 $\pm$ 1930 & 9500 $\pm$ 2300 & 9330 $\pm$ 2200 & $-$ & $-$ & $-$ \\
 & {\bf {\elect}(low)} & {\bf 8590 $\pm$ 380} & {\bf 8070 $\pm$ 270} & {\bf 9440 $\pm$ 190} & {\bf 9850 $\pm$ 440} & {\bf 8070 $\pm$ 770} & {\bf 10890 $\pm$ 1180} \\
 & {\foiii} & 7410 $\pm$ 530 & 8050 $\pm$ 610 & 8720 $\pm$ 300 & 9090 $\pm$ 270 & $-$ & 12320 $\pm$ 940\\ 
 & {\bf {\elect}(high)} & {\bf 7410 $\pm$ 530} & {\bf 8050 $\pm$ 610} & {\bf 8720 $\pm$ 300} & {\bf  9090 $\pm$ 270 } & {\bf 7070 $\pm$ 1020$^{\rm b}$} & {\bf 12320 $\pm$ 940} \\
     \hline
\end{tabular} 
\begin{description} 
\item[$^{\rm a}$] Estimated from {\elect}({\foiii}) and equation 3 of \citet{estebanetal09}. 
\item[$^{\rm b}$] Estimated from {\elect}({\fnii}) and equation 3 of \citet{estebanetal09}. 
\end{description} 
\end{table*} 

\begin{figure*} 
\centering 
\includegraphics[scale=0.20]{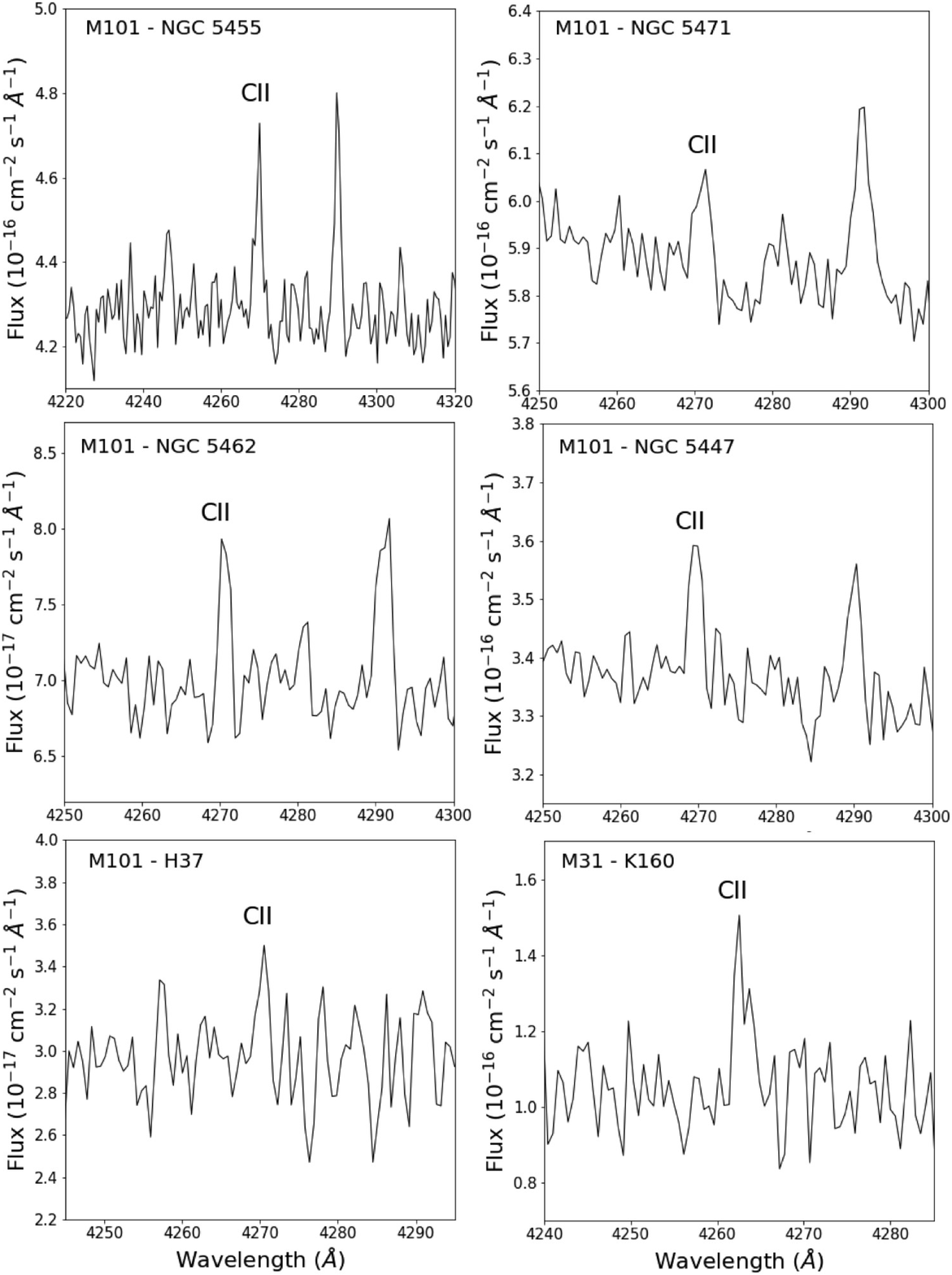} 
\caption{Section of the spectrum of the {\hii} regions NGC~5455, NGC~5471, NGC~5462, NGC~5447 and H37 of M101 and K160 of M31 showing the {\cii} 4267 \AA\ recombination line. } 
\label{fig:CII_lines} 
\end{figure*} 

\subsection{Abundances} 
\label{sec:abund} 
Using CELs, we have derived abundances of N$^+$, O$^+$, O$^{2+}$, Ne$^{2+}$, S$^+$, S$^{2+}$, Cl$^{2+}$, Ar$^{2+}$, Ar$^{3+}$ and Fe$^{2+}$. We assumed a two-zone scheme for 
calculating the ionic abundances. We used  \elect(low) for low ionization potential ions such as N$^+$, O$^+$, S$^+$ and Fe$^{2+}$ and \elect(high) for high ionization ones such as O$^{2+}$, Ne$^{2+}$ and 
Ar$^{3+}$. We used other temperature indicators for some particular ions due to the reasons outlined in Section~\ref{sec:M101_other} and following the prescriptions by \citet{dominguezguzmanetal19}. We used
 \elect({\fnii}) -- or {\elect}(low) when that indicator is not available -- for determining Cl$^{2+}$ abundances (see Section~\ref{sec:M101_other}), and the average of  \elect({\foiii}) and  \elect({\fnii}) 
 -- or {\elect}(high)  or  {\elect}({low) when one of those are not available -- for S$^{2+}$ and Ar$^{2+}$ (see Section~\ref{sec:M101_other}). 
 
The computations were made with {\sc pyneb} and using the atomic data listed in Table~\ref{tab:atomic}. We have measured 
several RLs of {\hei} in all the objects and of {\heii}, {\cii} and/or {\oii} in some of them. The He$^{+}$ abundance has been determined using {\sc pyneb},  \elect(high) and the effective 
recombination coefficient computations by \citet{porteretal12, porteretal13} for {\hei} lines, that include collisional and optical depth effects in the line intensities. The final adopted He$^{+}$ 
abundance is the weighted average of the ratios obtained from the individual brightest {\hei} lines measured in each object. We detect the {\heii} 4686 \AA\ line in NGC~5471 and H681, two {\hii} regions located in the 
external disc of M101. We have determined the He$^{2+}$ abundance with {\sc pyneb} using \elect(high) and the effective recombination coefficient calculated by 
\citet{storeyhummer95}. In both objects, the He$^{2+}$/H$^+$ ratios are very small and do not contribute significantly to the total He abundance. We have measured the {\cii} 4267~{\AA} line in the 
{\hii} regions NGC~5455, NGC~5471, NGC~5462, NGC~5447 and H37 of M101 and K160 of M31. The quality of the detections of the {\cii} 4267~{\AA} line can be noted in 
Figure~\ref{fig:CII_lines}. We compute the C$^{2+}$/H$^+$ ratios from the measured flux of the \cii\ 4267 \AA\ RL, {\elect}(high), and the \cii\ effective recombination coefficients calculated by 
\citet{daveyetal00} for case B. The  RLs of multiplet 1 of {\oii} have been only detected in NGC~5455 and NGC~5471. We have determined the O$^{2+}$ abundance from these lines making use of 
the effective recombination coefficients  for case B and assuming LS coupling calculated by \citet{storey94} and considering {\elect}(high). The relative populations of the {\oii} levels do not follow 
local thermodynamical  equilibrium (LTE)  for densities {\elecd} $<$ 10$^4$ {\cmc} \citep{ruizetal03}. We use the prescriptions of \citet{apeimbertpeimbert05} to calculate the appropriate corrections 
for the relative strengths between the individual \oii\ lines. 
Tables~\ref{tab:abund1} and \ref{tab:abund2} give the ionic abundances -- as well as their uncertainties -- for all the observed objects.

For the two objects where {\oii} RLs have been measured, NGC~5455 and NGC~5471, we were able to calculate the \ionic{O}{2+} abundances from both RLs and CELs. We have not considered the {\oii} 4638.86 and 4641.81 \AA\ lines for the calculations because they can be blended with {\niii} 4640.64 \AA\ and {\nii} 4643.06 \AA\ lines at the spectral resolution of our observations. 
It is a well-established fact that \ionic{O}{2+} abundances determined from RLs are always larger than those obtained from CELs, with differences typically ranging between 0.20 and 0.45 dex in {\hii} regions \citep[see][]{garciarojasesteban07,toribiosanciprianoetal17}. This phenomenon is the so-called abundance discrepancy, ADF, that in the case of \ionic{O}{2+}, {\adfo}, is defined as the difference of the O$^{2+}$/H$^+$ ratio determined from RLs with respect to CELs in logarithmic units \citep{liuetal00}. We obtain an 
ADF(\ionic{O}{2+}) = 0.13 $\pm$ 0.10 for NGC~5455  and of 0.19 $\pm$ 0.11 for  NGC~5471, both determinations lying in the lower range of typical values for {\hii} regions. As in other previous papers of our group, we have made the exercise of assuming the validity of the temperature fluctuations paradigm for explaining the abundance discrepancy. Doing this, we can estimate the {\tf} parameter that produces the agreement between the abundance of \ionic{O}{2+} determined from CELs and RLs, finding values of  
{\tf} = 0.015 $\pm$ 0.010 for NGC~5455 and {\tf} =  0.063 $\pm$ 0.008 for NGC~5471. For these two objects, Table~\ref{tab:abund1} lists the ionic abundances determined for the {\tf} values we have estimated  as well as the standard procedure of not considering temperature fluctuations, i.e.  assuming  {\tf}  = 0. For the rest of the objects -- for which we do not detect {\oii} RLs --  the calculations have only been made  assuming  {\tf}  = 0. The abundances for  {\tf} > 0 have been calculated following the formalism outlined by \cite{peimbertcostero69} and updated by \cite{apeimbertetal02}.

\begin{table*}
\centering
\caption{Ionic and total abundances$^{\rm a}$ for the {\hii} regions NGC~5462, NGC~5455,  H1216, NGC~5471, H37 and H219 of M101.}
\label{tab:abund1}
\begin{tabular}{lcccccccc}
\hline
& NGC~5462 & \multicolumn{2}{c}{NGC~5455} &  H11216 & \multicolumn{2}{c}{NGC~5471} &   H37 & H219  \\
&  {\tf} = 0.000 & {\tf} = 0.000 & {\tf} = 0.015 & {\tf} = 0.000 &   {\tf} = 0.000 & {\tf} = 0.063 & {\tf} = 0.000  &  {\tf} = 0.000 \\
&   &  & $\pm$  0.010 &  && $\pm$  0.008 &  & \\
 \hline
	\multicolumn{9}{c}{Ionic abundances from collisionally excited lines} \\
	N$^+$ & 6.67 $\pm$ 0.06 & 6.68 $\pm$ 0.02 & 6.75 $\pm$ 0.04 & 6.32 $\pm$ 0.02 & 5.71 $\pm$ 0.03 & 5.81 $\pm$ 0.04 & 6.36 $\pm$ 0.08 & 6.79 $\pm$ 0.04 \\
	O$^+$ & 7.59 $\pm$ 0.10 & 7.72 $\pm$ 0.02 & 7.80 $\pm$ 0.05 & 7.57 $\pm$ 0.04 & 7.13 $\pm$ 0.04 & 7.23 $\pm$ 0.04 & 7.79 $\pm$ 0.14 & 7.96 $\pm$ 0.08 \\
	O$^{2+}$ & 8.14 $\pm$ 0.01 & 8.17 $\pm$ 0.01 & 8.30 $\pm$ 0.09 &  8.06 $\pm$ 0.02 & 7.91 $\pm$ 0.01 &  8.10 $\pm$ 0.16 & 7.98 $\pm$ 0.02 & 7.85 $\pm$ 0.06 \\
	Ne$^{2+}$ & 7.38 $\pm$ 0.02 & 7.39 $\pm$ 0.03 &  7.53 $\pm$ 0.13 & 7.33 $\pm$ 0.03 & 7.26 $\pm$ 0.02 & 7.47 $\pm$ 0.24 & 7.40 $\pm$ 0.03 & 6.73 $\pm$ 0.14 \\
	S$^+$ &  5.81 $\pm$ 0.05 & 5.85 $\pm$ 0.01 & 5.92 $\pm$ 0.04 & 5.64 $\pm$ 0.02 & 5.33 $\pm$ 0.02 & 5.42 $\pm$ 0.03& 5.67 $\pm$ 0.07 & 5.95 $\pm$ 0.04 \\
	S$^{2+}$ & 6.61 $\pm$ 0.17 & 6.49 $\pm$ 0.04 & 6.63 $\pm$ 0.13 & 6.40 $\pm$ 0.04 & 6.05 $\pm$ 0.04 & 6.26 $\pm$ 0.25 & 6.26 $\pm$ 0.17 & 6.41 $\pm$ 0.13 \\
	Cl$^{2+}$ & 4.76 $\pm$ 0.15 & 4.70 $\pm$ 0.02 & 4.82 $\pm$ 0.11 & 4.51 $\pm$ 0.03 & 4.29 $\pm$ 0.04 & 4.47 $\pm$ 0.18 & 4.50 $\pm$ 0.12 & 4.76 $\pm$ 0.08 \\
	Ar$^{2+}$ & 5.91 $\pm$ 0.06 & 5.78 $\pm$ 0.02 & 5.89 $\pm$ 0.09 & 5.67 $\pm$ 0.02 & 5.36 $\pm$ 0.02 &  5.52 $\pm$ 0.15 & 5.43 $\pm$ 0.07 & 5.70 $\pm$ 0.06 \\
	Ar$^{3+}$ & $-$ & 4.58 $\pm$ 0.10 & 4.71 $\pm$ 0.14 & 4.59 $\pm$ 0.14 & 5.03 $\pm$ 0.02 & 5.22 $\pm$ 0.20 & $-$ & $-$ \\
	Fe$^{2+}$ & $-$ &  5.62 $\pm$ 0.02 & 5.75 $\pm$ 0.12 & 5.10 $\pm$ 0.04 & 5.07 $\pm$ 0.02 & 5.27 $\pm$ 0.20 & $-$ & 7.22 $\pm$ 0.13\\
	\multicolumn{9}{c}{Ionic abundances from recombination lines} \\
	He$^+$ & 10.96 $\pm$ 0.01 & \multicolumn{2}{c}{10.91 $\pm$ 0.01} & 10.89 $\pm$ 0.01 & \multicolumn{2}{c}{10.84 $\pm$ 0.01} & 10.85 $\pm$ 0.01 & 10.88 $\pm$ 0.01\\
	He$^{2+}$ & $-$ & \multicolumn{2}{c}{$-$} & $-$ &  \multicolumn{2}{c}{8.73 $\pm$ 0.02} & $-$  & $-$ \\
	C$^{2+}$ & 8.07 $\pm$ 0.10 & \multicolumn{2}{c}{7.98 $\pm$ 0.10} & $-$ & \multicolumn{2}{c}{7.71 $\pm$ 0.11} & 7.89 $\pm$ 0.13 &  $-$ \\
	O$^{2+}$ & $-$ & \multicolumn{2}{c}{8.30 $\pm$ 0.10} & $-$ & \multicolumn{2}{c}{8.10 $\pm$ 0.11} & $-$ & $-$ \\
	\multicolumn{9}{c}{Total abundances}\\
	He & $-$ & \multicolumn{2}{c}{$-$} & $-$ & \multicolumn{2}{c}{10.84 $\pm$ 0.01} & $-$ & $-$ \\
	C & 8.13 $\pm$ 0.10 & \multicolumn{2}{c}{8.08 $\pm$ 0.11} & $-$ & \multicolumn{2}{c}{7.78 $\pm$ 0.11} & 8.03 $\pm$ 0.13 & $-$ \\	
	N &7.32 $\pm$ 0.10 & 7.28 $\pm$ 0.03 &  7.36 $\pm$ 0.09 & 6.94 $\pm$ 0.04 & 6.53 $\pm$ 0.05 & 6.68 $\pm$ 0.13 &  6.83 $\pm$ 0.14 & 7.12 $\pm$ 0.10 \\
	O & 8.25 $\pm$ 0.02 & 8.30 $\pm$ 0.01 & 8.42 $\pm$ 0.07 & 8.18 $\pm$ 0.02 & 7.98 $\pm$ 0.01 & 8.16 $\pm$ 0.14 &  8.19 $\pm$ 0.06 &  8.21 $\pm$ 0.05 \\
	Ne & 7.44 $\pm$ 0.04 & 7.49 $\pm$ 0.03 & 7.57 $\pm$ 0.13 & 7.41 $\pm$ 0.03 & 7.28 $\pm$ 0.03 & 7.49 $\pm$ 0.25 & 7.56 $\pm$ 0.06 & 7.00 $\pm$ 0.16 \\
	S & 6.80 $\pm$ 0.18 & 6.67 $\pm$ 0.04 & 6.77 $\pm$ 0.14 & 6.58 $\pm$ 0.05 & 6.36 $\pm$ 0.06 & 6.63 $\pm$ 0.24 & 6.39 $\pm$ 0.19 & 6.53 $\pm$ 0.14 \\
	Cl & 4.83 $\pm$ 0.15  & 4.76 $\pm$ 0.02 & 4.95 $\pm$ 0.11 & 4.57 $\pm$ 0.03 & 4.43 $\pm$ 0.04 & 4.62 $\pm$ 0.20 & 4.56 $\pm$ 0.12 & 4.85 $\pm$ 0.09 \\
	Ar & 5.94 $\pm$ 0.06 & 5.81 $\pm$ 0.10 & 5.90 $\pm$ 0.17 & 5.70 $\pm$ 0.14 & 5.53 $\pm$ 0.03 & 5.70 $\pm$ 0.25 & 5.46 $\pm$ 0.07 & 5.75 $\pm$ 0.06 \\
	Fe & $-$ & 6.11 $\pm$ 0.03 & 6.28 $\pm$ 0.15 & 5.62 $\pm$ 0.06 & 5.81 $\pm$ 0.05 & 6.08 $\pm$ 0.25 & $-$ & 7.43 $\pm$ 0.16 \\
 \hline
 \end{tabular}
 \begin{description}
 \item[$^{\rm a}$] In units of 12+log(\ionic{X}{+n}/\ionic{H}{+}) or 12+log(X/H).  
 \end{description}
 \end{table*}

\begin{table*}
\centering
\caption{Ionic and total abundances$^{\rm a}$ for the rest of  {\hii} regions of M101 and all of M31.}
\label{tab:abund2}
\begin{tabular}{lcccccc}
\hline
&  NGC~5447 & H681 &  H1118 & H1146 & SDS323 & BA289 \\
 \hline
	\multicolumn{7}{c}{Ionic abundances from collisionally excited lines} \\
	N$^+$ & 6.40 $\pm$ 0.15 & 6.07 $\pm$ 0.03 & 6.35 $\pm$ 0.04 & 6.49 $\pm$ 0.06 & 5.93 $\pm$ 0.07 & 7.82 $\pm$ 0.07 \\
	O$^+$ & 7.43 $\pm$ 0.23 & 7.39 $\pm$ 0.02 & 7.61 $\pm$ 0.04 & 7.57 $\pm$ 0.08 & 7.17 $\pm$ 0.05 & 8.47 $\pm$ 0.12 \\
	O$^{2+}$ & 8.24 $\pm$ 0.01 &  7.61 $\pm$ 0.02 & 7.90 $\pm$ 0.04 & 7.96 $\pm$ 0.08 & 7.37 $\pm$ 0.05 & 8.13 $\pm$ 0.22  \\
	Ne$^{2+}$ & 7.50 $\pm$ 0.03 & 6.93 $\pm$ 0.05 & 7.04 $\pm$ 0.08 & 7.43 $\pm$ 0.16 & 6.54 $\pm$ 0.10 & $-$  \\
	S$^+$ &  5.52 $\pm$ 0.12 & 5.60 $\pm$ 0.02 & 5.75 $\pm$ 0.03 & 5.75 $\pm$ 0.05 & 5.38 $\pm$ 0.04 & 6.64 $\pm$ 0.07 \\
	S$^{2+}$ & 6.40 $\pm$ 0.17 & 5.92 $\pm$ 0.11 & 6.46 $\pm$ 0.15 & 6.46 $\pm$ 0.21 & $-$ &  $-$ \\
	Cl$^{2+}$ & $-$ & 4.08 $\pm$ 0.14 & $-$ & $-$ & $-$ &  $-$ \\
	Ar$^{2+}$ & 5.72 $\pm$ 0.08 & 5.27 $\pm$ 0.04 & 5.65 $\pm$ 0.05 & 5.68 $\pm$ 0.09 & 4.98 $\pm$ 0.14 & 5.99 $\pm$ 0.14 \\
	\multicolumn{7}{c}{Ionic abundances from recombination lines} \\
	He$^+$ & 10.86 $\pm$ 0.01 & 10.83 $\pm$ 0.01 & 10.87 $\pm$ 0.03 & 10.87 $\pm$ 0.03 & 10.86 $\pm$ 0.04 & 10.76 $\pm$ 0.02 \\
	He$^{2+}$ & $-$ &  8.41 : & $-$ & $-$ & $-$ & $-$ \\
	C$^{2+}$ & 7.99 $\pm$ 0.10 &  $-$ & $-$ & $-$ & $-$ & $-$ \\
	\multicolumn{7}{c}{Total abundances}\\
	He & $-$ & $-$ & $-$ & $-$ & $-$ & $-$ \\
	C & 8.03 $\pm$ 0.11 & $-$ & $-$ & $-$ & $-$ & $-$ \\	
	N & 7.23 $\pm$ 0.24 & 6.52 $\pm$ 0.04 & 6.85 $\pm$ 0.05 & 7.06 $\pm$ 0.10 & 6.34 $\pm$ 0.10 & 8.06 $\pm$ 0.20 \\
	O & 8.30 $\pm$ 0.03 & 7.81 $\pm$ 0.02 & 8.08 $\pm$ 0.03 & 8.11 $\pm$ 0.06 & 7.59 $\pm$ 0.03 & 8.63 $\pm$ 0.11 \\
	Ne & 7.52 $\pm$ 0.05 & 7.03 $\pm$ 0.05 & 7.17 $\pm$ 0.09 & 7.52 $\pm$ 0.18 & 6.61 $\pm$ 0.10 & $-$ \\
	S & 6.70 $\pm$ 0.26 & 6.13 $\pm$ 0.11 & 6.59 $\pm$ 0.15 & 6.62 $\pm$ 0.22 & $-$ & $-$ \\
	Cl & $-$ &  4.14 $\pm$ 0.14 & $-$ & $-$ & $-$ & $-$ \\
	Ar & 5.77 $\pm$ 0.09 & 5.30 $\pm$ 0.04 & 5.68 $\pm$ 0.05 & 5.71 $\pm$ 0.09 & 5.01 $\pm$ 0.14 &  6.06 $\pm$ 0.15 \\
	\\
&  K703 & K160 & BA379 & BA371 & BA310 & BA374 \\
	\\
	\multicolumn{7}{c}{Ionic abundances from collisionally excited lines} \\
	N$^+$ & 7.39 $\pm$ 0.04 & 7.58 $\pm$ 0.03 & 6.96 $\pm$ 0.02 & 6.84 $\pm$ 0.04 & 7.48 $\pm$ 0.09 &  7.21 $\pm$ 0.08 \\
	O$^+$ & 8.10 $\pm$ 0.07 & 8.38 $\pm$ 0.08 & 7.92 $\pm$ 0.03 & 7.76 $\pm$ 0.06 & 8.20 $\pm$ 0.15 & 8.05 $\pm$ 0.13 \\
	O$^{2+}$ & 8.10 $\pm$ 0.09 & 7.92 $\pm$ 0.09 & 8.16 $\pm$ 0.04 & 8.23 $\pm$ 0.03 &  8.13 $\pm$ 0.22 & 7.50 $\pm$ 0.06 \\
	Ne$^{2+}$ & 7.13 $\pm$ 0.18 & 6.85 $\pm$ 0.17 & 7.43 $\pm$ 0.05 & 7.55 $\pm$ 0.09 & $-$ & 6.55 $\pm$ 0.13 \\
	S$^+$ &  6.27 $\pm$ 0.04 & 6.39 $\pm$ 0.03 &  5.83 $\pm$ 0.02 & 5.75 $\pm$ 0.04 &  6.35 $\pm$ 0.08 & 6.16 $\pm$ 0.07 \\
	S$^{2+}$ & 7.10 $\pm$ 0.22 & 6.83 $\pm$ 0.14 & 6.55 $\pm$ 0.08 & 6.60 $\pm$ 0.11 & $-$ & $-$ \\
	Cl$^{2+}$ & $-$ & 5.03 $\pm$ 0.06 & 4.71 $\pm$ 0.03 & 4.70 $\pm$ 0.06 & 5.09 $\pm$ 0.31 & 4.77 $\pm$ 0.20 \\
	Ar$^{2+}$ & 5.97 $\pm$ 0.06 & 6.01 $\pm$ 0.07 & 5.94 $\pm$ 0.03 & 5.88 $\pm$ 0.05 & 6.18 $\pm$ 0.23 & 5.68 $\pm$ 0.07 \\
	Fe$^{2+}$ & $-$ &  $-$ & 6.83 $\pm$ 0.10 & $-$ & $-$ & $-$  \\ 
	\multicolumn{7}{c}{Ionic abundances from recombination lines} \\
	He$^+$ & 10.89 $\pm$ 0.01 & 10.90 $\pm$ 0.01 & 10.91 $\pm$ 0.01 & 10.91 $\pm$ 0.01 & 10.96 $\pm$ 0.01 & 10.96 $\pm$ 0.02 \\
	C$^{2+}$ & $-$ & 8.63 $\pm$ 0.10 &  $-$ & $-$ & $-$ & $-$  \\
	\multicolumn{7}{c}{Total abundances}\\
	He & $-$ & $-$ & $-$ & $-$ & $-$ & $-$ \\
	C & $-$ & 8.84 $\pm$ 0.10 & $-$ & $-$ & $-$ & $-$ \\	
	N & 7.77 $\pm$ 0.09 & 7.78 $\pm$ 0.09 & 7.45 $\pm$ 0.04 & 7.46 $\pm$ 0.06 & 7.83 $\pm$ 0.20 &  7.38 $\pm$ 0.24 \\
	O & 8.40 $\pm$ 0.06 & 8.51 $\pm$ 0.05 &  8.36 $\pm$ 0.03 & 8.35 $\pm$ 0.03 & 8.47 $\pm$ 0.13 & 8.16 $\pm$ 0.10 \\
	Ne & 7.37 $\pm$ 0.20 & 7.31 $\pm$ 0.19 & 7.58 $\pm$ 0.07 & 7.64 $\pm$ 0.10 & $-$ & 7.04 $\pm$ 0.16 \\
	S & 7.00 $\pm$ 0.22 & 6.94 $\pm$ 0.15 & 6.66 $\pm$ 0.08 & 6.76 $\pm$ 0.12 & $-$ & $-$ \\
	Cl & $-$ & 5.17 $\pm$ 0.07 & 4.76 $\pm$ 0.03 & 4.75 $\pm$ 0.06 & $-$ & 4.93 $\pm$ 0.22 \\
	Ar & 5.92 $\pm$ 0.07 & 6.05 $\pm$ 0.06 & 5.97 $\pm$ 0.03 & 5.90 $\pm$ 0.05 & 6.23 $\pm$ 0.23 & 5.76 $\pm$ 0.09 \\	
	Fe & $-$ & $-$ & 7.20 $\pm$ 0.11 &	$-$ &  $-$ &  $-$ \\
 \hline
 \end{tabular}
 \begin{description}
 \item[$^{\rm a}$] Assuming {\tf} = 0.0 and in units of 12+log(\ionic{X}{+n}/\ionic{H}{+}) or 12+log(X/H).  
 \end{description}
 \end{table*}

In {\hii} regions, O is commonly the only element for which no ionization correction factor (hereafter ICF) is necessary to derive its total abundance. Therefore, the total O abundance can be calculated simply by adding the 
O$^+$/H$^+$ and O$^{2+}$/H$^+$ ratios. Although we detect {\heii} lines in the spectra of two of our {\hii} regions -- indicating that some \ionic{O}{3+} should be present in those objects -- their small 
\ionic{He}{2+}/ \ionic{H}{+} ratio implies that the contribution of \ionic{O}{3+} to the O/H ratio is much smaller than the uncertainties and clearly negligible. For the other elements, we have to assume ICFs to correct for the contribution of ionization stages not showing detectable emission lines in the optical range. 

To compute the total abundances of N, Ne, S, Cl and Ar from CELs, we adopted the ICFs scheme proposed by \citet{izotovetal06}  based on photoionization models of extragalactic {\hii} regions. These authors propose different ICFs depending on the element and the metallicity range of the nebulae. We used the expressions for "high" metallicity -- 12+log(O/H) obtained from CELs $\geq$ 8.2 -- for most of the {\hii} regions of both galaxies.  
For SDS323 we used the expressions for "intermediate" metallicity -- 12+log(O/H) obtained from CELs $\sim$ 7.6. For H11146, H1118, NGC~5471, H681 and BA374, with values of 12+log(O/H) between 7.6 and 8.2, we used linearly 
interpolated ICFs between "intermediate" and "high" metallicity cases. For Fe, we adopted the ICF scheme suggested by \citet{rodriguezrubin05} which is based on the computed Fe$^{2+}$/H$^+$ ratio and photoionization models (their equation 2).  

In the case of C, we used the mean value of several ICF(\ionic{C}{2+}) schemes based on photoionization models: (a) the relation obtained by \citet{garnettetal99} based on models firstly described by \citet{garnettetal95}; (b) the scheme  proposed by \citet{bergetal19} that includes dependence on the metallicity and the ionization parameter; and (c) the sets of ICF(\ionic{C}{2+}) calculated by \citet{medina19},  based on a subset of photoionization models by \cite{valeasarietal16}. The values of the C/H ratio given in tables~\ref{tab:abund1} and \ref{tab:abund2} correspond to the application of the mean of the ICF(\ionic{C}{2+}) values obtained with the three schemes for each object, except in the case of K160, whose O/H ratio exceeds the limit of applicability of the ICF(\ionic{C}{2+}) of \citet{bergetal19}. The uncertainty assigned to the C abundance corresponds to the combination of the observational error of the \ionic{C}{2+}/\ionic{H}{+} ratio and the dispersion of the different ICF(\ionic{C}{2+}) values used for the mean. The C/H ratios obtained with the ICF(\ionic{C}{2+}) of \citet{garnettetal95} are systematically larger -- about 0.06 $\pm$ 0.07 dex -- than those obtained with the scheme proposed by \citet{bergetal19}. The C/H ratios obtained with the ICF(\ionic{C}{2+}) of \citet{medina19} are more consistent with the results obtained with the \citet{bergetal19} relations, being only about 0.02 $\pm$ 0.02 dex larger. 

We have several objects in common with the previous study on abundances of {\hii} regions in M101 carried out by \citet{croxalletal16}: NGC~5462, NGC~5455, H1216, NGC~5471, H219, NGC~5447, H681 and H1146. Our 
O/H ratios are on average 0.10 dex lower than those computed by \citet{croxalletal16}, with a dispersion of 0.07 dex (see Sect~\ref{sec:M101_O} ). It is important to remark that \citet{croxalletal16} and us use the same atomic datasets for the \ionic{O}{+} and \ionic{O}{2+} ions. The reason of the systematic differences in abundance is that our {\elect} determinations for the objects are higher in almost all objects. On average, the {\elect} values used for deriving \ionic{O}{2+} and \ionic{O}{+} abundances are about 600 K and 400 K, respectively, higher than those obtained by \citet{croxalletal16}. In the case of M31, we have four objects in common with the previous spectroscopic observations by \citet{zuritabresolin12}: BA310, BA371, BA374 and BA379. The mean difference between our O/H ratios and those calculated by \citet{zuritabresolin12} for these four objects is virtually negligible, 0.01 dex.

\section{Radial abundance gradients in M101} 
\label{sec:M101} 

\subsection{Oxygen} 
\label{sec:M101_O} 

\begin{figure*} 
\centering 
\includegraphics[scale=0.74]{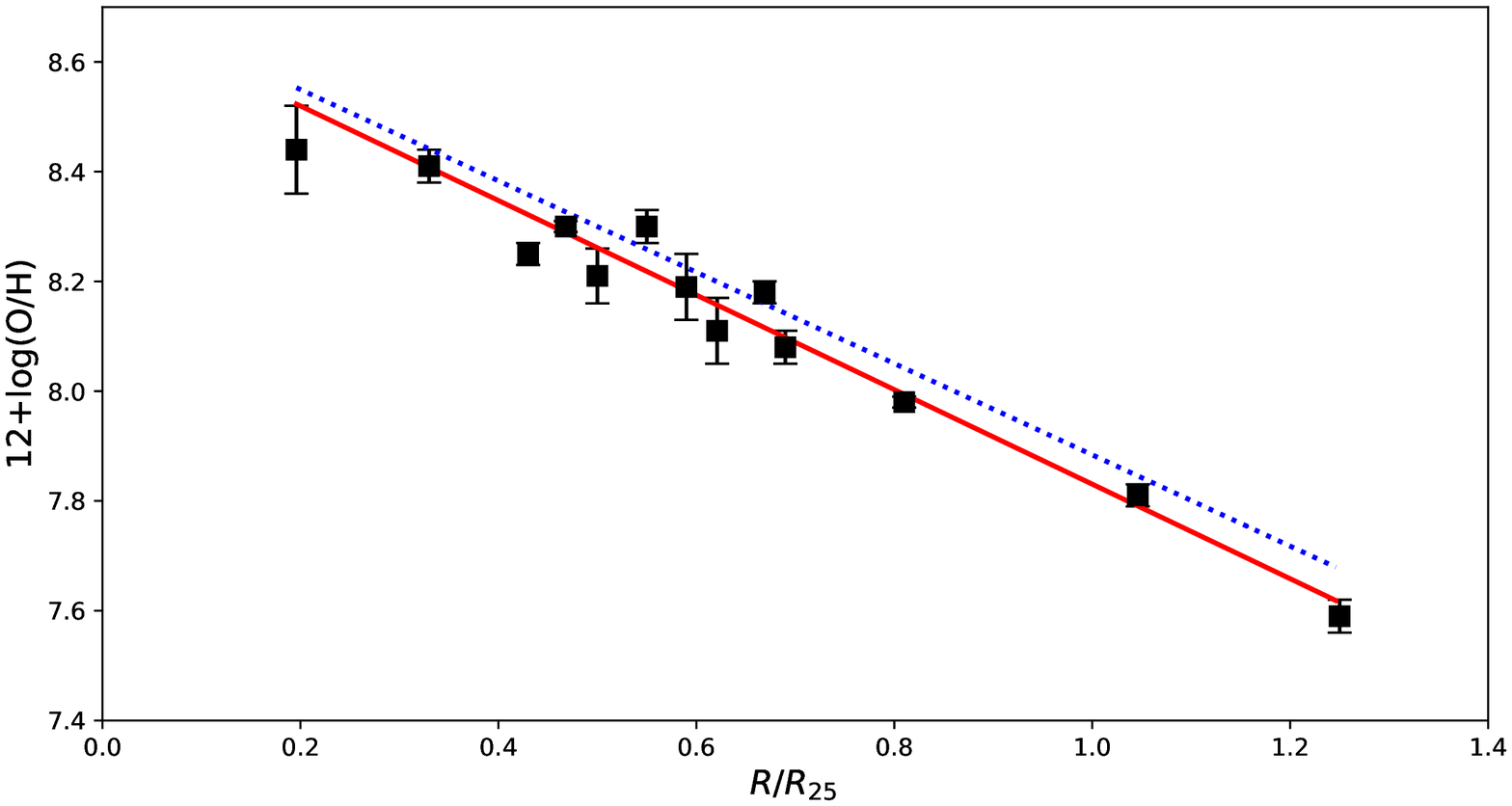} 
 \caption{Radial distribution of the O abundances calculated from CELs of the sample of  {\hii} regions in M101 (we have included data for two additional objects, see text for details) as a function of their fractional galactocentric distance ($R$/$R_\mathrm{25}$). The solid red line represents the least-squares fit to all the objects. The dotted blue line corresponds to the least-squares fit obtained by \citet{croxalletal16}.} 
 \label{fig:M101_Ograd} 
 \end{figure*} 

Fig.~\ref{fig:M101_Ograd} shows the radial distribution of the O abundances -- calculated from CELs -- of the sample of {\hii} regions in M101 as a function of their fractional galactocentric distance ($R$/$R_\mathrm{25}$). We have also included data for H1013 and NGC~5461 obtained by \citet{estebanetal09} for which we have recalculated the physical conditions and abundances following the same methodology and atomic data than for the rest of the objects of this paper. H1013 and NGC~5461 correspond to the two innermost objects shown in Fig.~\ref{fig:M101_Ograd}, located at $R$/$R_\mathrm{25}$ = 0.20 and 0.33 and with 12+log(O/H) = 8.44 and 8.41, respectively. We have performed a least-squares linear fit to the $R$/$R_\mathrm{25}$ ratio of the nebulae and their O abundance. The fit is indicated by a solid red line in Fig.~\ref{fig:M101_Ograd} and has the parameters given in the equation given below:
	\begin{equation} \label{eq:1} 12 + \log(\mathrm{O/H})_\mathrm{CELs} = 8.69(\pm 0.04) - 0.86(\pm 0.05) R/R_\mathrm{25}. \end{equation} 
For comparison, in Fig.~\ref{fig:M101_Ograd}, we have also included the least-squares fit obtained by \citet{croxalletal16}, whose slope and intercept are 
$-$0.83 $\pm$ 0.04 dex ($R$/$R_\mathrm{25}$)$^{-1}$ and 8.72 $\pm$ 0.02, respectively. Both fits are consistent within the errors. The scatter of the O abundances represented in 
Fig.~\ref{fig:M101_Ograd} with respect to the gradient fitting at the corresponding distance of the {\hii} regions is $\pm$0.04 dex. This is what \citet{croxalletal16} define as internal dispersion; they obtain a value of about 0.07 dex for their larger sample of 
{\hii} regions of  M101. \citet{croxalletal16} argue that the internal dispersion is larger than expected by the formal uncertainties of the data, not finding a clear cause for it but concluding its intrinsic nature. We think that this claim is debatable. In Section~\ref{sec:abund} we said that the mean difference in the O/H ratios obtained by us and \citet{croxalletal16} for the {\hii} regions of M101 in common is about 0.10 dex, a value larger than the "internal dispersion". 
It is very difficult to ascertain the reason of this systematics. One possibility is the unintentional bias introduced by the use of different methodologies of measuring weak lines \citep[see][]{rolastasinska94}. But 
other reasons are also arguable, such as anomalies in the relative flux calibration along the optical spectrum, or observations of different parts of objects with significant spatial variation of their physical properties, among others.  In Fig.~\ref{fig:M101_Ocomp} we plot the O/H ratios and O radial gradient obtained from our data and the O abundances we derive using the emission line ratios measured by \citet{croxalletal16} and \citet{kennicuttetal03} for the objects in common with our sample and applying our methodology and atomic data. In the figure, we can see that the different O abundances for the same object differ by values usually larger than the formal errorbars given in the papers. The average dispersion of the various determinations of the O/H ratio for a given object is about 0.04 dex, slightly smaller than the internal dispersion found by \citet{croxalletal16} but of the same order of the value estimated by us. This result suggests that uncertainties lower than about 0.04 or 0.05 dex for the O abundances may be somewhat underestimated and that the internal dispersion may not necessarily be related to chemical inhomogeneities across the disc of M101. 

\begin{figure} 
\centering 
\includegraphics[scale=0.37]{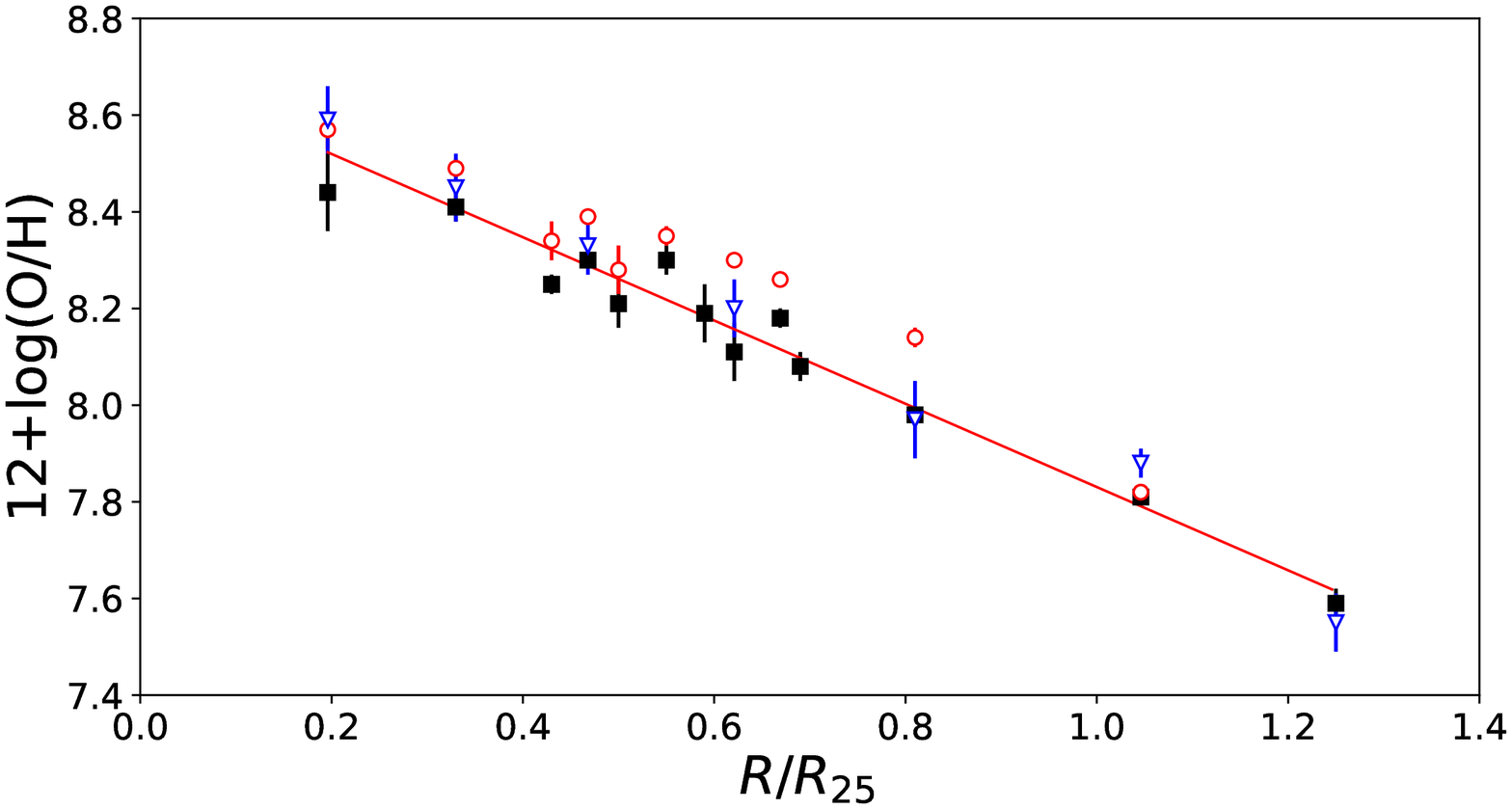} 
 \caption{Radial distribution of O abundances for the  {\hii} regions observed in M101 as a function of their fractional galactocentric distance ($R$/$R_\mathrm{25}$). The black squares and red line are our data and radial gradient shown in Fig.~\ref{fig:M101_Ograd}. The empty red circles and blue triangles show  the O abundances obtained using the emission line ratios measured by \citet{croxalletal16} and \citet{kennicuttetal03}, respectively, and applying our methodology and atomic data for the objects in common with our sample.} 
 \label{fig:M101_Ocomp} 
 \end{figure} 

As it was commented in Sect.~\ref{sec:abund}, we can determine the \ionic{O}{2+} abundance from {\oii} RLs in two objects of the sample, NGC~5455 and NGC~5471. \citet{estebanetal09} observed {\oii} RLs in additional {\hii} regions of M101: H1013, NGC~5461 and NGC~5447. This last object is included in our sample, but the {\oii} lines were not detected in the GTC spectra. We have recalculated the  \ionic{O}{2+} abundances of these three objects using the same methodology and atomic data and -- as indicated 
in Sect.~\ref{sec:abund} -- excluding the {\oii} 4638.86 and 4641.81 \AA\ lines. We have calculated the O/H ratio -- (O/H)$_\mathrm{RLs}$ -- using the \ionic{O}{2+} abundance from {\oii} RLs and the \ionic{O}{+} abundance determined from CELs and assuming {\ts} > 0.0 (these values are also included in Table~\ref{tab:abund1}). 
We obtain values of 12+log(O/H)$_\mathrm{RLs}$ of 8.75 $\pm$ 0.13, 8.49 $\pm$ 0.10 and 8.57 $\pm$ 0.11 for H1013, NGC~5461 and NGC~5447, respectively. With these three objects and NGC~5455 and NGC~5471 we can derive a radial O gradient with abundance determinations based on RLs (see Fig.~\ref{fig:M101_OgradRLs}), whose parameters are: 
	\begin{equation} \label{eq:2} 12 + \log(\mathrm{O/H})_\mathrm{RLs} = 8.81(\pm 0.27) - 0.75(\pm 0.55) R/R_\mathrm{25}. \end{equation}
The fit has very large uncertainties due to the large errors of the individual O/H ratio determinations and the small baseline of $R$/$R_\mathrm{25}$. The slope of the fit is not substantially different from the one obtained from CELs, but the fit shows a clear offset of about 0.2 dex (see Fig.~\ref{fig:M101_OgradRLs}), which is due to the abundance discrepancy factor (see Sect.~\ref{sec:abund}).  

\begin{figure} 
\centering 
\includegraphics[scale=0.37]{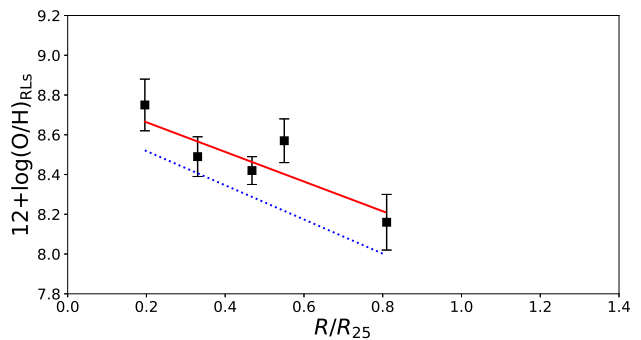} 
 \caption{Radial distribution of O abundances determined from {\oii} RLs for {\hii} regions in M101 as a function of their fractional galactocentric distance ($R$/$R_\mathrm{25}$). The solid red line represents the least-squares fit. The dotted blue line corresponds to the least-squares fit to the O abundances determined from CELs given by Eq.~\ref{eq:1}.} 
 \label{fig:M101_OgradRLs} 
 \end{figure} 
 
\begin{figure} 
\centering 
\includegraphics[scale=0.37]{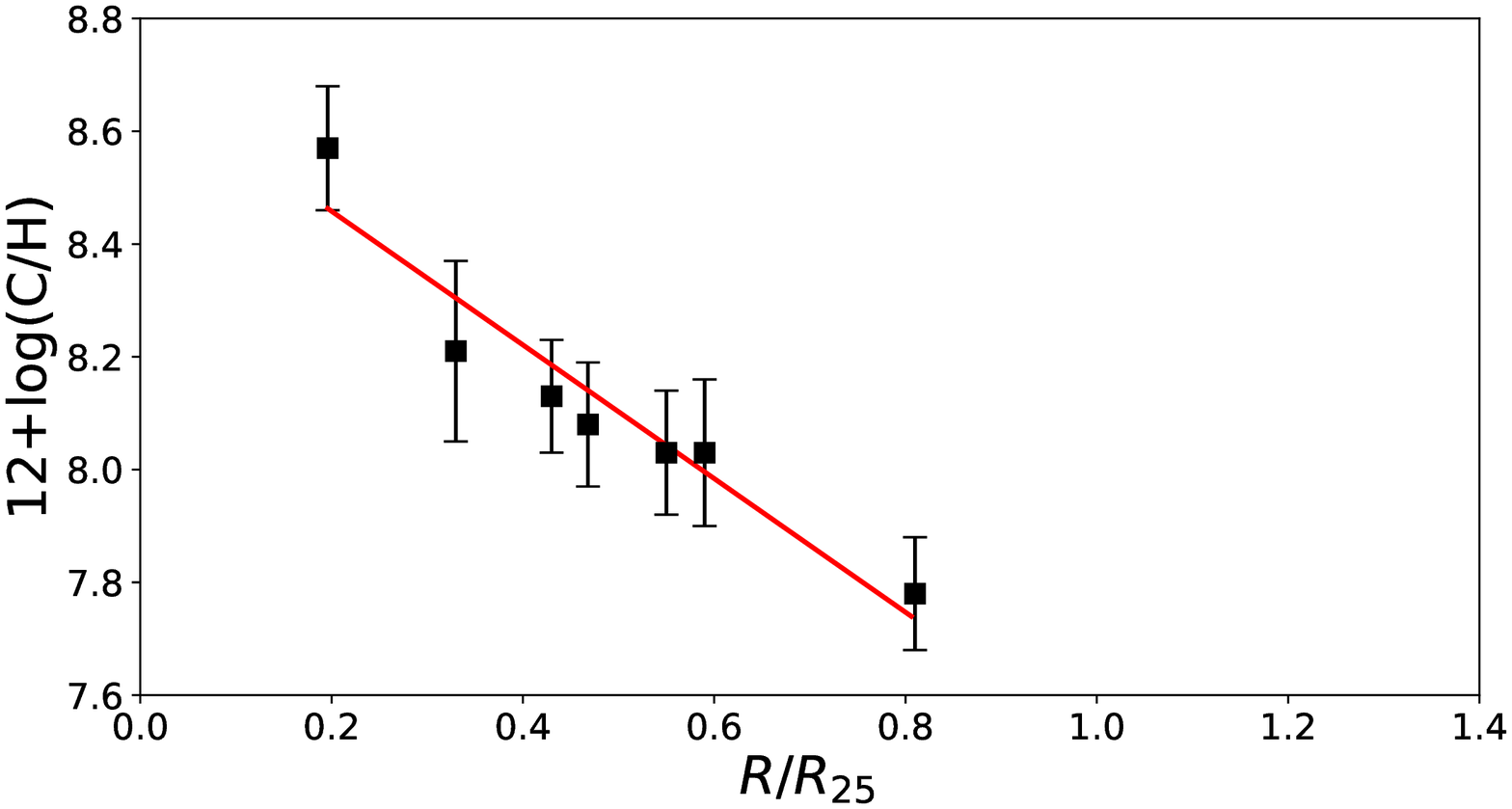} 
\includegraphics[scale=0.37]{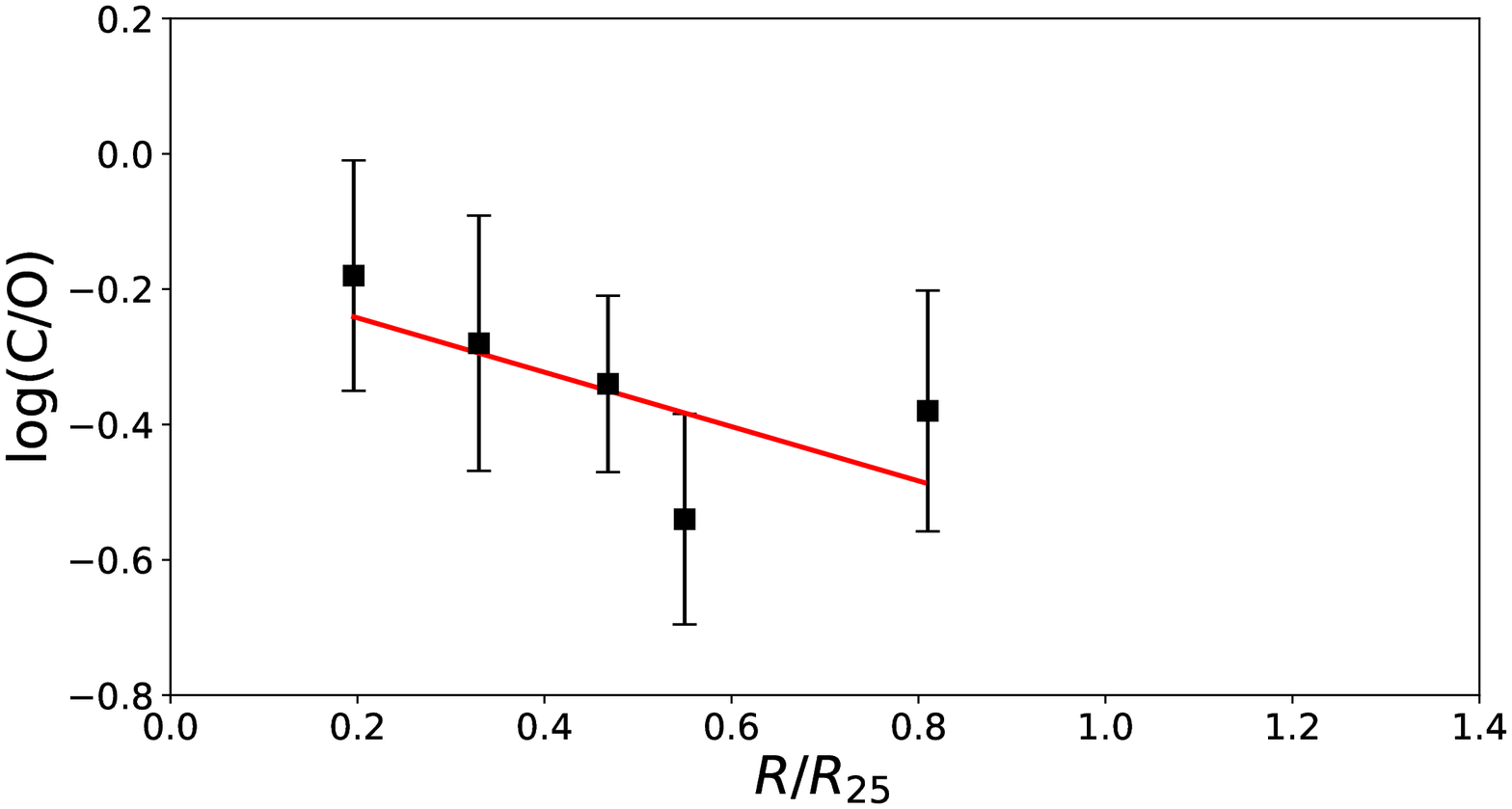} 
 \caption{Radial distribution of C abundances  (upper panel) and C/O ratios (lower panel) for {\hii} regions in M101 as a function of their fractional galactocentric distance ($R$/$R_\mathrm{25}$). The C and O abundances have been obtained from {\cii} and {\oii} RLs. The solid red line represents the least-squares fits. } 
 \label{fig:M101_Cgrad} 
 \end{figure} 
 
\begin{figure} 
\centering 
\includegraphics[scale=0.37]{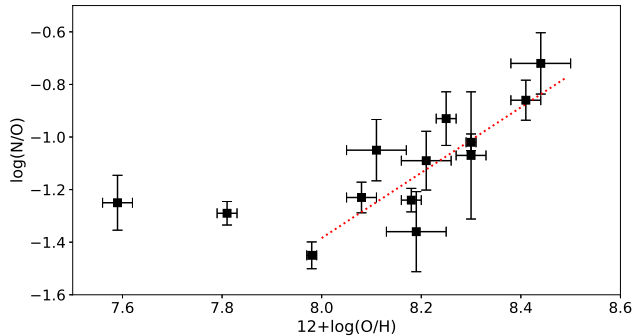} 
 \caption{N/O ratio as a function of 12+log(O/H) for {\hii} regions in M101. The dotted red line represents the linear  least-squares fits of the data with 12 + log(O/H) $\geq$ 8.0. } 
 \label{fig:M101_NO} 
 \end{figure} 
 
\begin{figure} 
\centering 
\includegraphics[scale=0.37]{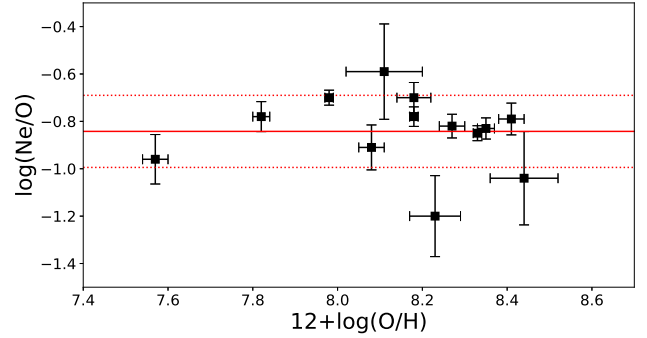} 
 \caption{Ne/O ratio as a function of 12+log(O/H) for {\hii} regions in M101. The continuous and dotted red lines represent the weighted mean and standard deviation of the values, respectively.} 
 \label{fig:M101_NeO} 
 \end{figure} 
 
\begin{figure*} 
\centering 
\includegraphics[scale=0.37]{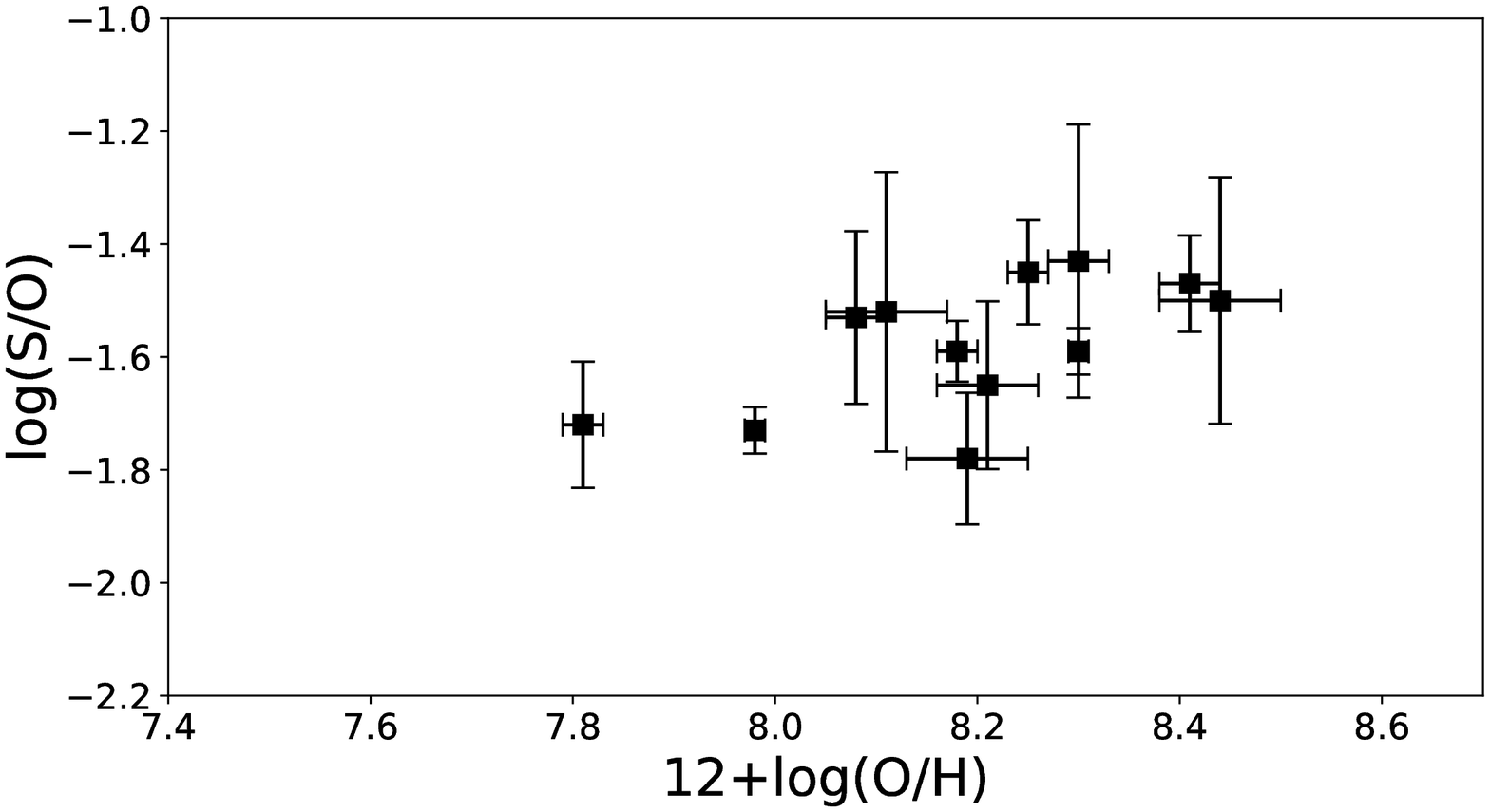} 
\includegraphics[scale=0.37]{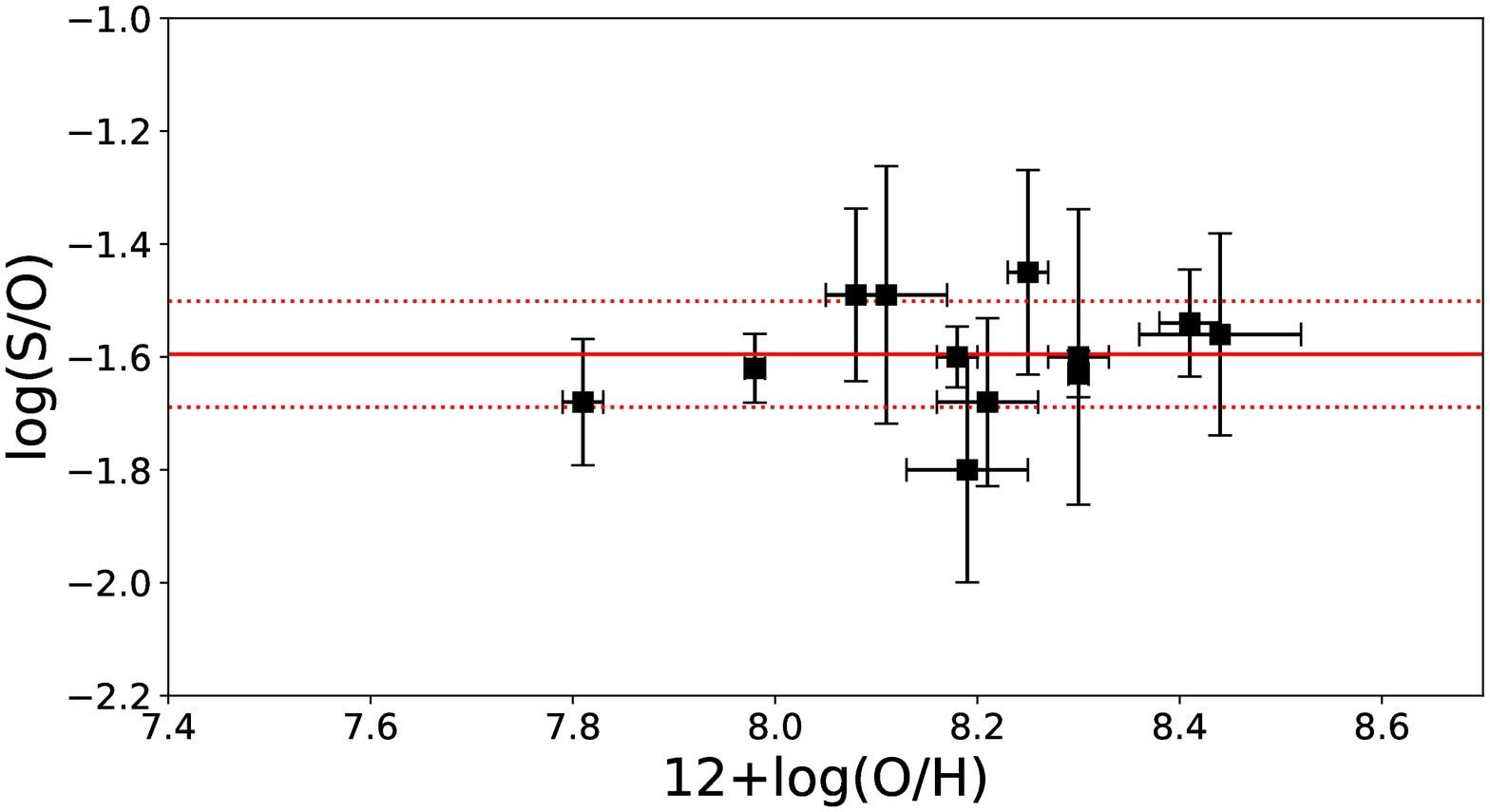} 
\\
\includegraphics[scale=0.37]{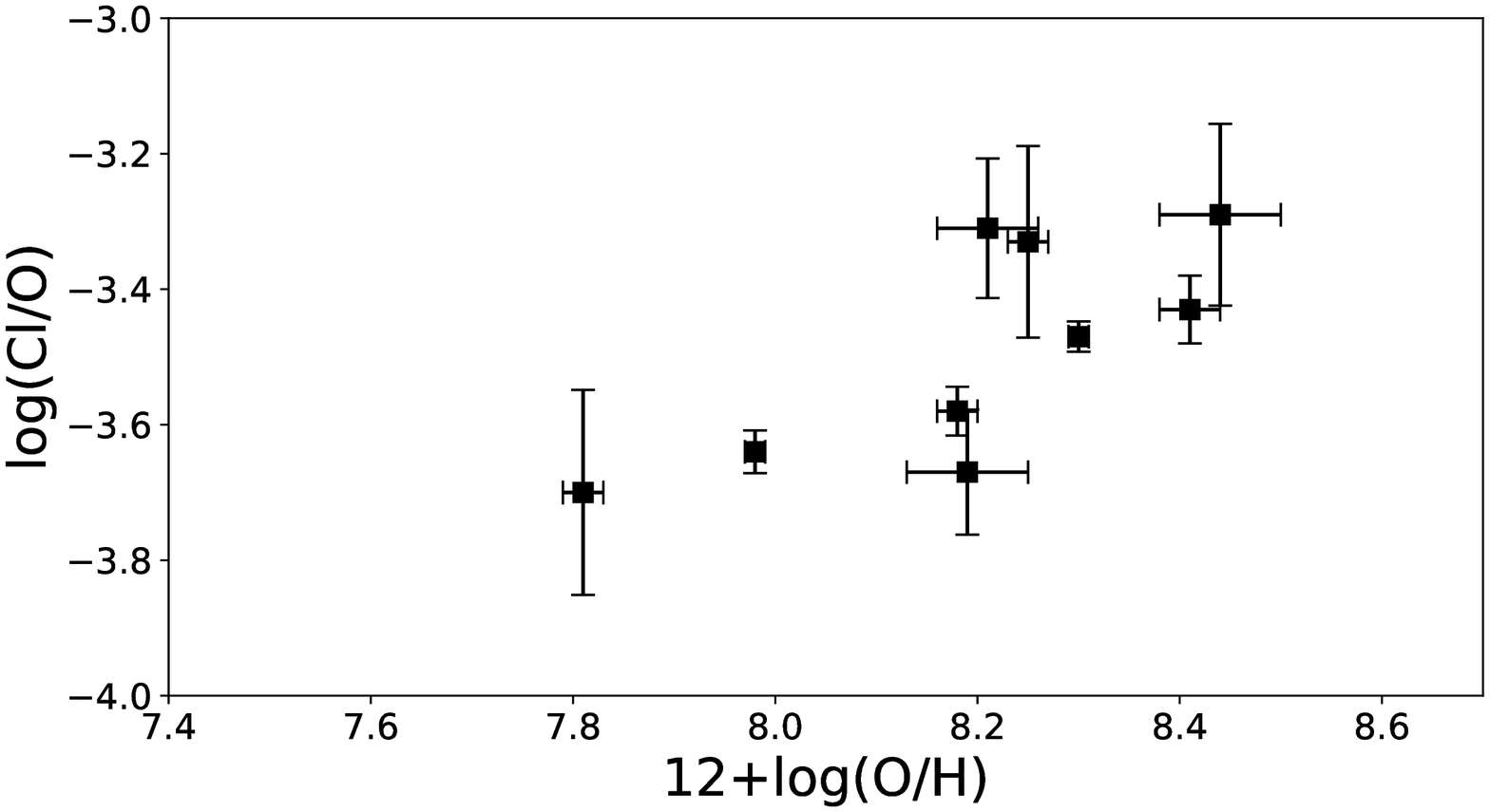} 
\includegraphics[scale=0.37]{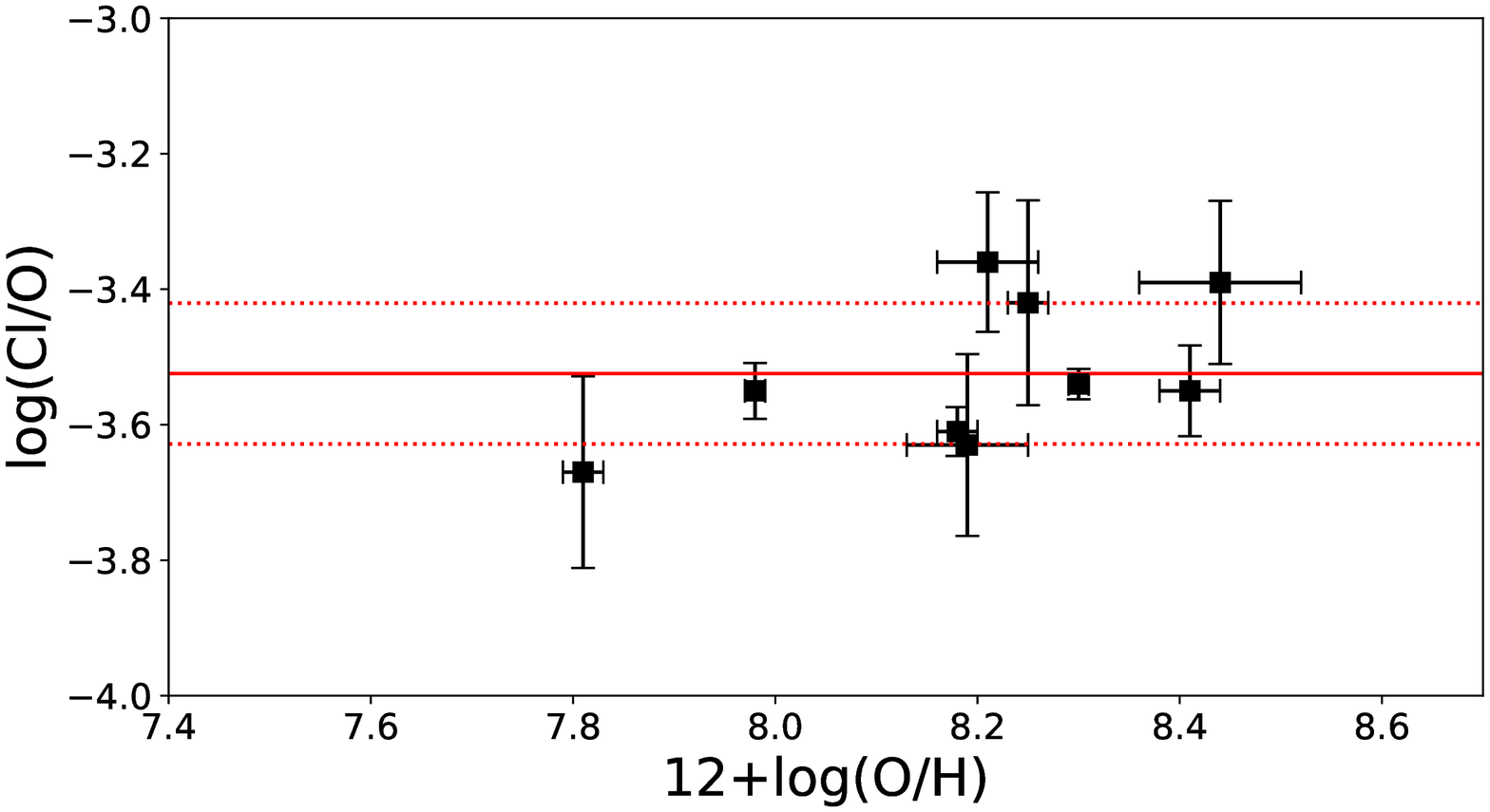}
\\
\includegraphics[scale=0.37]{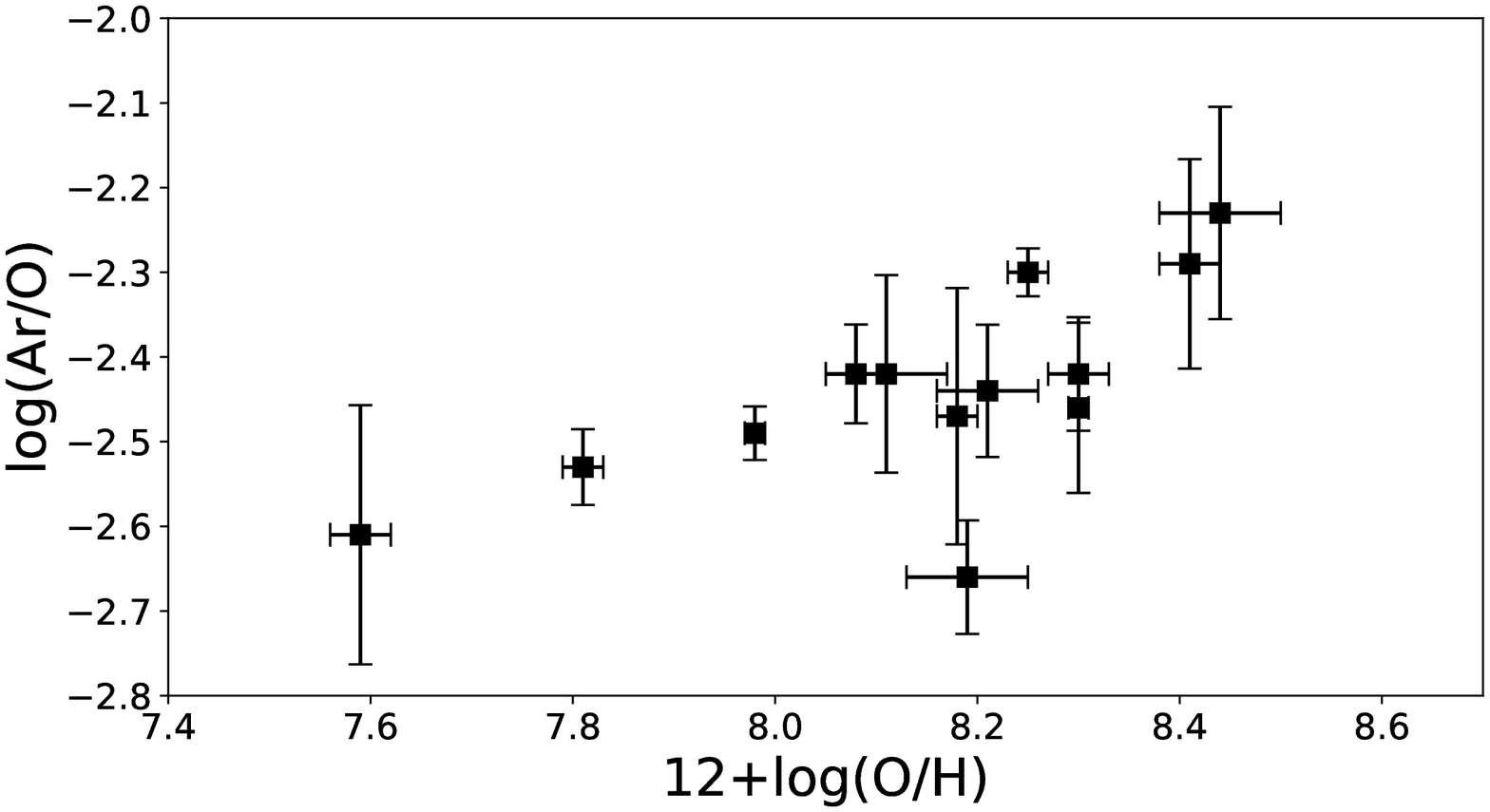} 
\includegraphics[scale=0.37]{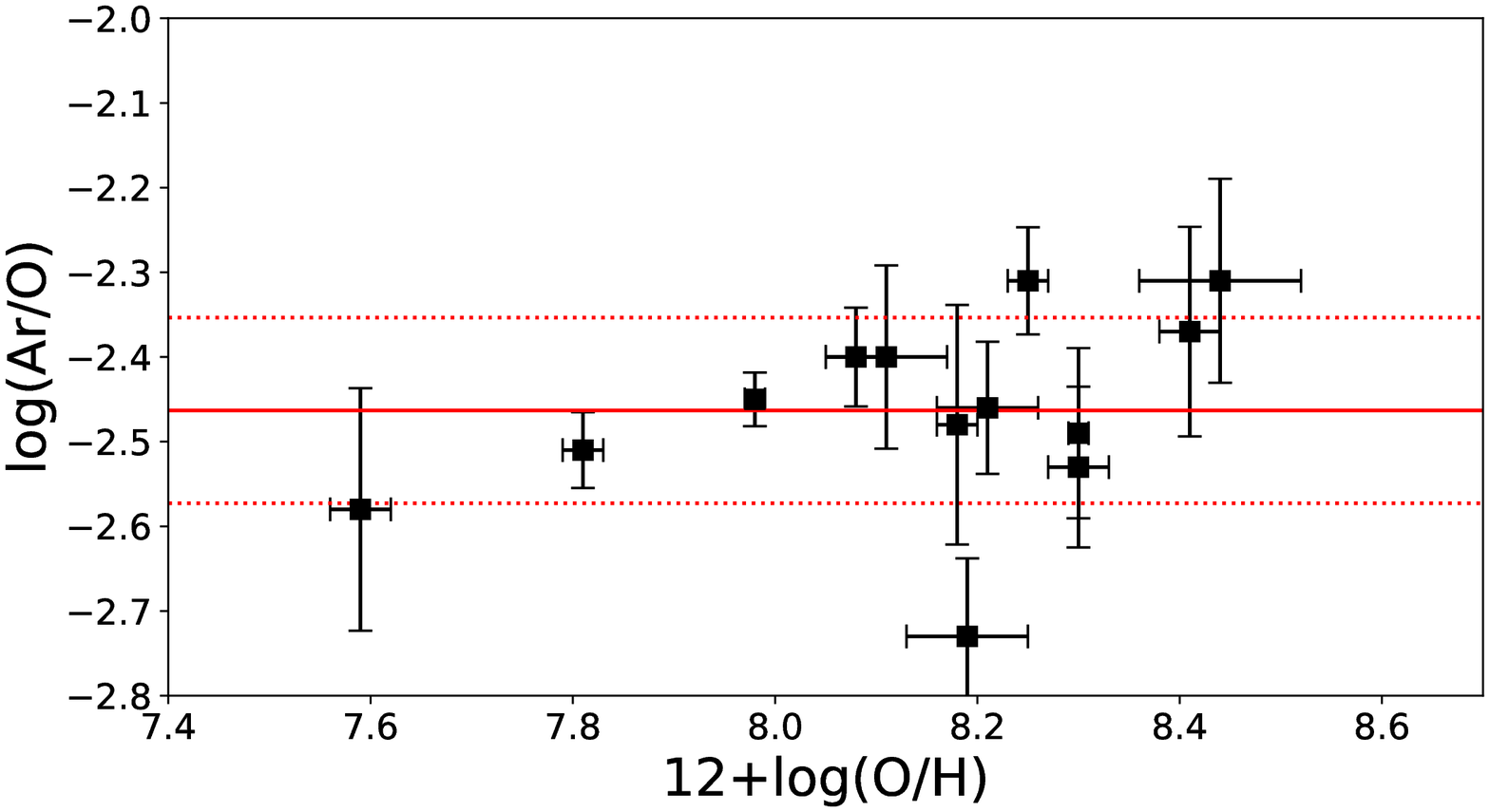} 
 \caption{Trends in the abundance ratios relative to oxygen as a function of 12+log(O/H) for {\hii} regions in M101. The left panels correspond to total abundances based on ionic abundances of \ionic{S}{2+},  \ionic{Cl}{2+} or  \ionic{Ar}{2+} determined using {\elect}(high). The right panels correspond to total abundances based on abundances of twice-ionized atoms following the prescriptions by \citet{dominguezguzmanetal19}: using {\elect}({\fnii}) -- or {\elect}(low) -- for determining \ionic{S}{2+} and the mean of {\elect}({\fnii}) and {\elect}({\foiii}) -- or {\elect}(low) and {\elect}(high) -- for 
 \ionic{Cl}{2+} and \ionic{Ar}{2+}. The continuous and dotted red lines in the right panels represent the weighted mean and standard deviation of the values, respectively.} 
 \label{fig:M101_XO} 
 \end{figure*} 

\subsection{Carbon} 
\label{sec:M101_C} 

 The radial C abundance gradient of M101 was studied by \citet{estebanetal09}, who determined the C/H ratio for H1013, NGC~5461 and NGC~5447 and use the previous -- but rather uncertain -- determination for NGC~5471 obtained by \citet{estebanetal02}. In this paper, we present C abundances for three new objects: NGC~5461, NGC~5455 and H37 as well as better determinations based on higher S/N ratio detections of the 
 {\cii} 4267 \AA\ line for NGC~5447 and NGC~5471. In total, we have C/H ratios for seven {\hii} regions of M101. We have recalculated the C abundances of H1013 and NGC~5461 following the methodology indicated in Sect.~\ref{sec:abund} using  the three ICF(\ionic{C}{2+}) schemes available in the literature. We obtain 8.58 $\pm$ 0.11 for H1013 and 8.21 $\pm$ 0.16 for NGC~5461. The least-squares linear fit to the $R$/$R_\mathrm{25}$ ratio of the objects and their C abundance gives the following radial gradient:
 	\begin{equation} \label{eq:3} 12 + \log(\mathrm{C/H}) = 8.70(\pm 0.10) - 1.19(\pm 0.19) R/R_\mathrm{25}; \end{equation}
\noindent whose slope is fairly similar to but more precise than the value of $-$1.32 $\pm$ 0.33 obtained by \citet{estebanetal09}. Fig.~\ref{fig:M101_Cgrad} shows the radial distribution of the C/H and C/O ratios. Eq.~\ref{eq:4} gives  the least-squares linear fit to the $R$/$R_\mathrm{25}$ and C/O ratios of the objects:
	 \begin{equation} \label{eq:4} \log(\mathrm{C/O}) = -0.16(\pm 0.24) - 0.40(\pm 0.46) R/R_\mathrm{25}. \end{equation}
	 Note that the O abundances used to derive the C/O ratios are those obtained from {\oii} RLs. Other spiral galaxies also show negative slopes of their C/O gradient \citep[e.g.][]{toribiosanciprianoetal16} reflecting the non-primary behavior of C enrichment. 

\subsection{Other elements} 
\label{sec:M101_other} 

We have also determined total abundances of N, Ne, S, Cl and Ar but their exact values rely on the ICFs assumed for each element. 

N/O is a very interesting abundance ratio because the bulk of both elements are produced by different stellar progenitors with different enrichment timescales. We have determined this abundance ratio using the O abundance calculated from CELs. Fig.~\ref{fig:M101_NO} shows the N/O ratio as a function of the O abundance for the sample of {\hii} regions of M101. We see an increase of N/O with 
O/H  when 12+log(O/H) $>$ 8.0 and a constant value at lower metallicities. This is a well-known behaviour, which  is related to the primary/secondary nature of N production and the anticorrelation between  the O yield and metallicity. In the flat regime, N is primary in origin and simply increases in lockstep with O. At higher metallicities, the combination of two effects produces the upturn in the N/O ratio. Firstly, the increase of the secondary production of N and, secondly, the decrease of the production of O by massive stars at higher metallicities \citep{henryetal00}. The  linear least-squares fits of the data represented in Fig.~\ref{fig:M101_NO} with 12 + log(O/H) $\geq$ 8.0 gives a slope of 1.24 $\pm$ 0.30; consistent with the value of $\sim$ 1.3 obtained by \citet{croxalletal16}. Our data give a mean value of log(N/O) = $-$1.29 $\pm$ 0.08 for the four objects with the lowest O/H ratios. We obtain the same value when assuming the standard approximation of N/O $\simeq$ \ionic{N}{+}/\ionic{O}{+} \citep{peimbertcostero69} instead of the ICF scheme by \citet{izotovetal06} for deriving the total N abundance. \citet{croxalletal16} obtain log(N/O) = $-$1.43 $\pm$ 0.11 for the plateau of the N/O {\it versus} O/H distribution. 

Fig.~\ref{fig:M101_NeO} shows the Ne/O ratio {\it versus} O abundance for the sample of {\hii} regions of M101. We see that the Ne/O trend is basically flat, with a mean value of log(Ne/O) = $-$0.84 $\pm$ 0.15. 
There are two objects with log(Ne/O) below $-$1.0 but with relatively large O/H ratios, H219 and H1013. In fact, H219 is outside the band defined by one standard deviation around the weighted mean of the Ne/O ratios of the rest of the objects. Those two nebulae might be part of the apparent low-Ne/O population of {\hii} regions found by \citet{croxalletal16} (see their 
figures 6 and 9). Those authors claim that such population is statistically confirmed and does not display unusual deficits in other abundance ratios such as N/O, S/O, or Ar/O, and this seems also to be the case according to our data (see Fig.~\ref{fig:M101_XO}). However, H219 and H1013 are precisely the nebulae of our sample showing the lowest \ionic{O}{2+}/\ionic{O}{+} ratios. Therefore, the apparent two populations in the Ne/O {\it versus} O/H diagram might be an artifact produced by the ICF schemes at low ionization degrees. The suitability of this explanation will be discussed in Arellano-C\'ordoba  et al. (in preparation). 

Fig.~\ref{fig:M101_XO} shows S/O, Cl/O and Ar/O ratios as a function of O/H. Stellar nucleosynthesis predicts that these distributions should be flat, as it has been corroborated in different 
observational works. In the left panels of Fig.~\ref{fig:M101_XO} we represent the S, Cl and Ar abundances determined using {\elect}(high) for deriving \ionic{S}{2+}, \ionic{Cl}{2+} and  \ionic{Ar}{2+}. This is the standard recipe when assuming the two-zone scheme for calculating ionic abundances. However, we can see that all the three abundance 
ratios show rather clear trends with O/H. This behaviour was not noted in the case of the Ne/O ratio (see Fig.~\ref{fig:M101_NeO}) and also when representing S/Ar or Cl/Ar {\it versus} O/H. 
We made the same plots using the ICF schemes by \citet{medina19} (see 
Section~\ref{sec:abund}) but the trends did not disappear. Although \citet{croxalletal16} claim that the distributions of S/O and Ar/O ratios with respect to O/H are flat, we think that their figures 6 and 7 might 
suggest some trend fairly similar to that shown in the left panels of our Fig.~\ref{fig:M101_XO}. \citet{kennicuttetal03} also claim the absence of any trend in their S/O {\it versus}  O/H plot (their figure 9) 
but again, this flatness is arguable to us. In a very recent work \citet{dominguezguzmanetal19} report exactly the same behaviour we find for the S/O, Cl/O and Ar/O (not in Ne/O) ratios, but for a sample of 37 Galactic and extragalactic {\hii} regions with high-quality spectra. \citet{dominguezguzmanetal19} propose that using {\elect}({\fnii}) instead of {\elect}(high) for 
determining \ionic{S}{2+} abundance and the mean of  {\elect}({\fnii}) and  {\elect}({\foiii}) instead of {\elect}(high) for \ionic{Cl}{2+} and  \ionic{Ar}{2+} the abundance ratios distributions become flatter and show a lower dispersion for a given O abundance. The results we obtain following \citet{dominguezguzmanetal19} can be seen in the right panels of Fig.~\ref{fig:M101_XO}. The S, Cl and Sr abundances included in tables~\ref{tab:abund1} and \ref{tab:abund2} are those obtained following these prescriptions. The average abundance ratios we obtain following these criteria are log(S/O) = $-$1.59$\pm$ 0.09, log(Cl/O) = $-$3.52 $\pm$ 0.10, log(Ar/O) = $-$2.46 $\pm$ 0.11.  It must be highlighted that, as we can see in Fig.~\ref{fig:M101_XO},  all the objects (except one in the Ar/O as a function of O/H diagram) are consistent within the errors with the band defined by one standard deviation around the weighted mean of the abundance ratios. The physical reason for the prescriptions proposed by \citet{dominguezguzmanetal19} does not seem to be related solely to the ionization potential of the ionic species involved. Perhaps the different shapes of the photoionization cross-sections   may be playing a role but, in any case this should be further explored with the help of photoionization models. 

\section{Radial abundance gradients in M31} 
\label{sec:M31} 

\begin{figure} 
\centering 
\includegraphics[scale=0.37]{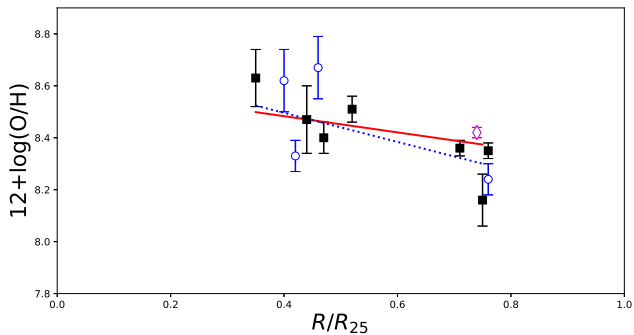} 
 \caption{Radial distribution of the O abundances calculated from CELs for {\hii} regions in M31 as a function of their fractional galactocentric distance ($R$/$R_\mathrm{25}$). Black squares represent our data; blue empty circles represent objects observed by \citet{zuritabresolin12}; the magenta empty diamond corresponds to K932, observed by \citet{estebanetal09}. The solid red line represents the least-squares fit to all the objects. The dotted blue line corresponds to the least-squares fit obtained by \citet{zuritabresolin12}.} 
 \label{fig:M31_Ograd} 
 \end{figure} 

\begin{figure} 
\centering 
\includegraphics[scale=0.37]{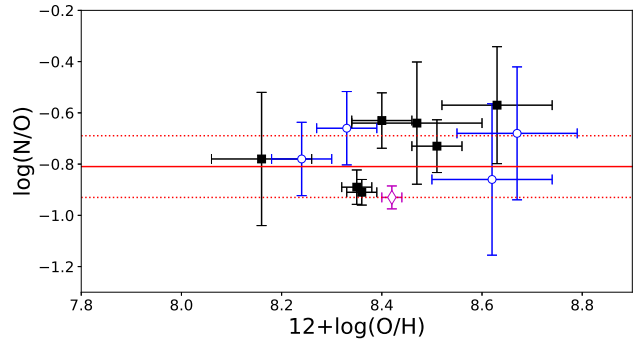} 
 \caption{Trend of the N/O ratio as a function of 12+log(O/H) for {\hii} regions in M31. The continuous and dotted red lines represent the weighted mean and standard dispersion of the values, respectively. Symbols are the same as in Fig~\ref{fig:M31_Ograd}. } 
 \label{fig:M31_NO} 
 \end{figure} 

\begin{figure*} 
\centering 
\includegraphics[scale=0.37]{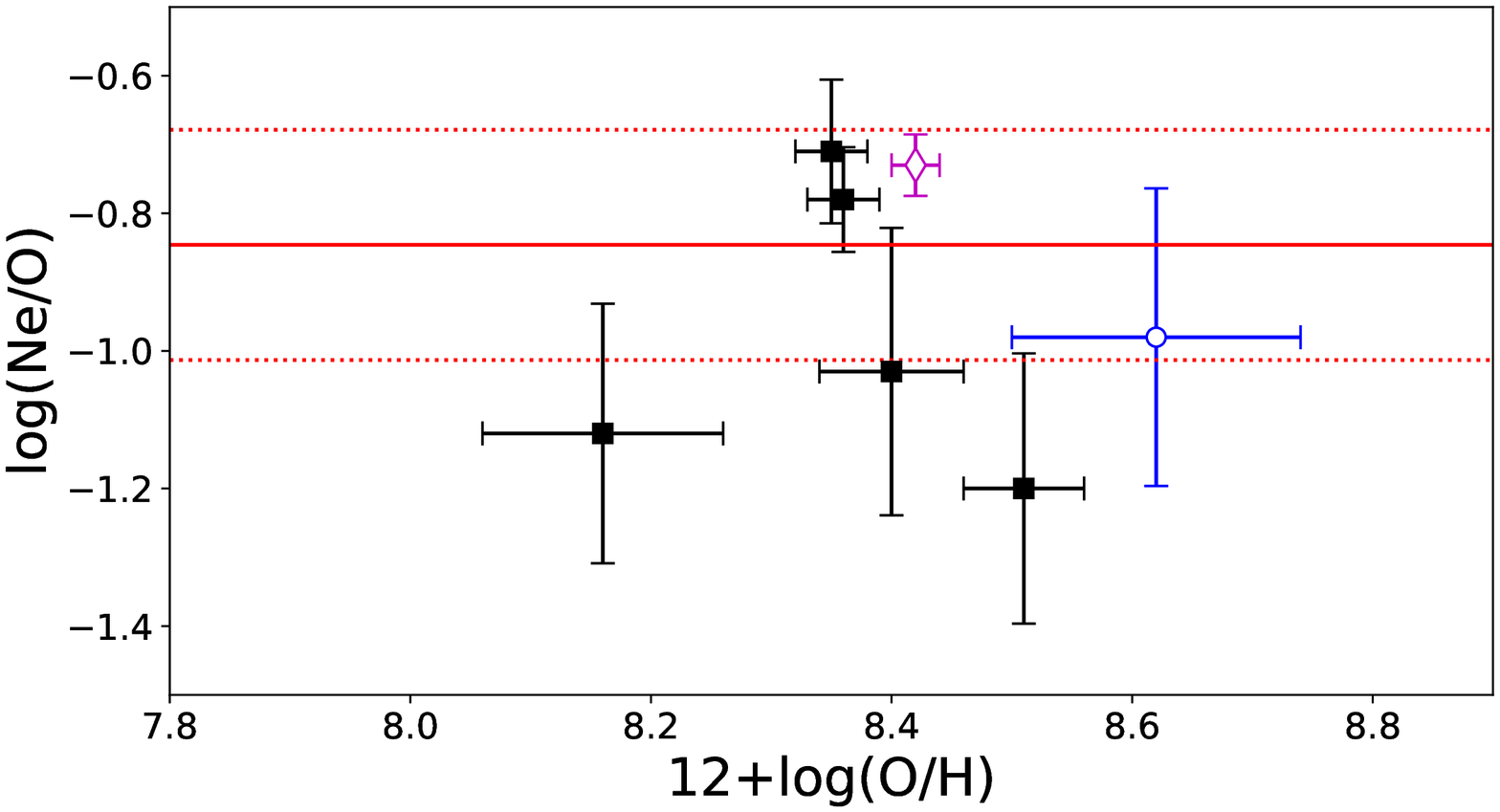} 
\includegraphics[scale=0.37]{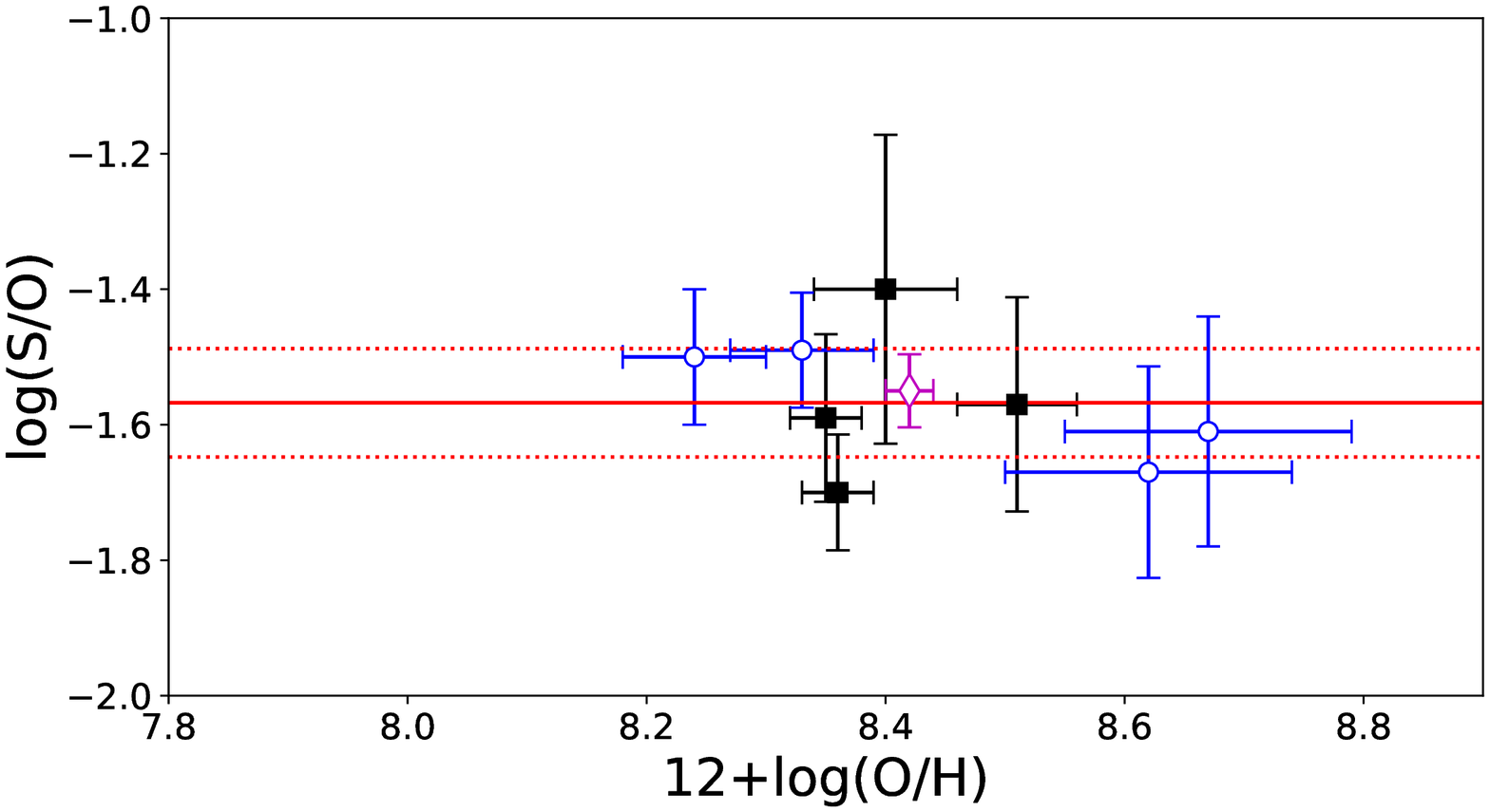} 
\\
\includegraphics[scale=0.37]{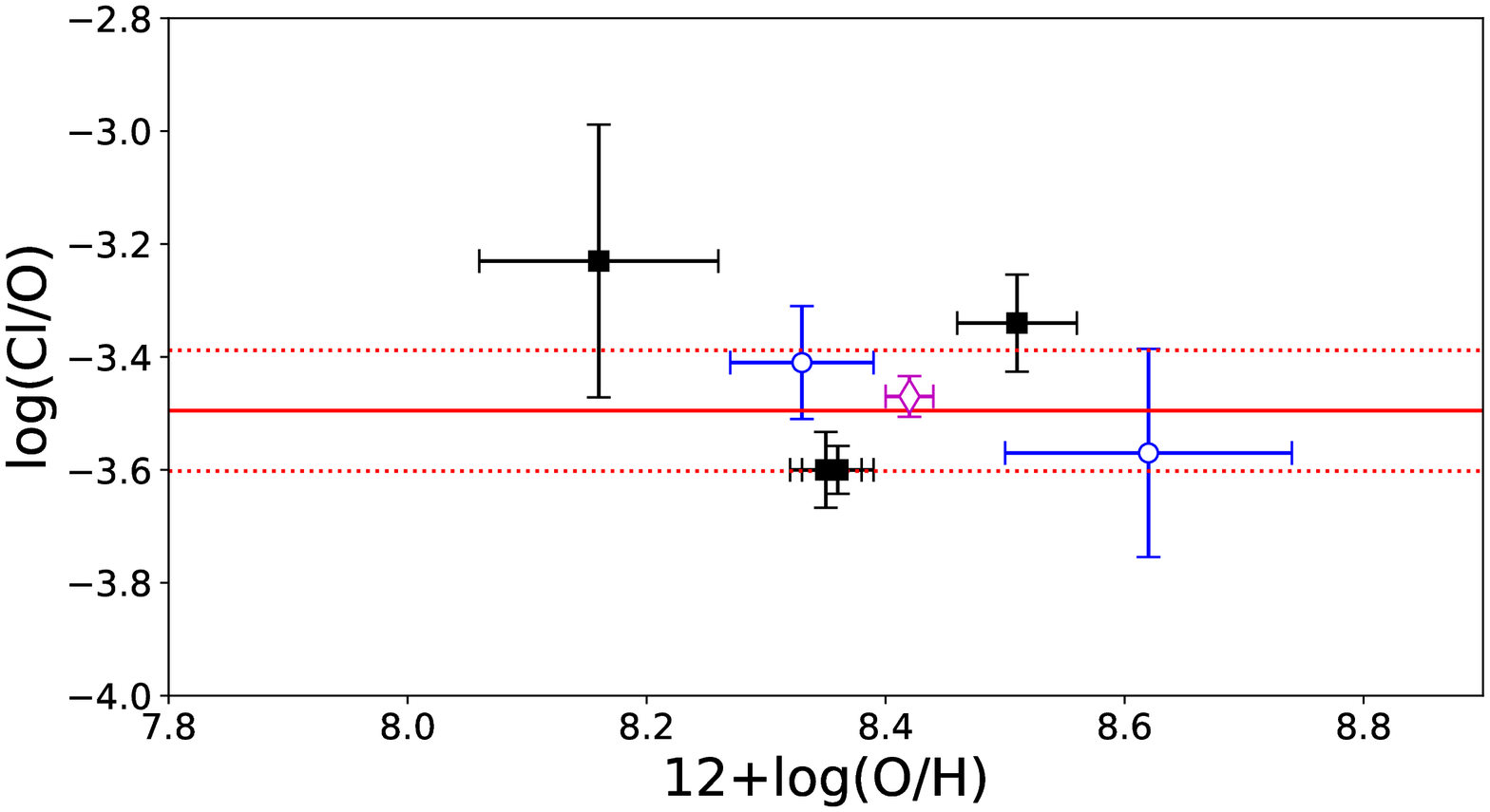} 
\includegraphics[scale=0.37]{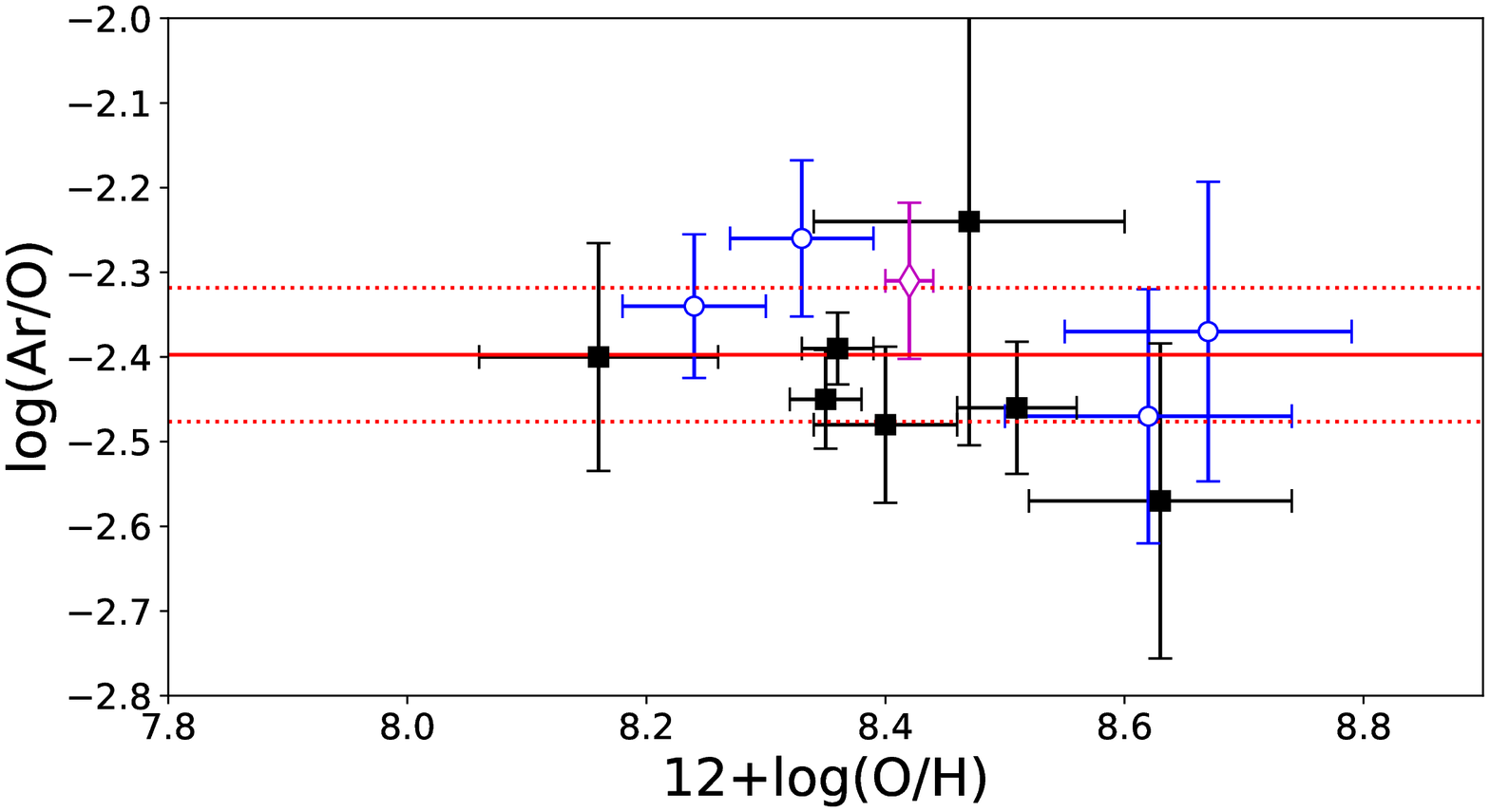} 
 \caption{Trends of abundance ratios relative to O as a function of 12+log(O/H) for {\hii} regions in M31. The total abundances are based on abundances of twice-ionized atoms following the prescriptions by \citet{dominguezguzmanetal19} (see text for details). The continuous and dotted red lines represent the weighted mean and standard deviation of the values, respectively. Symbols are the same as in Fig~\ref{fig:M31_Ograd}.}
 \label{fig:M31_XO} 
 \end{figure*} 

We have obtained spectra of seven {\hii} regions of M31. Four of them, BA310, BA371, BA374 and BA379 were also observed by \citet{zuritabresolin12}. The mean difference between the O abundance determined by us and by \citet{zuritabresolin12} for those objects in common is only 0.01 $\pm$ 0.05 dex. In Fig.~\ref{fig:M31_Ograd} we show the radial distribution of the O abundances -- calculated from CELs -- of our sample {\hii} regions in M31 as a 
function of their $R$/$R_\mathrm{25}$ ratio. We have also included data for BA312, BA362, BA419 and BA422 observed by \citet{zuritabresolin12} and K932 \citep[observed by][]{estebanetal09} for which we have recalculated the physical conditions and abundances following the same methodology and atomic data as for the rest of the objects of this paper. With these 12 {\hii} regions with direct determinations of {\elect}, we obtain the radial O abundance gradient given below:
	\begin{equation} \label{eq:6} 12 + \log(\mathrm{O/H})_\mathrm{CELs} = 8.61(\pm 0.15) - 0.31(\pm 0.21) R/R_\mathrm{25}; \end{equation} 
\noindent which is much flatter than the O gradient determined for M101 (see Eq.~\ref{eq:1}).  For comparison, in the figure, we have also included the least-squares fit obtained by \citet{zuritabresolin12}, whose slope and intercept are $-$0.56 $\pm$ 0.28 dex ($R$/$R_\mathrm{25}$)$^{-1}$ and 8.72 $\pm$ 0.18, respectively. Unfortunately, as in the case of \citet{zuritabresolin12} our observational points cover only two distinct quite narrow zones of galactocentric distances \citep[these zones can be easily seen in the combined GALEX+{\it Spitzer} image shown by][their figure 2]{zuritabresolin12}. It is clear 
that further {\elect}-based abundance determinations at different distances from the M31 center would be desirable, but the lack of bright {\hii} regions outside those star-forming annuli in the disc of M31 is a drawback for such a goal. The scatter of the O abundance of {\hii} regions represented in Fig.~\ref{fig:M31_Ograd} and the linear fit at their corresponding distance is $\pm$ 0.10 dex, larger than in the case of M101. We conclude that the presence of chemical inhomogeneities in the disc of M31 is more likely than in the case of M101. However, we have to take into account the fact that the mean uncertainty of the individual O/H determinations for the {\hii} regions of M31 is about 0.07 dex -- larger than in the case of M101 -- and the high inclination angle of M31 introduces an additional uncertainty in the estimation of the galactocentric distance of the {\hii} regions. 

Unfortunately, we did not detect {\oii} RLs in any of the {\hii} regions observed in M31. The only region where these lines have been reported is K932 \citep{estebanetal09}. However, as shown in Fig.~\ref{fig:CII_lines} and Table A5 the {\cii} 4267 \AA\ line was detected in K160. We found a rather high C/H abundance in this object (see Table~\ref{tab:abund2}). Although there are only two {\hii} regions with C/H ratio determinations available in M31, we can estimate a tentative radial gradient for 12+log(C/H), finding a slope of $-$2.00 $\pm$ 0.77 dex ($R$/$R_\mathrm{25}$)$^{-1}$, a much steeper -- but also far more uncertain -- C gradient than in the case of M101. 

In Fig.~\ref{fig:M31_NO} we plot the N/O ratio as a function of the O abundance for the {\hii} regions of M31. In contrast to what was found for M101 (Fig.~\ref{fig:M101_NO}), the N/O ratio in M31 shows a rather flat distribution with respect to O/H, 
with a mean log(N/O) = $-$0.81 $\pm$ 0.12.. This behaviour was also reported by \citet{zuritabresolin12} and will be further discussed in Sect.~\ref{sec:C_N_enrichment}. 

In Fig.~\ref{fig:M31_XO} we show the total abundances of Ne, S, Cl and Ar with respect to O as a function of the oxygen abundance for the {\hii} regions of M31. As in the case of M101, we have followed the prescriptions 
by \citet{dominguezguzmanetal19} for determining the total abundances. We have used {\elect}(high) for calculating \ionic{Ne}{2+}/\ionic{H}{+}, {\elect}({\fnii}) -- or {\elect}(low) -- for \ionic{Cl}{2+}/\ionic{H}{+} and the mean of  {\elect}({\fnii})  and  {\elect}({\foiii})  -- or {\elect}(low) and/or {\elect}(high) -- 
for \ionic{S}{2+}/\ionic{H}{+} and \ionic{Ar}{2+}/\ionic{H}{+}. The abundance ratios do not show clear trends and the average values -- as well as the dispersions --, log(Ne/O) = $-$0.85 $\pm$ 0.17, log(S/O) = $-$1.57 $\pm$ 0.08, log(Cl/O) = $-$3.50 $\pm$ 0.11 and log(Ar/O) = $-$2.40 $\pm$ 0.08 are almost identical to those found for M101. 

\section{Discussion} 
\label{sec:discussion} 

\subsection{About the normalization of radial gradients} 
\label{sec:normalization} 

\begin{table*} 
\caption{Comparison of O, C and N gradients for M101, M31 and other spiral galaxies.} 
\centering
\label{tab: comp_grad} 
\begin{tabular}{lccccccccccccc} 
\hline 
 &  &   &   &   &   \multicolumn{4}{c}{slope (dex ($R$/$R_{25})^{-1}$)} &  \multicolumn{4}{c}{slope (dex ($R$/$R_{e})^{-1}$)} & \\
 & Morphological & $R_{25}$ &  $R_{\mathrm e}$ &  & \multicolumn{2}{c}{O/H} & C/H & N/H & \multicolumn{2}{c}{O/H} & C/H & N/H & \\
Galaxy &  type & (kpc) & (kpc) & $M_{\mathrm V}$ & (CELs) & (RLs) & (RLs) & (CELs) & (CELs) & (RLs) & (RLs) & (CELs) & Reference \\
\hline                                                                                                                   
NGC~300 & Sc & 5.33 & 2.5 & $-$18.99 & $-$0.30 & $-$0.14 & $-$0.43 & $-$0.83 & $-$0.14 & $-$0.07 & $-$0.20 & $-$0.39 & 1 \\
M33 & SAcd & 6.85 & 3.7 & $-$19.41 & $-$0.36 & $-$0.33 & $-$0.61 & $-$0.82 & $-$0.19 & $-$0.18 & $-$0.33 & $-$0.44 & 1 \\
Milky Way & SBbc & 11.25 & 4.55 & $-$20.90 & $-$0.46 & $-$0.41 & $-$0.79 & $-$0.74 & $-$0.19 & $-$0.17 & $-$0.32 & $-$0.30 & 2  \\
M101 & SABc & 31.0 & 8.6 & $-$21.36 & $-$0.86 & $-$0.75 & $-$1.19 & $-$1.41 & $-$0.24 & $-$0.21 & $-$0.33 & $-$0.39 & 3 \\
M31 & SAb & 21.75 & 9.4 & $-$21.78 & $-$0.31 & $-$ & $-$2.00 & $-$1.15 & $-$0.13& $-$ & $-$0.86 & $-$0.49 &3 \\
\hline           
           \multicolumn{14}{l}{1 --  \citet{toribiosanciprianoetal16}.} \\
           \multicolumn{14}{l}{2 --  \citet{estebangarciarojas18}.} \\
           \multicolumn{14}{l}{3 --  This work.} \\
  
  \end{tabular} 
\end{table*}

\begin{figure*} 
\centering 
\includegraphics[scale=0.37]{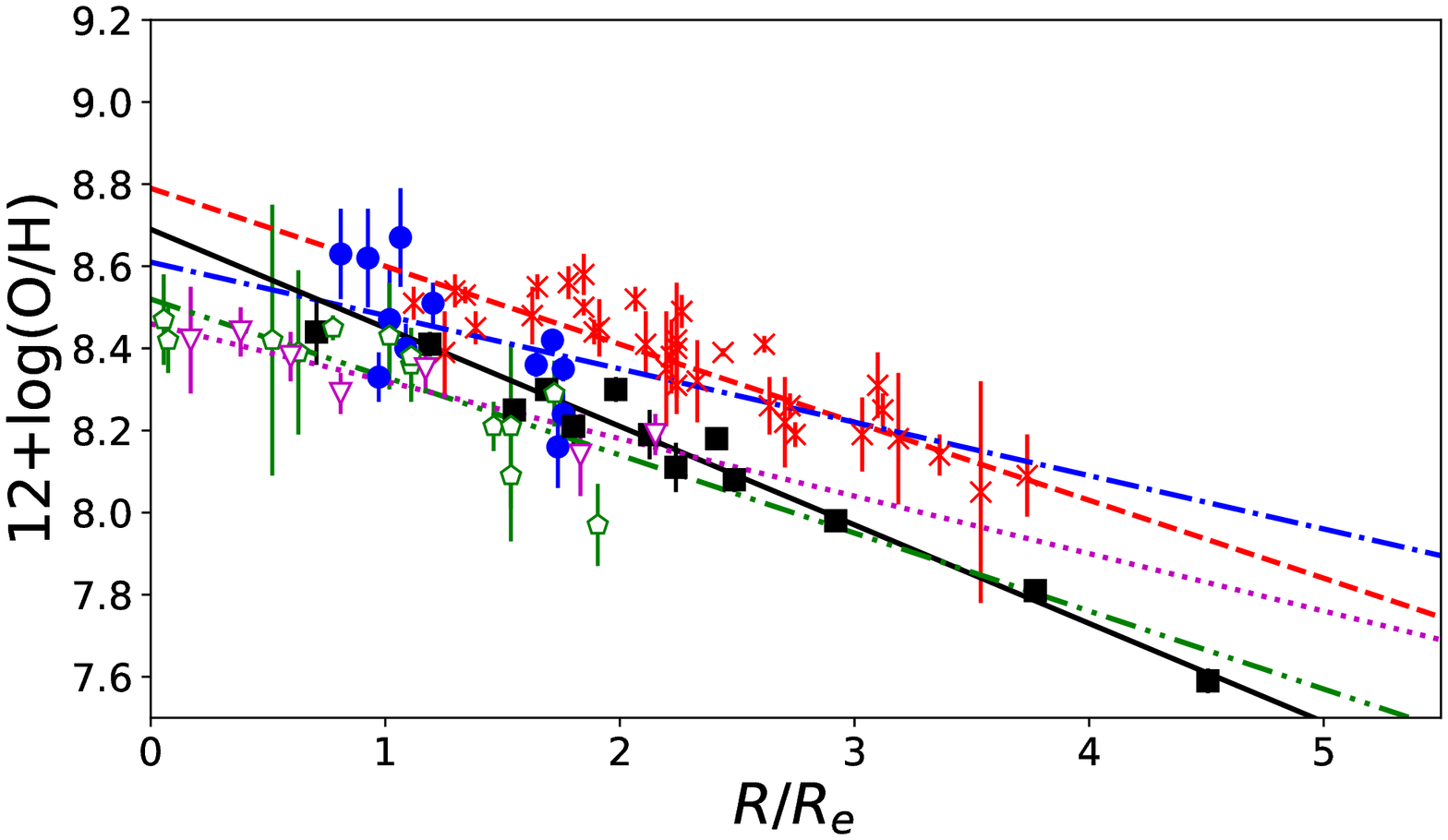} 
\includegraphics[scale=0.37]{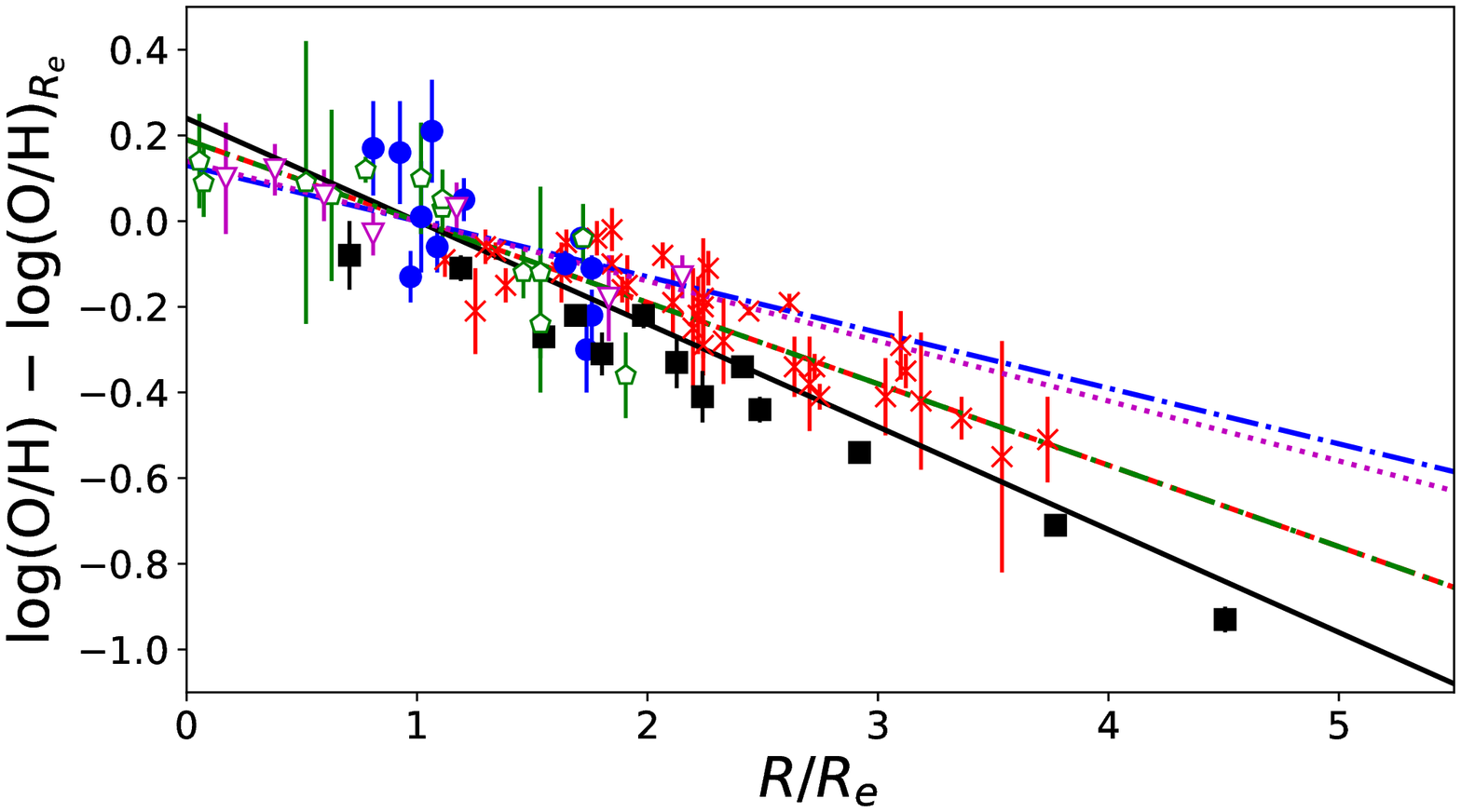} 
\\
\includegraphics[scale=0.37]{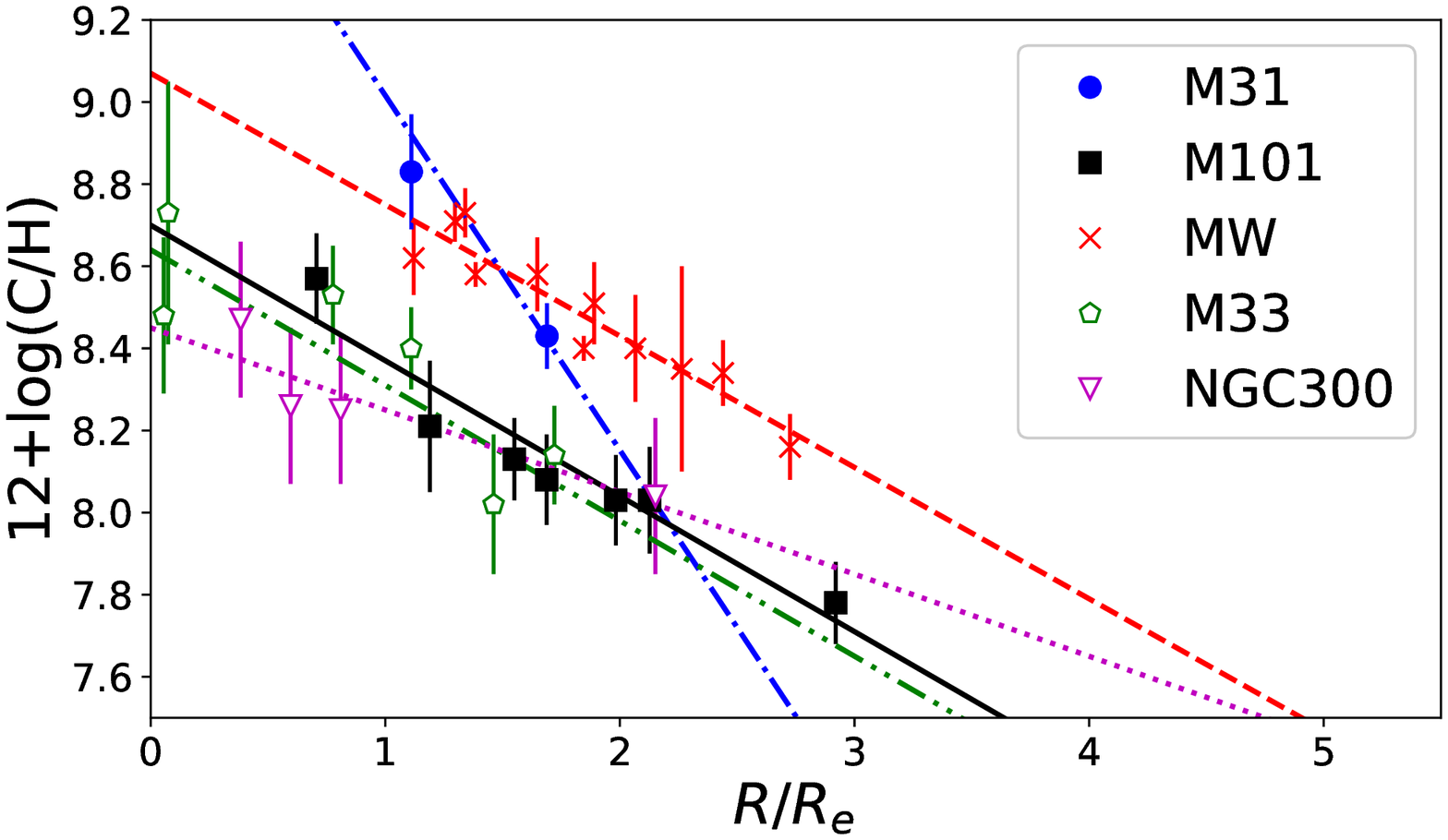} 
\includegraphics[scale=0.37]{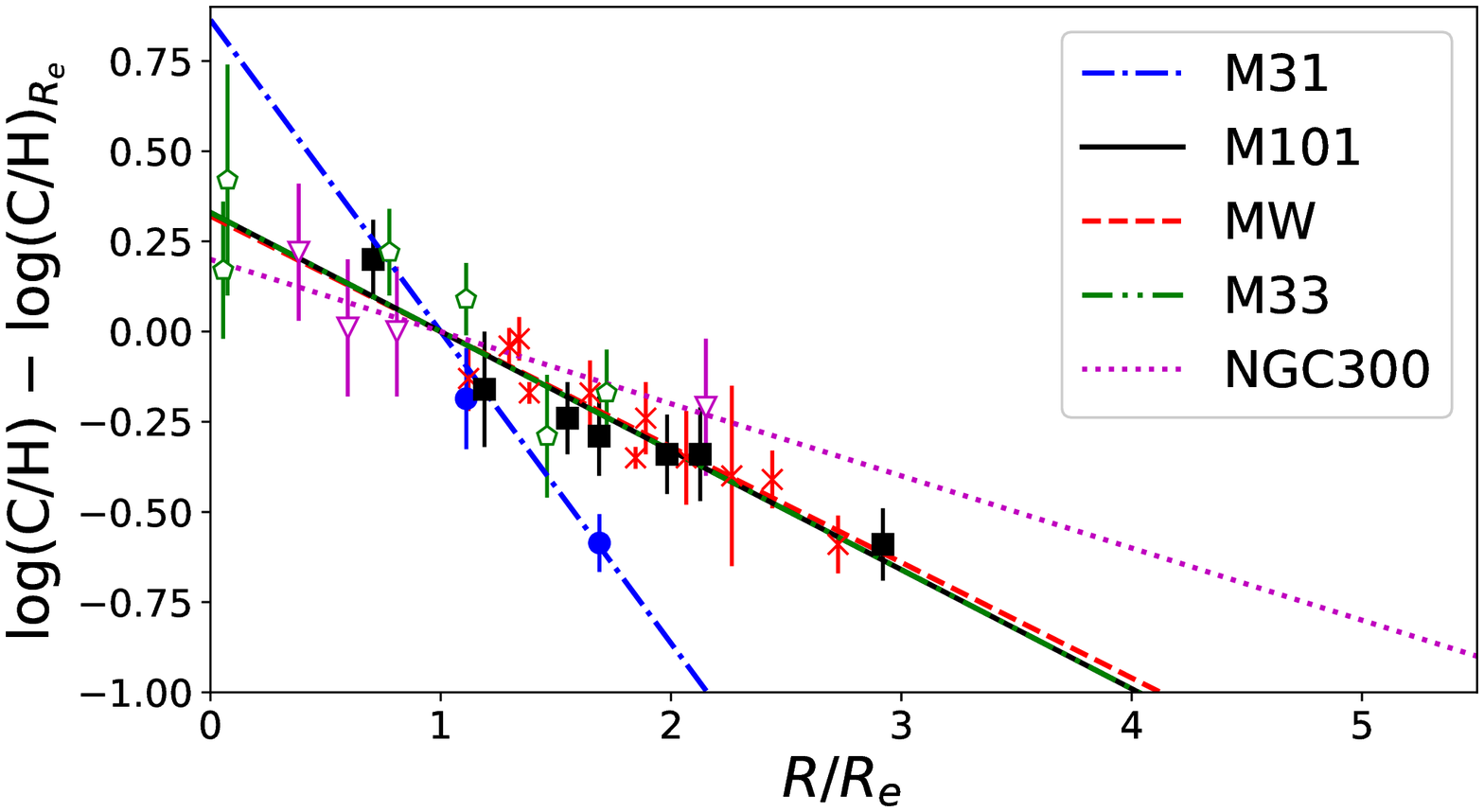}
\\
\includegraphics[scale=0.37]{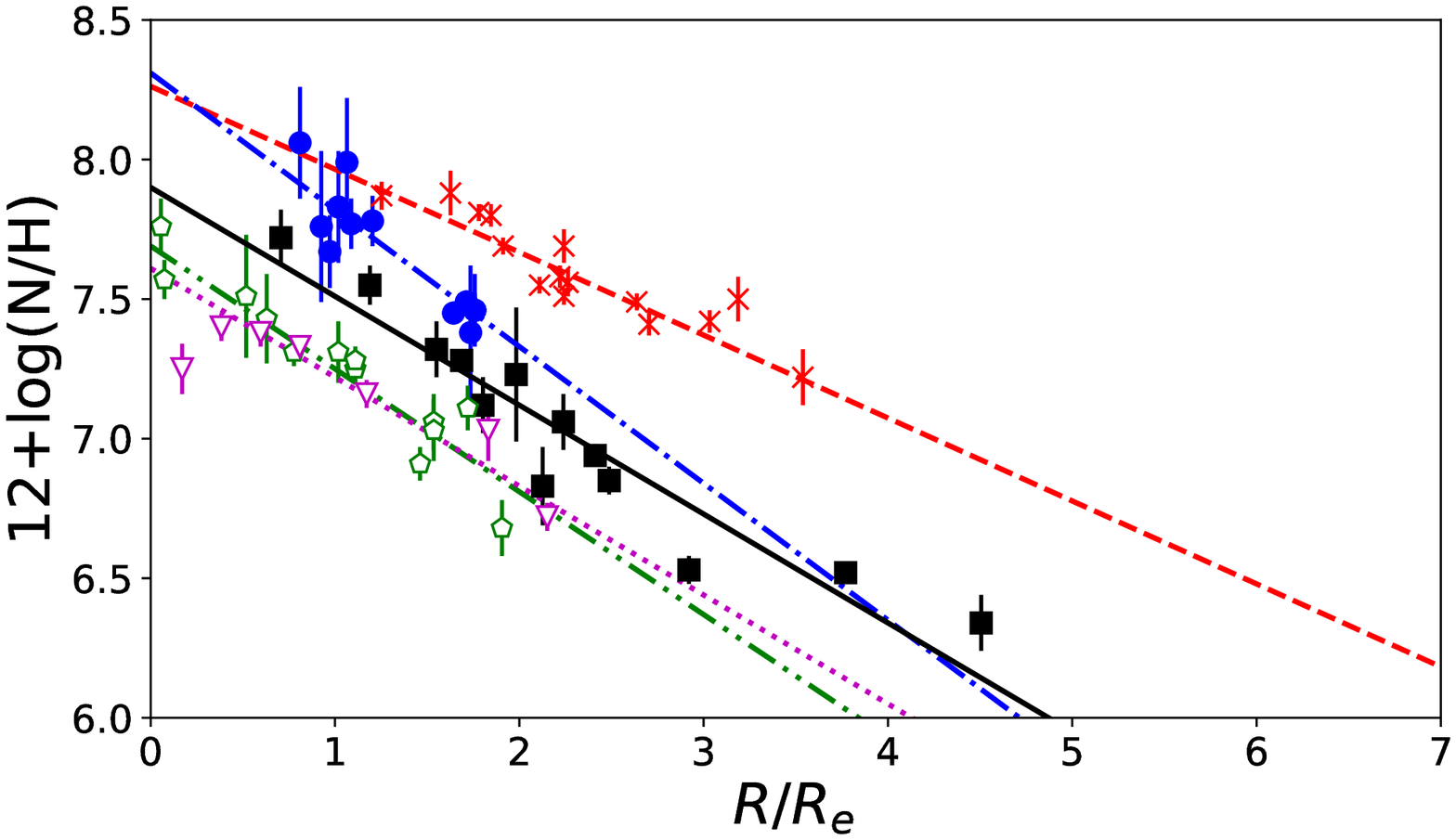} 
\includegraphics[scale=0.37]{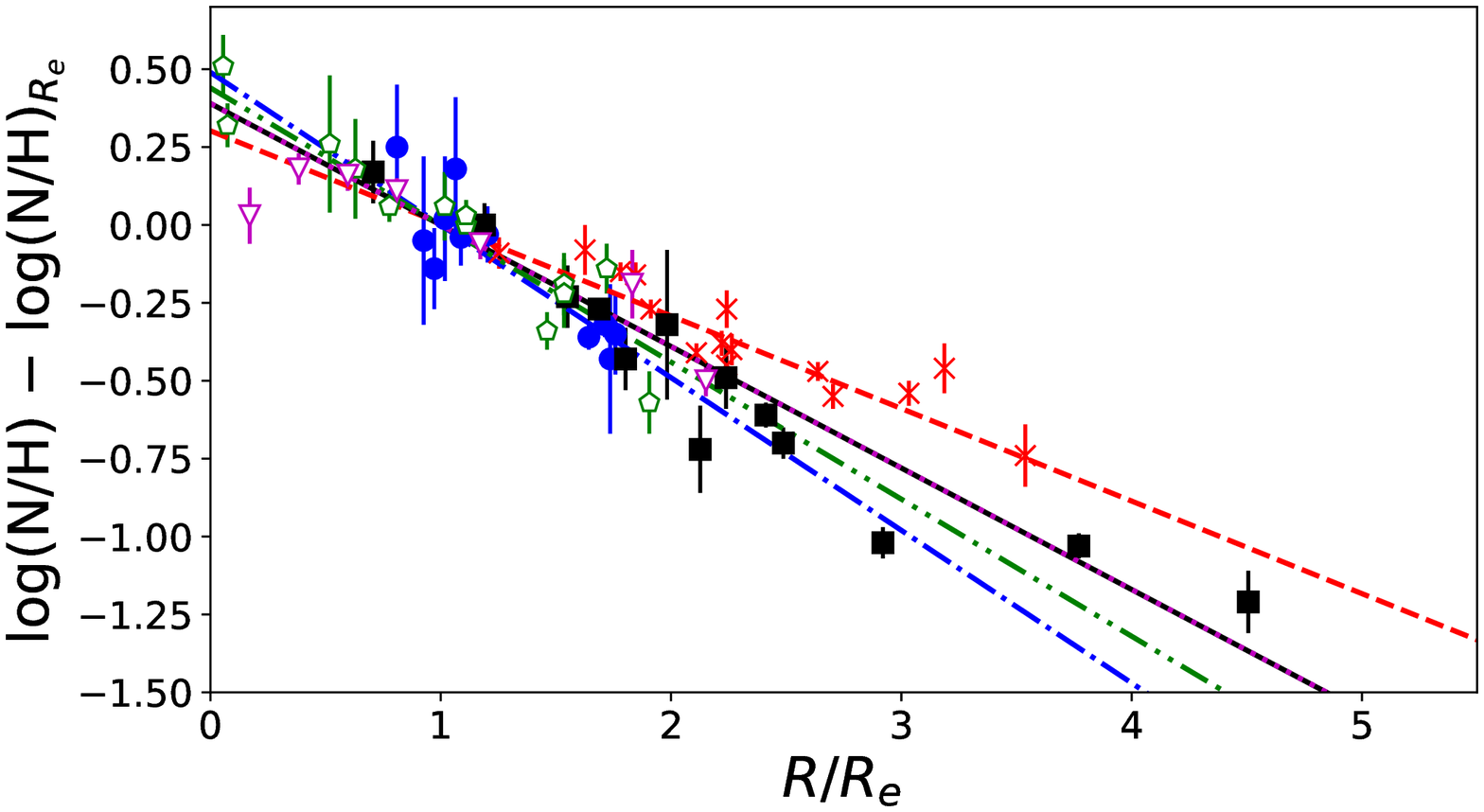} 
 \caption{From top to bottom: radial distribution of O, C, and N abundances as a function of their fractional galactocentric distance with respect to the disc effective radius ($R/R_\mathrm{e}$) for {\hii} regions in M31 (blue circles and dotted-dashed 
 line, this work); M101 (black squares and solid line, this work); Milky Way  \citep[red crosses and dashed line,][]{estebangarciarojas18}; 
 M33 \citep[green double-dotted and dashed line and empty pentagons,][]{toribiosanciprianoetal16}; and NGC~300 \citep[magenta triangles and dotted line,][]{toribiosanciprianoetal16}. The right panels represent the same but with O abundances normalized to its value at the corresponding $R_\mathrm{e}$ of each galaxy.} 
 \label{fig:5gals_mosaic} 
 \end{figure*} 

 \citet{toribiosanciprianoetal16} found that the radial C/H abundance gradients are steeper than those of O/H in spiral galaxies. In addition, they found a correlation between the slope of the C/H gradient 
and the galactic absolute magnitude $M_{\mathrm V}$ when the abundance is represented as a function of $R$/$R_{25}$. In Table~\ref{tab: comp_grad} we show the slope of the O, C and N gradients with respect to $R/R_\mathrm{25}$ and  $R/R_\mathrm{e}$ for M101 and 
M31, the Milky Way \citep{estebangarciarojas18} and other nearby spiral galaxies: M33 and NGC~300 \citep{toribiosanciprianoetal16}. In the table, we also give the morphological type, $R_\mathrm{25}$, the effective radius  ($R_{\mathrm e}$) and the absolute $V$-band magnitude ($M_{\mathrm V}$) of the galaxies. The values of 
$R_{\mathrm e}$ have been taken from Zurita, Florido  et al. (in preparation), who determine it from the radial scale length of the disc ($R_{\mathrm d}$) using the relation: 
$R_{\mathrm e}$ = 1.678 $\times$ $R_{\mathrm d}$ \citep[e.g.][]{sanchezetal14} valid under the assumption of an exponential disc. In the case of M33, M101 and NGC~300, we have used the $R_{\mathrm d}$ values calculated by Zurita, Florido  {\it et al.} from the fitting of the $r$-band surface brightness profiles. Those authors recommend to use the value of $R_{\mathrm d}$  obtained  by \citet{licquiaetal16} for the Milky Way and the one obtained by \citet{courteauetal11} from the $I$-band light profile for M31. 
The comparison between radial abundance gradients of different galaxies has been done normalizing the radial galactocentric distance to either $R_{25}$ or $R_{\mathrm e}$. In most published works, the normalization parameter is  $R_{25}$  \citep[e.g.][] {rosalesortegaetal11, croxalletal16, toribiosanciprianoetal16}, not only for the abundance gradients but also to study the radial profiles of other main galaxy properties as, for example, dust, gas, stars or star formation rate 
\citep[e.g.][]{casasolaetal17}.
\citet{diaz89} argued that $R_{\mathrm e}$ is the most appropriate parameter for normalizing the abundance gradients because it is related to the mass concentration and reflects the recycling rate of the gas in different galaxies 
\citep[see][]{sanchezetal12}. Using the data presented in Table~\ref{tab: comp_grad} for the five galaxies, we find that the relative standard deviation  -- the ratio of the standard deviation to the mean -- of the gradient slopes becomes smaller -- almost a factor of two lower -- when  
$R_{\mathrm e}$ is used  as normalization parameter except in the case of C, for which using $R/R_\mathrm{25}$ or  $R/R_\mathrm{e}$ gives the same relative standard deviation. However, removing the slope of the C/H gradient of M31 -- which is the most uncertain one because it relies only on two objects -- the relative standard deviation  using $R_{\mathrm e}$ becomes equally smaller for all chemical elements. In the following, we will use $R_\mathrm{e}$ as our preferred normalizing parameter for the comparison of radial abundance gradients. 

Fig.~\ref{fig:5gals_mosaic} shows the radial O, C and N abundance gradients of M31, M101, the Milky Way, M33 and NGC~300 with respect to $R/R_\mathrm{e}$. In the left panel, we can see that the {\hii} regions of the Milky Way (red crosses) show  
 O, C and N abundances larger than those of the remaining galaxies for the same $R/R_\mathrm{e}$ ratio. This indicates that the Milky Way seems to have an average metallicity at $R_{\mathrm e}$ larger than the more massive and 
luminous galaxies M31 or M101, in contradiction with what we would expect from the mass-metallicity relation. This apparent metallicity excess of the Milky Way would be alleviated if the true 
$R_{\mathrm e}$ of the Milky Way is larger than the value determined from the $R_{\mathrm d}$ proposed by \citet{licquiaetal16}. In fact, a $R_{\mathrm e}$ about twice larger would reconcile the values of O, C and N with respect to $R/R_\mathrm{e}$ of the Milky Way with those of the other galaxies. Although this might seem simply an argument of convenience, a larger true $R_{\mathrm e}$ is a possibility that certainly cannot be ruled out. The different 
measurements of $R_{\mathrm d}$ over the last few decades have provided a fairly wide range of values, between 1 and 5 kpc, although the latest determinations tend towards values smaller than 2.6 kpc 
\citep[e.g.][]{sackett97, amoresetal17}. On the other hand, as \citet{licquiaetal16} have pointed out, small values of $R_{\mathrm d}$ cause the Milky Way to be located outside the three-dimensional luminosity-velocity-radius scaling relation. Values of  $R_{\mathrm d}$ about 5 kpc -- twice as large as the one we assume -- would be necessary for solving that anomaly. Finally, another remarkable feature in the left panel of Fig. ~\ref{fig:5gals_mosaic} is that the C abundances in {\hii} regions of M31 are considerably higher than in M101, M33 or NGC~300 for the same $R/R_\mathrm{e}$ ratio.

In the right panels of Fig.~\ref{fig:5gals_mosaic} we represent the abundances normalized to the ones corresponding to $R_{\mathrm e}$ in order to wash 
out the effect of the different mean metallicities of the galaxies. Using this representation, it is easier to note that the gradient slopes are in general fairly similar, especially in the cases of O and N. We see a larger dispersion for the C/H gradients, but this is mostly due to the rather steeper slope of M31, which -- as has been said before -- shows the largest uncertainty. Another fact we can note in the central right panel of  Fig.~\ref{fig:5gals_mosaic} -- and Table~\ref{tab: comp_grad} -- is the lack of correlation between the slope of C/H gradient and the $M_V$
of the galaxies in contrast with what was found by  \citet{toribiosanciprianoetal16} when using $R_{25}$ as a normalization parameter.  

\begin{figure} 
\centering 
\includegraphics[scale=0.37]{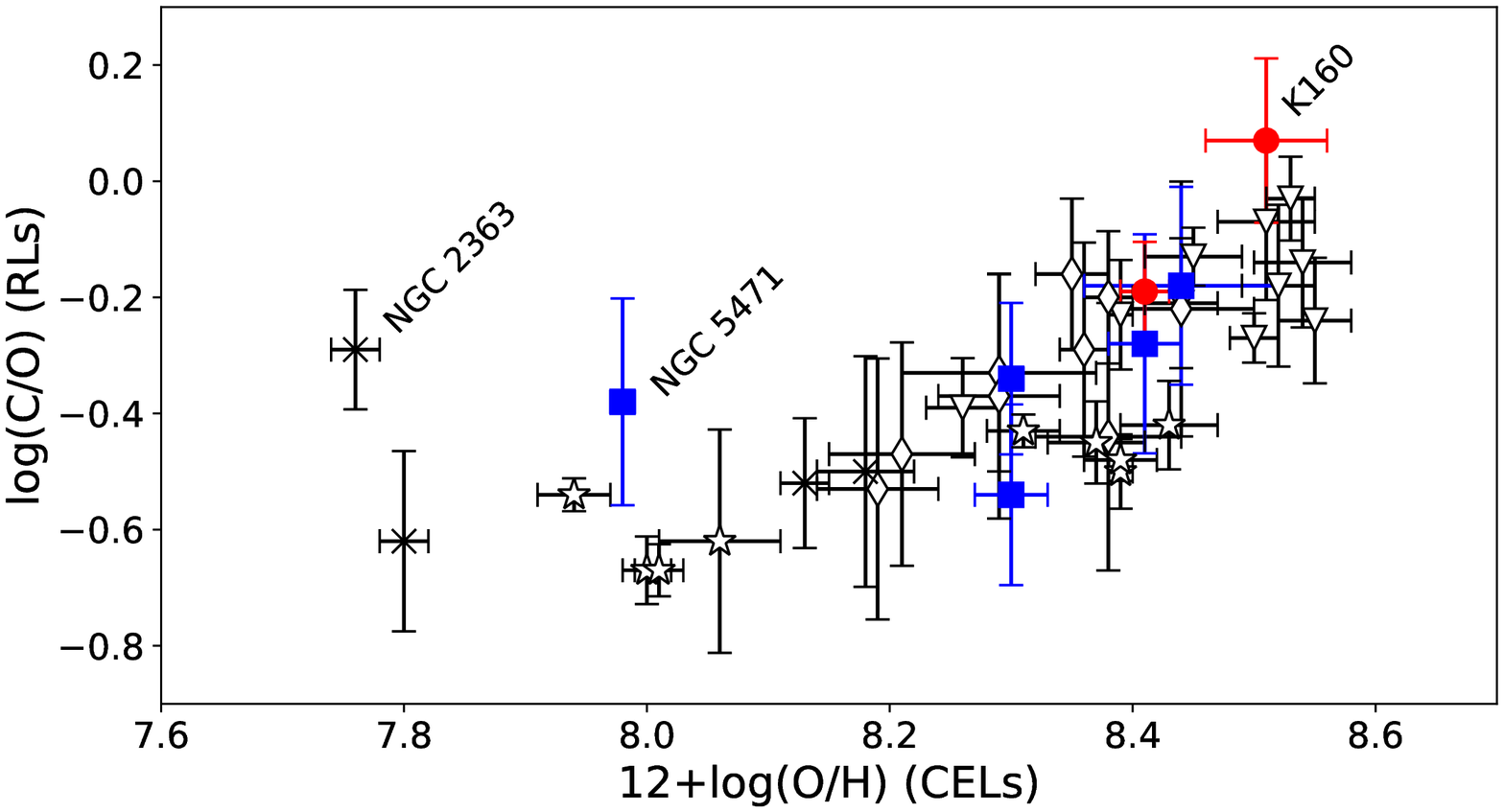} 
 \caption{Trend of log(C/O) determined from RLs as a function of 12+log(O/H) determined from CELs for {\hii} regions in M31 (red circles), M101 (blue squares) and in other spiral and dwarf galaxies. Empty triangles 
 correspond to {\hii} regions of the Milky Way \citep{garciarojasesteban07, estebanetal13, estebanetal17}; empty diamonds correspond to {\hii} regions of the spiral galaxies M33, NGC~300 
and NGC2403 \citep{toribiosanciprianoetal16, estebanetal09}; empty stars represent objects of the Magellanic Clouds \citep{toribiosanciprianoetal17} and crosses indicate the position of {\hii} galaxies or blue compact dwarves 
\citep{lopezsanchezetal07, estebanetal09, estebanetal14}. The peculiar positions of NGC~2363 and NGC~5471 are indicated (see text for details).} 
 \label{fig:CO_O} 
 \end{figure} 

\begin{figure} 
\centering 
\includegraphics[scale=0.37]{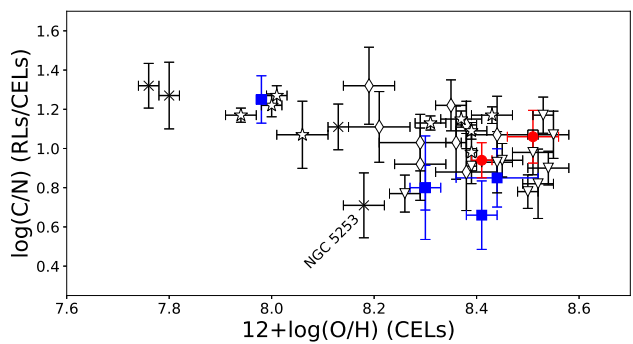} 
 \caption{Log(C/N) {\it versus} 12+log(O/H) determined from CELs for {\hii} regions in M31, M101 and other spiral and dwarf galaxies. Symbols are the same as in Fig.~\ref{fig:CO_O}. C abundances are determined from RLs and N abundances from CELs.} 
 \label{fig:CN_O} 
 \end{figure} 

\begin{figure} 
\centering 
\includegraphics[scale=0.37]{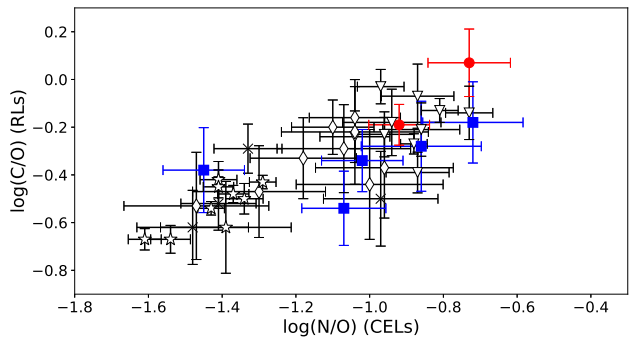} 
 \caption{Log(C/O) determined from RLs {\it versus} log(N/O) determined from CELs for {\hii} regions in M31, M101 and other spiral and dwarf galaxies. Symbols are the same as in Fig.~\ref{fig:CO_O}.} 
 \label{fig:CO_NO} 
 \end{figure} 

\subsection{About the C and N enrichment} 
\label{sec:C_N_enrichment} 

Fig.~\ref{fig:CO_O} shows the C/O {\it versus} O/H ratios for the {\hii} regions of M101 and M31 and others from the literature. The C/O ratio has been derived from C and O abundances determined from RLs and the O/H 
ratio represented in the abscissae is derived from CELs. This is a new version of a plot presented in \citet{estebanetal14}  and \citet{toribiosanciprianoetal16, toribiosanciprianoetal17}. The figure shows a well-known correlation originally found by \citet{garnettetal95, garnettetal99} that demonstrates the different enrichment time-scales of C and O.  Red circles represent our data for M31 and 
blue squares for M101, the rest of the symbols correspond to other spiral and dwarf galaxies (see the caption for a description 
 of the symbols and references). We can see that the C/O ratios of the {\hii} regions  of M31 and M101 follow the general C/O {\it versus}  O/H trend, with K160 the object with the largest C/O ratio. \citet{bergetal16, bergetal19} obtain C/O ratios from UV CELs for a 
sample of metal-poor dwarf galaxies,  covering a range of 12+log(O/H) from 7.4 to 8.0 and finding an average value of log(C/O) = $-$0.71 $\pm$ 0.17. Those authors suggest that the large scatter of C/O is due to the different star 
formation efficiencies, burst duration and supernova feedback of the dwarf galaxies located in the low-metallicity regime. In this context, the position of NGC~2363 reported by \citet{estebanetal14} (the point with the lower O/H 
ratio in Fig.~\ref{fig:CO_O}), with a C/O ratio of $-$0.29 dex, might be interpreted as a star-forming burst at its latest moments of evolution, when the C/O has increased to its largest value because the O production 
has exhausted but C is being ejected by slower-evolving stars. It is interesting to note the position of NGC~5471 in the C/O {\it versus}  O/H diagram. This is the {\hii} region 
belonging to a large spiral galaxy (M101) with the lowest O abundance for which we have a C/O ratio derived from RLs. Its C/O ratio of $-$0.39 is larger than most of the {\hii} regions of the Small Magellanic Cloud -- with similar O/H ratios -- and also larger than the aforementioned mean value of the C/O ratio of metal-poor dwarf galaxies determined by \citet{bergetal19}. However, the position of NGC~5471 in Fig.~\ref{fig:CO_O} is compatible with the results of chemical evolution models of the Milky Way -- assuming that our galaxy is representative of a prototypical large spiral galaxy --  presented by \citet{estebanetal13} for {\hii} regions located beyond the isophotal radius, that predict values of C/O of about $-$0.40 dex \citep[see][their figure 6]{estebanetal14}. 

Fig.~\ref{fig:CN_O} shows the C/N {\it versus} O/H ratios for the {\hii} regions of M101 and M31 and other galaxies from the literature. The C abundances have been derived from 
RLs and N and O abundances from CELs. The C/N ratios should only be considered as indicative because of the abundance discrepancy problem. Actually, the C/N ratios would be smaller in the case where both lines are determined using the same kind of lines, optical RLs or UV CELs \citep[e.g.][]{garciarojasesteban07, estebanetal16b}. Fig.~\ref{fig:CN_O} shows a rather large dispersion of the C/N ratio for 12+log(O/H) $>$ 8.1 but a remarkably smaller one for lower metallicity objects. From the figure, one can guess some trend of a slightly  increasing C/N at lower metallicities. This contrasts with the lack of correlation found by \citet{garnettetal99} or the opposite tendency reported by \citet{kobulnickyskillman98} using UV CELs.  \citet{bergetal16, bergetal19} also find  a flat C/N {\it versus} O/H trend for metal-poor dwarf galaxies but with a very large dispersion, log(C/N) = 0.75 $\pm$ 0.20. The log(C/N) values we find for objects with 12+log(O/H) $\geq$ 8.0 are between 
1.2 and 1.4, substantially larger than the C/N ratios obtained by \citet{bergetal16, bergetal19} from UV CELs and this difference should be an effect of the abundance discrepancy problem. In fact, \citet[][their figure 5]{toribiosanciprianoetal16} reported that the difference between the \ionic{C}{2+}/\ionic{H}{+} ratio 
determined from RLs and that calculated with UV CELs -- the so called abundance discrepancy (AD) -- increases as the O abundance decreases. The amount of such AD is consistent with the observed C/
N trend. The object at 12+log(O/H) $\sim$ 8.2 with the lowest C/N ratio in Fig.~\ref{fig:CN_O} corresponds to the zone of NGC~5253 where \citet{lopezsanchezetal07} reported  localized N and He pollution produced by massive Wolf-Rayet stars.  In Fig.~\ref{fig:CN_O}, we can also see that the C/N ratio of NGC~5471 is in complete agreement with the C/N ratios of {\hii} regions of the SMC having similar O abundance.

Fig.~\ref{fig:CO_NO} plots the C/O ratio determined from RLs for the objects included in figures~\ref{fig:CO_O} and \ref{fig:CN_O} as a function of their N/O ratio determined from CELs.  The figure  shows that the C/O and N/O abundance ratios are tightly correlated, with a slope close to 45$^\circ$. As reported by \citet{estebanetal14}, this trend is in agreement with the predictions of  the chemical evolution models for the Milky Way presented by \citet{estebanetal13}, that reproduce the position of \hii\ regions in the discs of spiral galaxies and the slope of the correlation. Figures~\ref{fig:CN_O} and \ref{fig:CO_NO} 
indicate the coupling between C and N enrichments in \hii\ regions. \citet{estebanetal14} propose that, despite the fact that the nucleosynthesis processes of both elements 
are different, their enrichment time-scales result to be rather similar due to the cumulative-temporal contribution of the C and N yields by stars of different masses. According to \citet{carigietal05}, at high metallicities, 12+log(O/H) $\geq$ 8.5, the main contributors to C and N should be massive stars. However, at intermediate metallicities,  8.1 $\leq$ 12+log(O/H) $\leq$ 8.5, the main contributors to C are low metallicity low-mass stars, but for N the main producers are massive stars and intermediate-mass stars of intermediate and low metallicity, respectively. Based on their results for low-metallicity dwarf galaxies, \citet{bergetal19} also highlight the similar behaviour shown by the C and N enrichment. They propose that the significant scatter of the C/O and N/O ratios indicate their production by stars of different average masses.  

\begin{figure} 
\centering 
\includegraphics[scale=0.37]{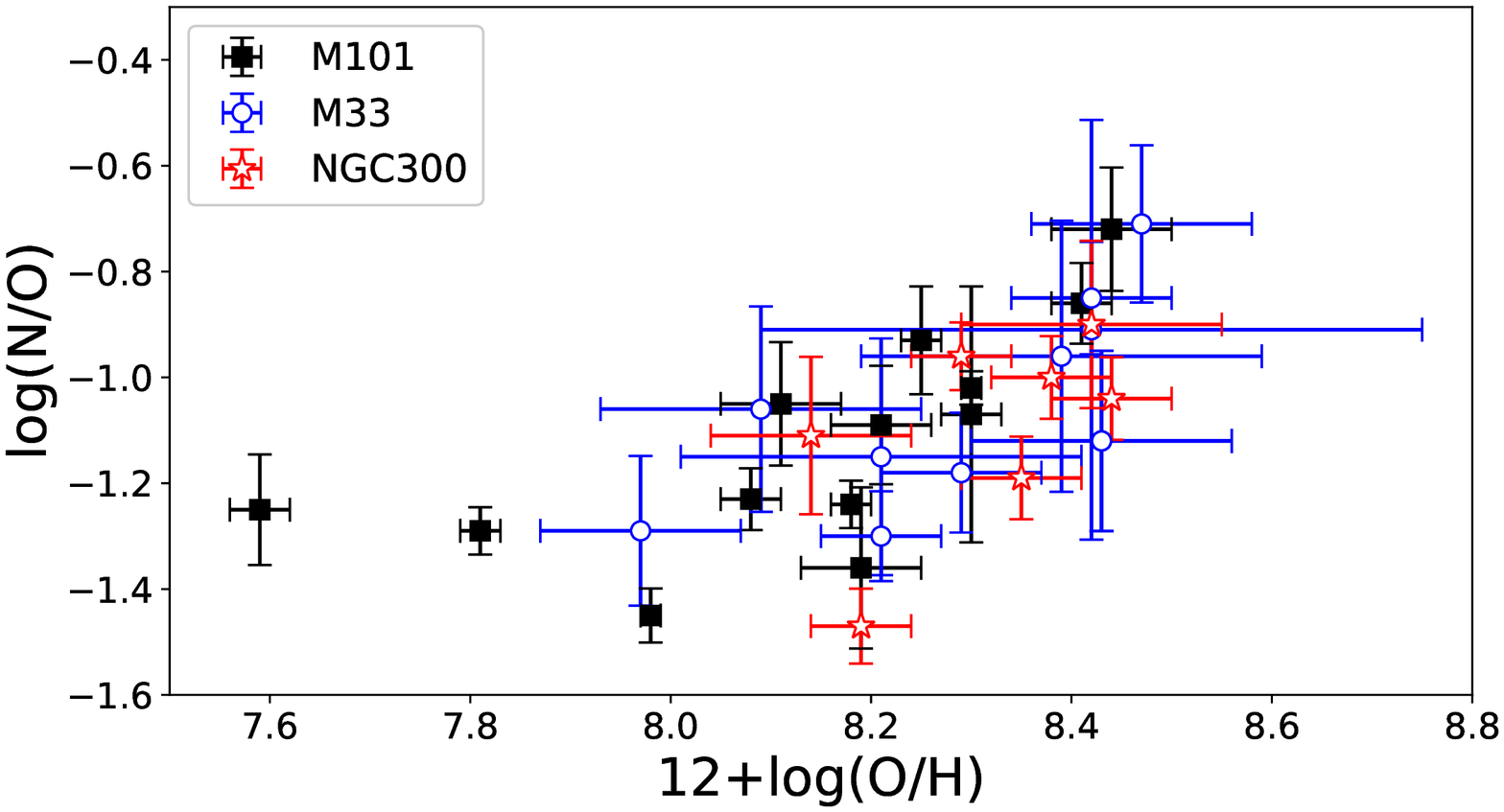} 
\includegraphics[scale=0.37]{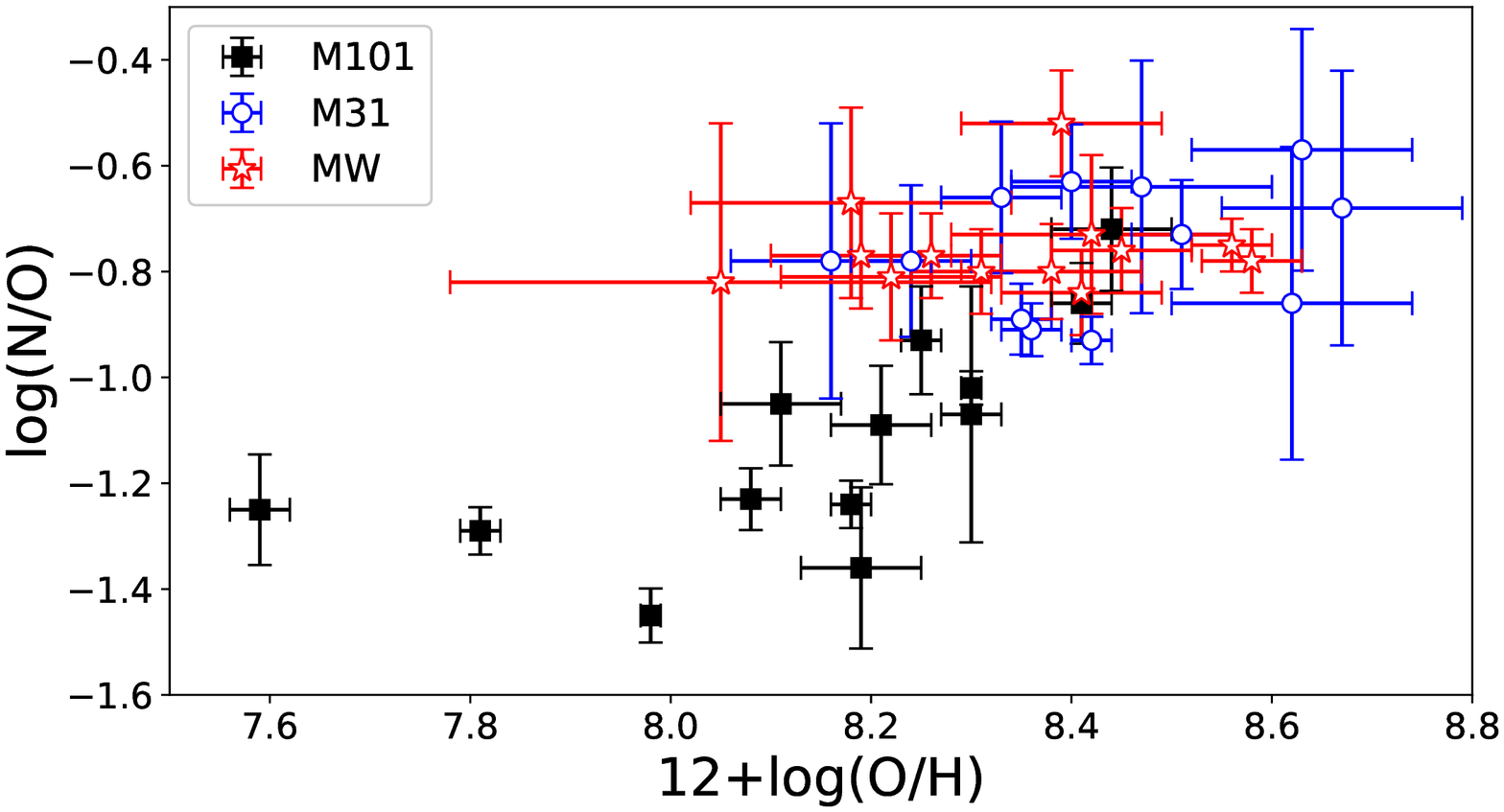} 
 \caption{Log(N/O) {\it versus} 12+log(O/H) determined from CELs for {\hii} regions in several spiral galaxies. Upper panel: data for M101 (this work) and for M33 and NGC~300 \citep{toribiosanciprianoetal16}. Lower panel: 
  data for M101 and M31 (this work) and for the Milky Way \citep{estebangarciarojas18}.} 
 \label{fig:NO_O_3gals} 
 \end{figure} 
As we can see in figures~\ref{fig:M101_NO} and \ref{fig:M31_NO} and commented in Sect.~\ref{sec:M31}, the trend of N/O as a function of O/H is very different in M101 and M31. In the case of M101, it follows the typical relation, i.e. constant N/O for lower metallicities and increasing N/O for growing O/H due to the production of secondary N for large metallicities \citep[e.g.][]{perezmonteroetal16, belfioreetal17} but in the case of M31 the N/O 
ratio is virtually flat. In the upper panel of Fig.~\ref{fig:NO_O_3gals} we compare the N/O {\it versus} O/H trend for M101 with the data for {\hii} regions of M33 and NGC~300 obtained by \citet{toribiosanciprianoetal16} from 
{\elect}-based abundances. The upper panel of Fig.~\ref{fig:NO_O_3gals} indicates that the three galaxies follow the standard N/O {\it versus} O/H trend. In fact, this is reflected by the 
presence of negative linear radial N/O gradients in them. 

A negative radial N/O gradient is a common property in spiral galaxies. There are several studies dedicated to determine and analyze N/O gradients in large samples of spirals. However, these extensive studies use strong-line methods for estimating abundances because they usually lack direct determinations of {\elect}. \citet{pilyuginetal04}  using published spectra of {\hii} regions of 54 spiral galaxies and applying the $P$-method for estimating the O abundance, found negative slopes of the N/O gradient in virtually all their sample galaxies. This is also the general result of \citet{belfioreetal17}, who use resolved spectroscopic data from the SDSS IV MaNGA survey for 550 nearby galaxies. However, \citet{perezmonteroetal16} using spectra of {\hii} regions in a sample of 350 spiral galaxies of the CALIFA survey, find that about 4-10\% of their galaxies  display a flat or even a positive radial N/O gradient. The results shown in 
Fig.~\ref{fig:M31_NO} and the lower panel of Fig.~\ref{fig:NO_O_3gals}, illustrate the fact that, contrary to what happens in most spiral galaxies, M31 and the Milky Way show a rather constant N/O ratio across their discs. \citet{perezmonteroetal16}  indicate that there are no differences between the average integrated properties of their whole sample and the subset of galaxies with flat or inverted N/O gradient, finding no special feature that could explain this odd behavior. The relation between the N/O ratio and metallicity is a subject of intense debate \citep[e.g.][]{mccalletal85, thuanetal95, henryetal00, izotovetal06, vincenzoetal16, vincenzokobayashi18}.  On the other hand, nucleosynthesis models produce very different stellar yields of N depending on the assumptions made in the stellar evolution models \citep[e.g.][]{woosleyweaver95,  marigoetal96, marigoetal98, portinarietal98,  hirschietal05, kobayashietal11, chieffilimongi13}. Therefore, the presence of a flat N/O versus O/H relation in a galactic 
disc has a difficult and non-unique explanation. In their paper about the N and O abundance gradients in the Milky Way, \citet{estebangarciarojas18} pointed out that the constant N/O ratio would indicate that the bulk of N should have a primary origin, and this would be also the situation in M31. Hydrodynamical cosmological simulations of galaxies by \citet{vincenzokobayashi18} using the stellar yield set by \citet{kobayashietal11} find almost flat trends in N/O {\it versus} O/H diagrams when ignoring the contribution of failed supernovae. Finally, the chemical evolution models by \citet{vincenzoetal16} highlight the importance of the star-formation efficiency (SFE) on fixing the value of the N/O ratio for a given O abundance. According to these models, a lower SFE results in a lower production of oxygen per unit time by massive stars decreasing the metallicity of the corresponding regions. If there is a delay in the production of nitrogen, the N/O ratio will increase. Therefore, radial changes of SFE may be another possibility to modulate the behaviour of the N/O {\it versus} O/H diagram for a given galaxy.

\section{Conclusions} 
\label{sec:conclusions} 

We present deep spectra of 18 {\hii} regions in the nearby massive spiral galaxies M101 and M31. The data have been obtained with the OSIRIS spectrograph attached to the 10.4~m {\it Gran Telescopio Canarias} telescope. We have obtained direct determinations of the electron temperature in all the nebulae. We detect the pure recombination line of {\cii} 4267 \AA\ in one {\hii} region of M31 and in five of M101, including a good quality measurement of this line for NGC5471, a giant {\hii} region in the outskirts of M101. We derive the C/H ratio for those {\hii} regions and determine the radial abundance gradient of this element in both galaxies. We also determine the radial abundance gradients of O as well as those of the N/O, Ne/O, S/O, Cl/O and Ar/O ratios. As in other spiral galaxies, the C/H gradients are steeper than those of O/H giving rise to a negative slope of the C/O gradient, a fact that reflects the non-primary behavior of C enrichment. 

The dispersion of the O abundance of the {\hii} regions of M101 with respect to the computed radial gradient is $\pm$0.04 dex, of the order of the uncertainties of the individual O/H ratio determinations, indicating the absence of significant chemical inhomogeneities across the disc of M101. Conversely, for M31 we find a higher internal dispersion, of about $\pm$0.10 dex. Although, in principle, we cannot discard the presence of some chemical inhomogeneities, we have to consider that (a) 
the individual O/H ratio determinations of the {\hii} regions of M31 have an average uncertainty of 0.07 dex, considerably larger than for M101, and (b) the high inclination angle of M31 makes the determination of galactocentric distances for the {\hii} regions of this galaxy more uncertain. 

Contrary to what is expected, we find trends in the S/O, Cl/O and Ar/O ratios as a function of O/H in M101 when using the standard methods for deriving S, Cl and Ar abundances. 
These trends are reduced when applying the prescription by \citet{dominguezguzmanetal19}, using {\elect}({\fnii}) instead of {\elect}(high) for determining \ionic{S}{2+}/\ionic{H}{+} and the mean of  {\elect}({\fnii}) and  {\elect}({\foiii}) instead of {\elect}(high) for \ionic{Cl}{2+} and  \ionic{Ar}{2+}. The final average values of the Ne/O, S/O, Cl/O and Ar/O ratios are almost identical for M101 and M31. 

Contrary to what is found for M101 and most other spiral galaxies, the distribution of the N/O ratio with respect to O/H is rather flat in M31. It should be noted that this is also the behavior found in the
Milky Way by several authors suggesting that the bulk of N should have a primary origin in both galaxies, although other explanations are possible.

The use of  $R_{\mathrm e}$ as normalization parameter of the radial distance provides lower relative standard deviations when comparing the gradient slopes of the different galaxies. The comparison of the radial gradients with respect to $R/R_\mathrm{e}$ of M101, M31, M33, NGC~300 and the Milky Way suggests 
an apparent metallicity excess in the Milky Way. This odd behavior might be solved if the true $R_{\mathrm e}$ is about two times larger than the most recent determinations. 

Using different diagrams involving C, N and O abundances, we explore the coupling between C and N enrichments in \hii\ regions. We find that, despite the fact that the nucleosynthesis processes of C and N 
are different, their enrichment time-scales result to be rather similar due to the cumulative-temporal contribution of the C and N yields by stars of different masses. 

\section*{Acknowledgements} We are grateful to Almudena Zurita, Estrella Florido  and Leticia Carigi for their useful comments. This paper is based on observations made with the Gran Telescopio Canarias (GTC), installed in the Spanish Observatorio del Roque de los Muchachos of the Instituto de Astrof\'isica de Canarias, in the island of La Palma, Spain. We acknowledge support from the State Research Agency (AEI) of the Spanish Ministry of Science, Innovation and Universities (MCIU) and the European Regional Development Fund (FEDER) under grant with reference AYA2015-65205-P. CEL acknowledges the joint sponsorship by the Fulbright Program and the Spanish Ministry of Education during a stay in the Institute of Astronomy of the University of Hawaii at Manoa, where part of this work was carried out. JGR acknowledges support from an Advanced Fellowship from the Severo Ochoa excellence program (SEV-2015-0548). 




\bibliographystyle{mnras} 
\bibliography{cesar_bibliography} 


\section*{Supporting information}

Supplementary data are available at MNRAS online.
\hfill \break \\
\noindent \textbf{Appendix A:} Line intensity ratios of the {\hii} regions}
\hfill \break \\
Please note: Oxford University Press is not responsible for the content or functionality of any supporting materials supplied by the authors. Any queries (other than missing material) should be directed to the corresponding author for the article.


\bsp	
\label{lastpage} 
\end{document}